\documentclass[printer]{aa}

\usepackage{txfonts,textcomp}
\usepackage{amsfonts}
\usepackage{placeins}
\usepackage{tablefootnote}
\usepackage{graphicx}
\usepackage{lscape}
\usepackage{multirow}
\usepackage{dirtytalk}
\usepackage[nopostdot,acronym,xindy,indexonlyfirst=true,stylemods=longbooktabs,automake]{glossaries-extra}
\makeglossaries
\newacronym{ac}{AC}{across scan}
\newacronym{aen}{AEN}{astrometric excess noise}
\newacronym{al}{AL}{along scan}
\newacronym{bd}{BD}{brown dwarf}
\newacronym{cu8}{CU8}{Coordination unit 8}
\newacronym{dec}{DEC}{declination}
\newacronym{dof}{DOF}{degree of freedom}
\newacronym{dpac}{DPAC}{Data processing and analysis consortium}
\newacronym{dr}{DR}{data release}
\newacronym{flame}{FLAME}{final luminosity age and mass estimate}
\newacronym{fov}{FoV}{field of view}
\newacronym{fp}{FP}{false-positives}
\newacronym{gaston}{GASTON}{\textit{Gaia} astrometric noise simulation to derive orbital inclination}
\newacronym{gost}{GOST}{\textit{Gaia} observation forecast tool}
\newacronym{g3}{GDR3}{third \textit{Gaia} data release}
\newacronym{hg}{HG}{\textsc{Hipparcos}--\textit{Gaia}}
\newacronym{hg3}{HG}{\textsc{Hipparcos} -- \textit{Gaia} DR3}
\newacronym{iads}{IADs}{intermediate astrometric data}
\newacronym{icrs}{ICRS}{International celestial reference system}
\newacronym{ipd}{IPD}{Image parameter determination}
\newacronym{lr}{LR}{likelihood ratio}
\newacronym{lsf}{LSF}{line spread function}
\newacronym{mad}{MAD}{median absolute deviation}
\newacronym{mcmc}{MCMC}{Markov-chains Monte-Carlo}
\newacronym{ms}{MS}{main-sequence}
\newacronym{mse}{UEVA}{unbiased estimator of variance a posteriori}
\newacronym{pa}{PA}{position angle}
\newacronym{pdf}{PDF}{probability density function}
\newacronym{pma}{PMa}{proper motion anomaly}
\newacronym{pmex}{GaiaPMEX}{\textit{Gaia} DR3 proper motion anomaly and astrometric noise excess}
\newacronym{psf}{PSF}{point spread function}
\newacronym{ra}{RA}{right ascension}
\newacronym{rse}{RSE}{regression standard error}
\newacronym{rms}{RMS}{root mean square}
\newacronym{rss}{RSS}{residuals sum of square}
\newacronym{ruwe}{\texttt{ruwe}}{renormalized unit weight error}
\newacronym{rv}{RV}{radial velocity}
\newacronym{sma}{sma}{semi-major axis}
\newacronym{snr}{S/N}{signal-to-noise ratio}
\newacronym{uwe}{UWE}{unit weight error}
\newacronym{wc}{WC}{window class}
\newacronym{wfs}{WFS}{wavefront sensor}
\newacronym{5p}{5p}{five parameters}
\newacronym{6p}{6p}{six parameters}

\usepackage[breaklinks]{hyperref}

\makeatletter
\renewcommand*\aa@pageof{, page \thepage{} of \pageref*{LastPage}}
\makeatother

\begin{document}

\title{Searching for substellar companion candidates with \textit{Gaia}}
\subtitle{I. Introducing the GaiaPMEX tool}
\author{F. Kiefer\inst{1} \and A.-M. Lagrange\inst{1}  \and P. Rubini\inst{2} \and F. Philipot\inst{1}}

\institute{
\label{inst:1}LESIA, Observatoire de Paris, Universit\'e PSL, CNRS, Sorbonne Universit\'e, Universit\'e de Paris, 5 place Jules Janssen, 92195 Meudon, France\thanks{Please send any request to flavien.kiefer@obspm.fr} \and
\label{inst:2}Pixyl, 5 av du Grand Sablon 38700 La Tronche
}
        
   \date{Received 31/07/2024 ; accepted 03/09/2024}

  
  \abstract 
{The \textit{Gaia} mission is expected to yield the detection of several thousands of exoplanets, perhaps at least doubling the number of known exoplanets. However, only 72 candidates have been reported with the publication of the \textit{Gaia} third data release, or \glsxtrlong{g3} (\glsxtrshort{g3}). Although a greater harvest of exoplanets is expected to occur with the publication of the astrometric time series in the DR4 at the eve of 2026, the \glsxtrshort{g3} is already a precious database that can be used to search for exoplanets beyond 1\,au.
}
{With this objective, we characterized multiple systems by exploiting two astrometric signatures derived from the \glsxtrshort{g3} astrometric solution of bright sources with $G<16$. We have the \glsxtrlong{pma}, or \glsxtrshort{pma}, for sources also observed with \textsc{Hipparcos} and the excess of residuals present in the \glsxtrlong{ruwe} (\glsxtrshort{ruwe}) and the \glsxtrlong{aen} (\glsxtrshort{aen}). These astrometric signatures give an accurate measurement of the astrometric motion of a source seen with \textit{Gaia}, even in the presence of non-negligible calibration and measurement noises.}
{We introduce a tool called \glsxtrlong{pmex}, or \glsxtrshort{pmex} for short, that is able for a given source to model the astrometric signatures that are hidden within the \glsxtrshort{pma}, \glsxtrshort{ruwe}, and \glsxtrshort{aen} by a photocenter orbit due to a companion with a certain mass and relative \glsxtrlong{sma} to the primary star (\glsxtrshort{sma}). \glsxtrshort{pmex} calculates a confidence map of the possible companion's mass and \glsxtrshort{sma}, given the actual measurements from \glsxtrshort{g3}, and \textsc{Hipparcos}, when available. This tool allowed us to determine for any source of interest if it may be a binary (or planetary) system and the possible companion's mass and \glsxtrshort{sma}. }
{We find that the astrometric signatures can allow for identification of stellar binaries and hint toward companions with a mass in the planetary domain. The constraints on mass are, as expected, degenerate, but when allowed, coupling the use of \glsxtrshort{pma} and \glsxtrshort{ruwe} or \glsxtrshort{aen}, they may significantly narrow the space of solutions.}
{Thanks to combining \textit{Gaia} and \textsc{Hipparcos}, planets are expected to be most frequently found within 1--10\,au from their star, at the scale of Earth-to-Saturn orbits. In this range of \glsxtrshort{sma}, exoplanets with a mass down to 0.1\,M$_{\rm J}$ are more favorably detected around M-dwarfs closer than 10\,pc to Earth. Some fraction, if not all, of companions identified with \glsxtrshort{pmex} may be characterized in the future using the astrometric time series that will be published in the forthcoming DR4.}


   \maketitle
%

\section{Introduction}

Finding and characterizing exoplanets has become one of the most active areas in astronomy. So far, most exoplanets have been found by the transit and the \glsxtrlong{rv} (\glsxtrshort{rv}\footnote{All acronyms used are summarized and indexed in Appendix~\ref{sec:acronyms}.}) techniques, as seen in the few publicly available exoplanet catalogs. Notably, \textit{Gaia} absolute astrometry is expected to identify (tens of) thousands of new exoplanets and brown dwarfs (\glsxtrshort{bd}) in the near future~\citep{Perryman2014,Sahlmann2015,Holl2022,Arenou2023,Holl2023}. 

The current number of exoplanet candidates identified with \glsxtrlong{g3} (\glsxtrshort{g3}) astrometry\,\citep[72;][]{Arenou2023} is still much below expectations. Therefore, a major challenge is to exploit the \textit{Gaia} data currently made public in the online catalogs in its most recent \glsxtrlong{dr} (\glsxtrshort{dr}3; ~\citealt{Gaia2021}) to detect unknown exoplanet candidates, as nicely illustrated with the discovery of AF\,Lep\,b~\citep{Mesa2023,Franson2023,DeRosa2023}. Incidentally, \textit{Gaia}'s astrometry can also help validate (or reject) candidate exoplanets detected by other means (\glsxtrshort{rv}, transit, imaging) and further characterize them~\citep{Kiefer2019a,Kiefer2019b,Kervella2019,Brandt2019,Kiefer2021,Dalal2021,Brandt2021,Feng2021,Kervella2022,Feng2022,Xiao2023,Philipot2023a,Philipot2023b} or aid in assessing the existence of a companion (possibly supplementary) of a given star or set of stars of interest.  

With this objective, we set up a tool called \glsxtrshort{pmex} for \glsxtrlong{pmex} based on the original works of~\citet{Kiefer2019a,Kervella2019,Kiefer2019b,Kiefer2021,Kervella2022} that allows for determination of the mass of possible candidate companions and their relative \glsxtrlong{sma} in relation to their primary star (abbreviated to \glsxtrshort{sma} hereafter) from consideration of, individually or in combination, the constraints from the \glsxtrlong{pma} (hereafter \glsxtrshort{pma}; \citealt{Kervella2019,Brandt2021,Kervella2022}), the astrometric excess noise (\glsxtrshort{aen}; see \citealt{Kiefer2019a,Kiefer2019b,Kiefer2021}), and the \glsxtrlong{ruwe} (\glsxtrshort{ruwe}; see~\citealt{Lindegren2018,Lindegren2021}). This tool models, within a Bayesian framework, the observed \glsxtrshort{aen}, \glsxtrshort{ruwe}, and \glsxtrshort{pma} through simulated outcomes of \textit{Gaia}'s observations of a source if it had a companion of a given mass and \glsxtrshort{sma}. It leads to a 2D confidence map of the companion mass and \glsxtrshort{sma}. Introducing this tool is the purpose of the present paper; a series of further papers will report the results of its application on other systems.

In Sect.~\ref{sec:observables}, we recall the definitions of \glsxtrshort{aen}, \glsxtrshort{ruwe}, and \glsxtrshort{pma}. In Sect.~\ref{sec:noise_proxy}, we describe our reverse-engineering method to determine the noise levels of \textit{Gaia}'s observations of individual sources. In Sect.~\ref{sec:simu}, we explain the modeling of any star's orbital motion due to a companion and the simulations of \textit{Gaia} astrometric measurements of that star. In Sect.~\ref{sec:astro_sig} we define the \glsxtrshort{pma}, \glsxtrshort{ruwe}, and \glsxtrshort{aen} astrometric signatures. In Sect.~\ref{sec:PMEX}, we present the \glsxtrshort{pmex} tool in detail. In Sect.~\ref{sec:examples}, we show illustrative examples of the application of \glsxtrshort{pmex} on a few chosen sources. Finally, in Sect.~\ref{sec:discussion}, we discuss the perspectives opened by the application of this tool regarding the detection of exoplanets and brown dwarfs using \textit{Gaia}.

\section{Astrometric excess noise, RUWE, and proper motion anomaly}
\label{sec:observables}
\subsection{The astrometric excess noise}\label{sec:AEN}
The \glsxtrshort{aen} of a source, as introduced in~\citet{Gaia2016}, is the excess of scatter in the residuals of \glsxtrlong{al} angle measurements compared to the astrometric displacement of the source modeled as a single-star, that includes position, linear proper motion and parallactic motion. At each epoch of transit of a source along one of the detectors, there is a specific scan direction, the \glsxtrlong{al} direction (\glsxtrshort{al}), along which the source image is moving during the rotation of the spacecraft. The position of the source on the detector can be determined in 2D, since there is also an \glsxtrlong{ac} (or \glsxtrshort{ac}) direction, but it is much less precisely measured along the \glsxtrshort{ac} than along the \glsxtrshort{al} direction. Therefore, in all \textit{Gaia} data releases, only the \glsxtrshort{al} angles are used as astrometric measurements to determine the main astrometric data of a source~\citep{Lindegren2016,Lindegren2018,Lindegren2021}.

In the \glsxtrshort{g3}, as in previous releases, the process of fitting the astrometric data is iterative. At each iteration, individual errors, $\sigma_{\rm AL}$, of \glsxtrshort{al} angle measurements performed during a transit of a star on the detector are estimated or updated and then used to calculate a $\chi^2$. Since \glsxtrshort{dr}2~\citep{Lindegren2018}, a spacecraft attitude excess noise $\sigma_{\rm att}$ is quadratically added to $\sigma_{\rm AL}$ in the calculation of the $\chi^2$. Its amplitude is typically about 0.076\,mas, while individual measurement errors are within 0.05-0.15\,mas~\citep{Lindegren2021}. Both form a \say{formal error} $\sigma_{\rm formal}$=$\sqrt{\sigma_{\rm att}^2 + \sigma_{\rm AL}^2}$. We give more details and estimation of their variations with respect to the magnitude, color, \glsxtrlong{ra} (\glsxtrshort{ra}) and \glsxtrlong{dec} (\glsxtrshort{dec}) of targets in Sect.~\ref{sec:formal_error}. Their time series will only be known upon the publication of the DR4. The monitoring of the residuals root-mean-square shows that the measurement and excess attitude errors are constant most of the time, with rare deviations (see Fig. A.2 in~\citealt{Lindegren2021}). We thus assume in the following that the attitude excess noise of a time series for any given target remains relatively constant in time. With this assumption, the $\chi^2$, as it appears in the archives (namely \verb+astrometric_chi2_al+), written here $\chi^2_{\rm astro}$ is
\begin{equation}\label{eq:chi2_archive}
\chi^2_{\rm astro}  = \frac{\sum_{\ell=1,N} R^2_\ell}{\sigma_{\rm att}^2 + \sigma_{\rm AL}^2}, 
\end{equation} 
where $R_\ell$ are the residuals of the $N$ astrometric measurements (\verb+astrometric_n_good_obs_AL+) after subtraction of the fitted model. 
If some additional calibration noise -- that is, a non-subtracted residual instrumental jitter beyond the attitude excess noise -- or real astrometric signal were to be present, it would not be accounted for in the formal errors used to calculate the $\chi^2$ and the reduced $\chi^2$ would be larger than 1. Deviations of the reduced $\chi^2$ beyond $1$ are accounted for in the \glsxtrshort{aen} (\verb+astrometric_excess_noise+). To calculate the final uncertainties of fitted parameters of a given target, the \glsxtrshort{aen} is quadratically added to the formal error of any astrometric measurements such as to impose a reduced $\chi^2$ of 1. Still assuming that the errors are uniform along the time series, the \glsxtrshort{aen} is related to the \glsxtrlong{rss} through
\begin{equation}\label{eq:AEN}
\text{AEN}^2 +   \sigma_{\rm att}^2 + \sigma_{\rm AL}^2 = \frac{\sum_{i=1,N} R^2_i}{N-5},
\end{equation} 
counting $N-5$ \glsxtrlong{dof} (\glsxtrshort{dof}), with five parameters fit to the astrometry. The exact definition of the \glsxtrshort{aen} involves possibly non-uniform errors and it is fixed iteratively during the reduction. Its value might thus slightly deviate from this definition. 
The level of the additional calibration noise still present in the data, not accounted for in the formal error of Eq.~\ref{eq:chi2_archive} but contributing to the \glsxtrshort{aen} in Eq.~\ref{eq:AEN}, strongly depends on the magnitude and the color of the observed targets~\citep{Lindegren2016,Lindegren2018,Lindegren2021}. We invented a method to estimate it for any source from the whole \textit{Gaia} catalog of bright sources with magnitude $G<16$, as thoroughly explained in Sect.~\ref{sec:jitter}.

The identification of many zero-valued \glsxtrshort{aen} for sources dimmer than $G$=13 led us to become aware of an issue with the estimation of the calibration noise in the \glsxtrshort{g3}'s reduction. When the $\chi^2$ was smaller or equal to the 95th--percentile of the $\chi^2$ distribution with $N_{\rm \glsxtrshort{dof}}$ \glsxtrlong{dof}, that is, when the reduced $\chi^2$ was smaller than $1+\frac{N_{\rm \glsxtrshort{dof}}}{\sqrt{2\,N_{\rm \glsxtrshort{dof}}}}$, the \glsxtrshort{aen} was almost always fixed to zero in the archives~\citep{Lindegren2012}. For sources dimmer than $G$=13, the attitude excess noise, common to all sources observed at the same epoch on the detector, overestimates the calibration noise and thus the format error to compute the $\chi^2$ (Lindegren, priv. comm.). This led to an \glsxtrshort{aen} wrongly fixed to zero for many sources beyond $G$=13, thus erasing any information on supplementary signals. Below $G$=13 this problem did not arise, because the calibration noise was conversely underestimated by the attitude excess noise, leading always to strictly positive values of the \glsxtrshort{aen}. Our present understanding is that the \glsxtrshort{aen} can be used as a binarity indicator and even used to characterise orbital motion, as long as the calibration noise and the attitude excess noise are both well known, and that the zero-valued \glsxtrshort{aen} are discarded. The \glsxtrlong{ruwe}, discussed in the next section, being directly proportional to the reduced $\chi^2$ will be less problematic in this regard because it is not cut off below some value.

\subsection{The renormalized unit weight error}
\label{sec:RUWE}
An alternative to overcome the above issue is to use the renormalized unit weight error, or \glsxtrshort{ruwe}, instead of the \glsxtrshort{aen}. By definition~\citep{Lindegren2018},
\begin{align}\label{eq:ruwe_chi2}
  {\rm \glsxtrshort{ruwe}} = \frac{1}{u_0} \times \sqrt{\frac{\chi^2_{\rm astro}}{N-5}},
\end{align}
where $u_0$ is a factor that depends on magnitude and color. It can be determined from the \glsxtrshort{g3} database values of $\chi_{\rm astro}^2$ or \verb+astrometric_chi2_al+), \verb+ruwe+ and number points $N$ or \verb+astrometric_n_good_obs_al+. 
With the approximate Eqs.~\ref{eq:chi2_archive} and~\ref{eq:AEN}, the \glsxtrshort{ruwe} and the \glsxtrshort{aen} are directly associated:
\begin{align}\label{eq:ruwe}
  {\rm \glsxtrshort{ruwe}} \approx \frac{1}{u_0} \times \sqrt{\frac{\text{AEN}^2 +   \sigma_{\rm att}^2 + \sigma_{\rm AL}^2}{ \sigma_{\rm att}^2 + \sigma_{\rm AL}^2}}.
\end{align}

\noindent
The \glsxtrshort{ruwe} is a unit-less scalar, but by the use of this formula, it could be conveniently transformed to an \glsxtrshort{aen}. With a unit of angle -- expressed in milli-arcsecond (mas) in the catalog -- the \glsxtrshort{aen} is directly commensurate to any possible astrometric motion -- in au if divided by the parallax. 
A large value ($>$1.4) for a source is often accepted as indicating binarity. In many cases, this is indeed true, but it is nevertheless a misinterpretation of the DR3's documentation, rather cautiously indicating that well-behaved sources (single or not), that is, for which the five-parameter fit gives a reasonably good fit, should have \glsxtrshort{ruwe}<1.4. We noticed, indeed, that the deviation of the \glsxtrshort{ruwe} above 1 in the \glsxtrshort{g3} catalog is sometimes unreliable as a binarity indicator. This is most frequent for sources whose \textit{Gaia} data were fit using six parameters (\texttt{astrometric\_params\_solved}=95). The case of the star $\beta$\,Pictoris is an excellent counter-example, with a \glsxtrshort{ruwe} of 3.07, that, we show in Sect.~\ref{sec:betapic}, can be explained by noise only, for this very bright star.

\subsection{The proper motion anomaly}\label{sec:PMa}

The \glsxtrshort{pma}, as initially introduced in~\citet{Kervella2019}, is the proper motion offset between the \textsc{Hipparcos}-\textit{Gaia} average proper motion (with a baseline $\sim$24.5\,years), and the \glsxtrshort{g3} fitted linear proper motion (with a baseline of $36$ months). It thus measures an acceleration of the primary star due to the presence of a long-period secondary companion. 
The most recent measurements of \glsxtrshort{pma} can be found in~\citet{Kervella2022} as well as in~\citet{Brandt2021} with a different treatment of the global reference frames matching between \glsxtrshort{g3} and the \textsc{Hipparcos} \glsxtrlong{icrs} (\glsxtrshort{icrs} for short). In brief, noting $\mu$ the 2D proper motion, with index \glsxtrshort{hg} for \say{\glsxtrlong{hg}}, 
\begin{equation}
    \mu_{\rm \glsxtrshort{pma}} = \mu_{\rm \glsxtrshort{g3}} - \mu_{\rm \glsxtrshort{hg}}.
\end{equation}

\noindent
The non-linear perspective acceleration is assumed to be corrected in $\mu_{\rm \glsxtrshort{hg}}$. In this sense, $\mu_{\rm \glsxtrshort{hg}}$ is the average 3-D \textsc{Hipparcos}--\textit{Gaia} linear proper motion projected on the tangent plane at \glsxtrshort{g3} epoch. Moreover, the effect of perspective acceleration is taken into account in the \glsxtrshort{g3} astrometric solution, and $\mu_{\rm \glsxtrshort{g3}}$ is thus already the proper motion of the star in the cartesian tangent plane. With these definitions, we can thus consider that $\mu_{\rm \glsxtrshort{pma}}$ is the projected tangential PMa as measured from a reference frame co-moving with the system's barycenter. 

\begin{figure}
    \centering
    \includegraphics[width=89.3mm]{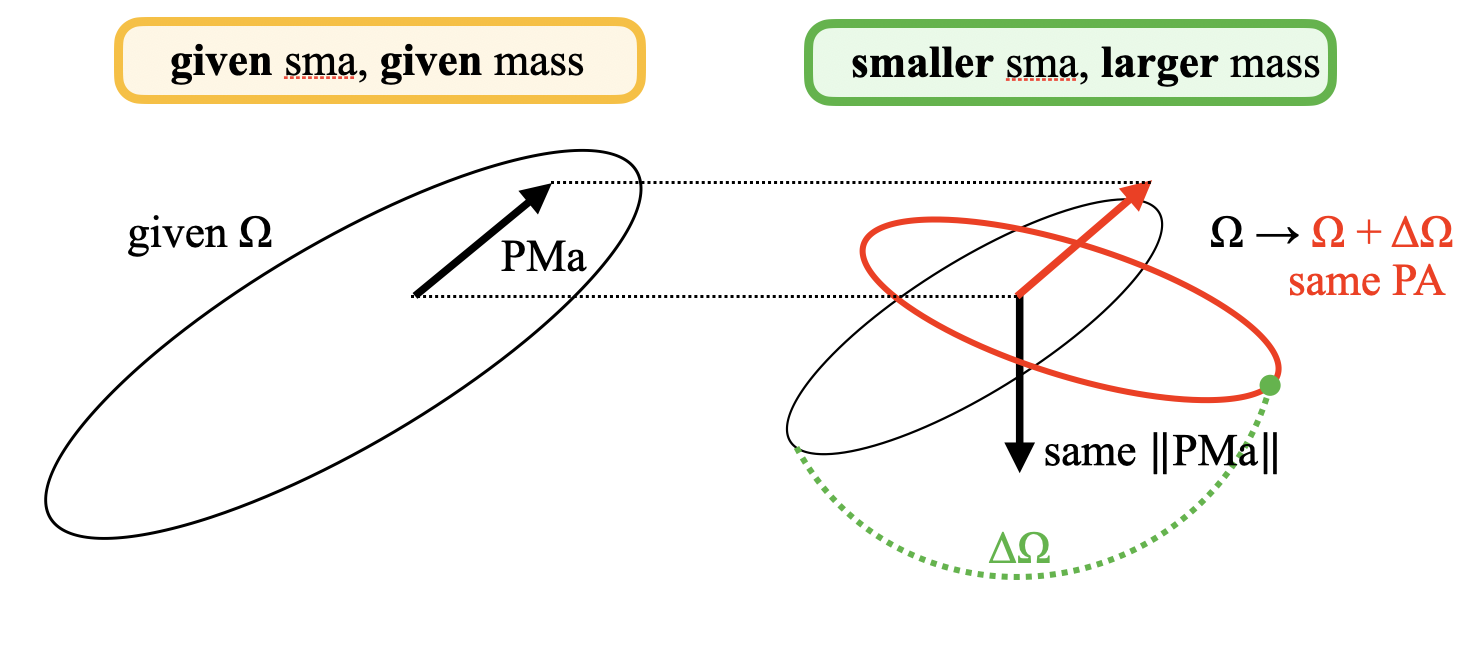}
    \caption{Illustration of the equality of \glsxtrshort{pma} modulo $\Omega$ between two systems with the same central star and a companion on a long-period orbit but with different values of sma and mass. For a given sma and a given mass of the companion (left panel) the PMa is directed toward the companion. There exists a smaller sma and a larger mass for which the $\Vert{\rm \glsxtrshort{pma}}\Vert$ is the same (right panel) but the orientation at equal $\Omega$ is different. Nevertheless, it is possible to align the \glsxtrshort{pma} on the same \glsxtrlong{pa} (\glsxtrshort{pa}) by rotating the system by some $\Delta\Omega$.}
    \label{fig:PMa_illustrated}
\end{figure}

As illustrated in Fig.~\ref{fig:PMa_illustrated}, for any mass and \glsxtrshort{sma}, there always exists a longitude of ascending node $\Omega$ that fits any \glsxtrshort{pma} position angle, the mass and \glsxtrshort{sma} can only be constrained from $\left\Vert \text{\glsxtrshort{pma}}\right\Vert$. When referring to \glsxtrshort{pma} in the rest of the text, we thus always refer to $\left\Vert \text{\glsxtrshort{pma}}\right\Vert$, that is:
\begin{equation}
    {\rm \glsxtrshort{pma}} = \left\Vert\mu_{\rm \glsxtrshort{g3}} - \mu_{\rm \glsxtrshort{hg}}\right\Vert
\end{equation}.

\noindent
Even though the \glsxtrshort{pma} measures a variability in the proper motion of a star, noise in the astrometric measurement may induce a non-zero \glsxtrshort{pma}. We assess the significance of the \glsxtrshort{pma} in Sect.~\ref{sec:significance_PMa}. It turns out that this is different than comparing the value of the \glsxtrshort{pma} to its error bar that is calculated from the published measurement errors from \textit{Gaia} and \textsc{Hipparcos}. 

\section{Noises and errors in \textit{Gaia} observations}
\label{sec:noise_proxy}

One the issues with interpreting correctly the \glsxtrshort{aen}, \glsxtrshort{ruwe} and \glsxtrshort{pma} as indicators of binarity and even measurements of companion's properties, is our  ignorance, a priori, of the noise budget in those quantities. Indeed, measurement noise and instrumental calibration noise participate at a certain degree in the excess of residuals beyond "formal error" (see Sect.~\ref{sec:formal_error} for a definition), as well as in any excess of proper motion fitted to noisy astrometric data. To complicate the task further, the level of those noises and error in \textit{Gaia} data for any given source is not published and thus unknown to the community. As a prerequisite to the functioning of \glsxtrshort{pmex}, whose goal is to model the astrometric motion beyond noise in \glsxtrshort{aen}, \glsxtrshort{ruwe} and \glsxtrshort{pma}, we thus present, in the following sections, a method that we developed to determine the noises and error levels in \textit{Gaia} data for any source with $G$$<$16. 

\subsection{The formal error}
\label{sec:formal_error}

What we call the formal error, $\sigma_{\rm formal}$, is the unknown error that appears in the denominator of $\chi^2_{\rm astro}$ in Eq.~\ref{eq:chi2_archive}, that is, $\sqrt{\sigma^2_{\rm att}+\sigma^2_{\rm \glsxtrshort{al}}}$. Combining this equation with Eq.~\ref{eq:AEN} led us to express a simple approximation of $\chi^2_{\rm astro}$ with respect to \glsxtrshort{aen}:
\begin{equation}\label{eq:chi2_errors}
\chi^2_{\rm astro} = (N-5) \times \frac{\text{AEN}^2 +   \sigma_{\rm att}^2 + \sigma_{\rm AL}^2}{\sigma_{\rm att}^2 + \sigma_{\rm AL}^2}.
\end{equation} 

\noindent
The formal error could thus be guessed by inverting this formula for all the sources observed in the \glsxtrshort{g3} that have an \glsxtrshort{aen} not compatible with 0\,mas, that is, with an \verb+astrometric_excess_noise_sig+\,$\ge$\,2~\citep{Lindegren2012}:
\begin{equation} \label{eq:formal_error}
\sigma_{\rm formal} = \text{AEN} \, \left(\frac{\chi^2_{\rm astro}}{N-5} - 1 \right)^{-1/2}.
\end{equation} 

\noindent
This estimate of the typical errors used in the $\chi^2_{\rm astro}$ for several million sources allowed us to study the impact of magnitude, color, \glsxtrshort{ra} and \glsxtrshort{dec} on \glsxtrshort{g3} astrometric errors, and, more specifically, as we show in Sects.~\ref{sec:attitude_error} and~\ref{sec:AL_error} of the attitude excess noise and the \glsxtrshort{al} measurement errors. 
We adopted the bins defined in Table~\ref{tab:bins_params}. To adapt to more rapid variations of the errors between magnitudes of 10.5 and 13.5, we adopted a smaller bin size $\sim$0.1 between 10.5 and 12.5. Moreover, a strong discontinuity in the errors occurs at $G$=13. It is related to the change in \glsxtrlong{wc} (or \glsxtrshort{wc}) from $G<13$ (WC0) to $G>13$ (WC1). It goes with a different level of charge transfer inefficiency (CTI) that is increasing in WC0 up to $G=13$, but strongly decreasing in WC1~\citep{Lindegren2021}. Because of this, we had to adopt an even smaller bin size of 0.05 between 12.5 and 13.5 $G$-mag.

\begin{table}[hbt]
    \centering
    \caption{Bins used for $G$ magnitude, $Bp-Rp$ color, \glsxtrshort{ra}, and \glsxtrshort{dec}.}
    \begin{tabular}{lcc}
        parameters & bounds & bin sizes \\
        \hline
        \multirow{4}{*}{$G$--mag}& $[1;10.5]$ & 0.25 \\ & $[10.5;12.5]$ & 0.1 \\ & $[12.5;13.5]$ & 0.05 \\ & $[13.5;16]$ & 0.25 \\ 
        \hline
        $Bp-Rp$ & $[-3;9]$ & 0.25 \\ 
        \hline
        \glsxtrshort{ra} ($^\circ$) & [0;360] & 6 \\
        \hline
        \glsxtrshort{dec} ($^\circ$) & [-90;90] & 3 \\
        \hline
    \end{tabular}
    \label{tab:bins_params}
\end{table}

\begin{figure*}[hbt]
    \centering
    \includegraphics[width=178.6mm,clip=true,trim=100 0 90 0]{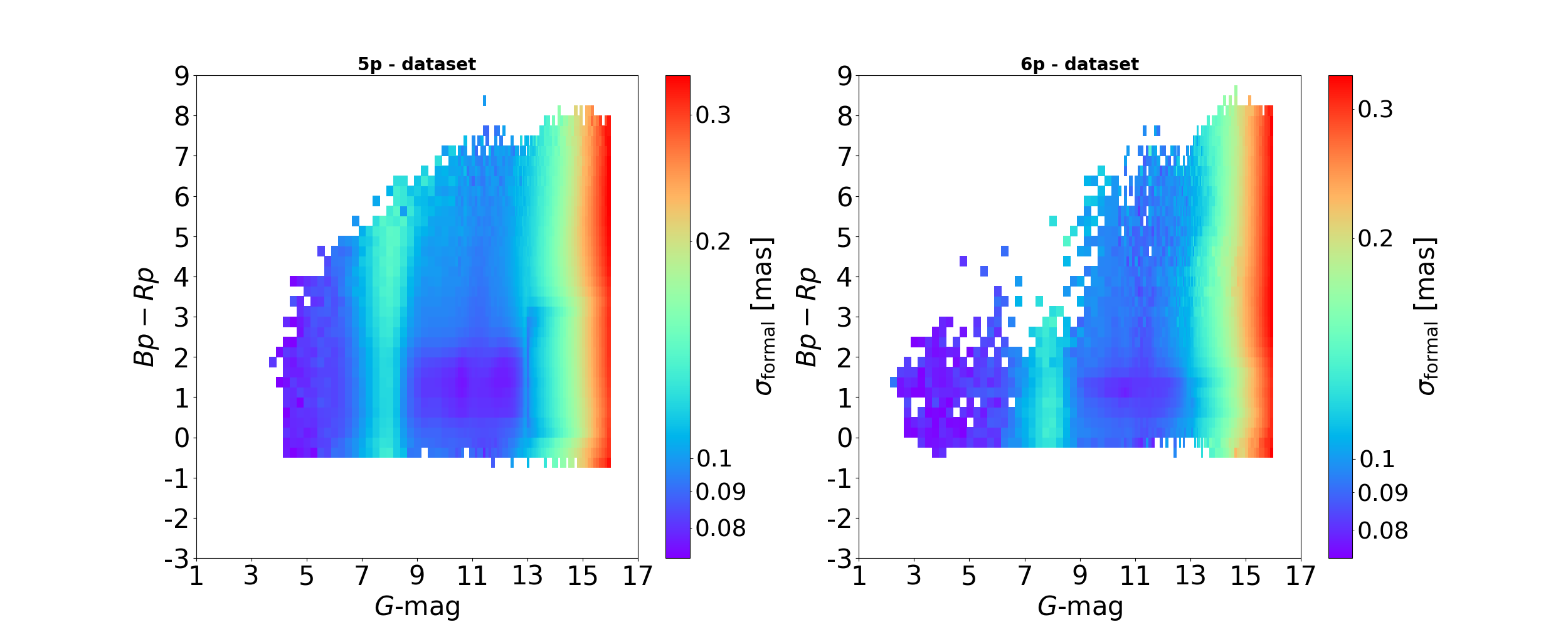}
    \includegraphics[width=178.6mm,clip=true,trim=100 0 90 0]{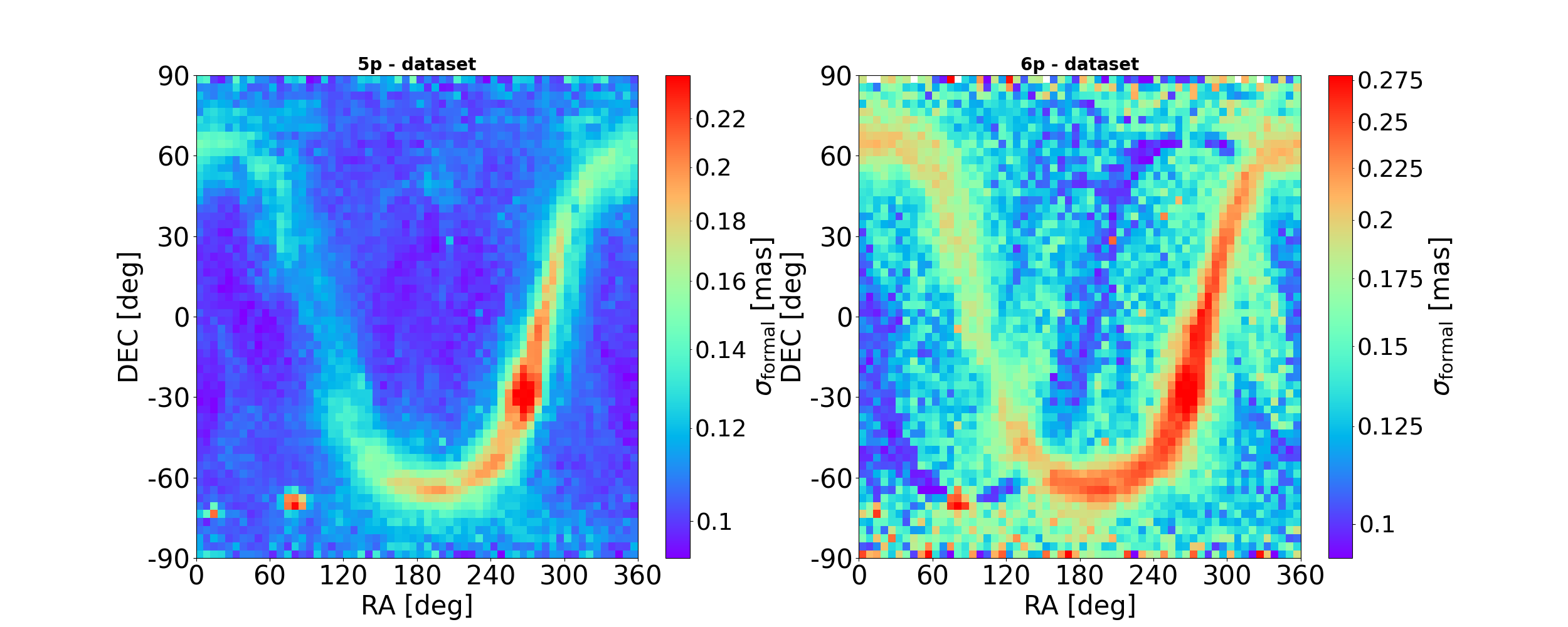}
    \caption{Median formal error distribution $\sigma_{\rm formal}$ with respect to magnitude and color (top) and \glsxtrshort{ra} and \glsxtrshort{dec} (bottom) in the \glsxtrshort{g3} database of sources brigther than $G$=16.}
    \label{fig:formal_error}
\end{figure*}

In each magnitude-color or \glsxtrshort{ra}-\glsxtrshort{dec} bin, we calculate the median formal error of all sources in these bins, respectively $\sigma_{\rm formal}(\text{mag},\text{color})$ and $\sigma_{\rm formal}(\text{\glsxtrshort{ra}},\text{\glsxtrshort{dec}})$. This gives the relationship between formal error and magnitude \& color or \glsxtrshort{ra} \& \glsxtrshort{dec}. Figure~\ref{fig:formal_error} shows the variations of the median formal error for the \textit{Gaia} sources with \glsxtrshort{aen}$>$0\,mas and brighter than $G$=16, with respect to those parameters. We consider separately the sources whose data were fit by a \glsxtrlong{5p} model (\verb+astrometric_params_solved+ = 31), hereafter called '\glsxtrshort{5p}' dataset, and those whose data -- astrometry plus photometry -- were fit by a \glsxtrlong{6p} model (\verb+astrometric_params_solved+ = 95) that includes an astrometric estimate of the effective wavenumber, or pseudocolor, $\nu_{\rm eff}$, hereafter called '\glsxtrshort{6p}' dataset. The sources whose data were only fit by a two-parameter model were not considered. Interestingly, this shows that the most crowded regions of the Galaxy have a larger error on average, as well as the sources with a $G$-mag of about 7--9. This latter dependence on magnitude agrees well with Fig. A.1 of~\citet{Lindegren2021}.

\subsection{Attitude excess noise}
\label{sec:attitude_error}
The \textit{Gaia} spacecraft attitude is modeled during the data reduction. It converts a rigid solid-body motion within the \glsxtrshort{icrs} reference frame into \textit{Gaia}'s own reference frame where the CCDs are fixed~\citep{Lindegren2021}. It thus models part of the path followed by any source along the detectors during a transit. This model suffers from time-dependent attitude excess noise, due, for example, to micro-clanks, calibration errors, etc., and has a typical level of 76\,$\mu$as on average~\citep{Lindegren2021}. The attitude excess noise varies with time but at a given epoch all stars observed share a common attitude excess noise (Lindegren, \textit{priv. comm.}). Depending on the magnitude and the color, $\sigma_{\rm att}$ tends to over/under estimate the calibration noise (see also Sects.~\ref{sec:AEN} and~\ref{sec:jitter}).

The time-dependency of formal errors or attitude excess noise is not available. We can only assume that for any source, those errors are relatively constant (see, e.g., the Fig. A.3 in~\citealt{Lindegren2021} for an example with time-dependent attitude excess noise variations). Nonetheless, for a given source with specific \glsxtrshort{ra} \& \glsxtrshort{dec} direction, the attitude excess noise is probed at more-or-less regularly spaced epochs because of the scanning law of the spacecraft. Sources in different directions might thus probe disjoint sets of attitude excess noise values, and the mean attitude excess noise might thus depend on the \glsxtrshort{ra} \& \glsxtrshort{dec} direction. Being fixed, by construction, for all stars observed at the same epoch, the attitude excess noise do not dependent on magnitude or color. 

For any source in the \glsxtrshort{g3} database that has \glsxtrshort{aen}$>$0\,mas, quadratically removing the \glsxtrshort{al} measurement error from the formal error leads to the attitude excess noise. To do this computation, we need to know the \glsxtrshort{al} measurement error for any source. By conversely quadratically subtracting the attitude excess noise from the formal error one in fact can estimate the \glsxtrshort{al} measurement error. At any bin of magnitude \& color, $\sigma_{\rm formal}(\text{mag},\text{color})$ is the median formal error among all sources in that bin, distributed on all directions of the sky. We thus expect that, at any magnitude \& color, the median $\sigma_{\rm att}$ is close to 76\,$\mu$as. This led to a first estimation of the \glsxtrshort{al} measurement error, with respect to the magnitude and the color of the source, by applying
\begin{equation}
\sigma_{\rm AL} (\text{mag},\text{color}) = \sqrt{\left[\sigma_{\rm formal}(\text{mag},\text{color})\right]^2 - 0.076^2}
\end{equation}

\noindent
This estimation is refined in Sect.~\ref{sec:AL_error}. The \glsxtrshort{al} measurement error of a given source depends mainly on the optical properties associated with a CCD measurement of its \glsxtrlong{psf} (\glsxtrshort{psf}) on the detector, thus related to the magnitude and the color of the source. 
Linearly interpolating through this magnitude-color relationship, we can estimate the \glsxtrshort{al} measurement error for any source of given magnitude and color (within available convex hull), or $\sigma_{\rm AL, mc}$.
Our best guess of the attitude excess noise for any source can then be obtained by quadratically subtracting this $\sigma_{\rm AL, mc}$ from the formal error:
\begin{equation}
\sigma_{\rm att} = \sqrt{\sigma_{\rm formal}^2 - {\sigma^2_{\rm AL, mc}}}
\end{equation}

\begin{figure*}[hbt]
    \centering
    \includegraphics[width=178.6mm,clip=true,trim=100 0 90 0]{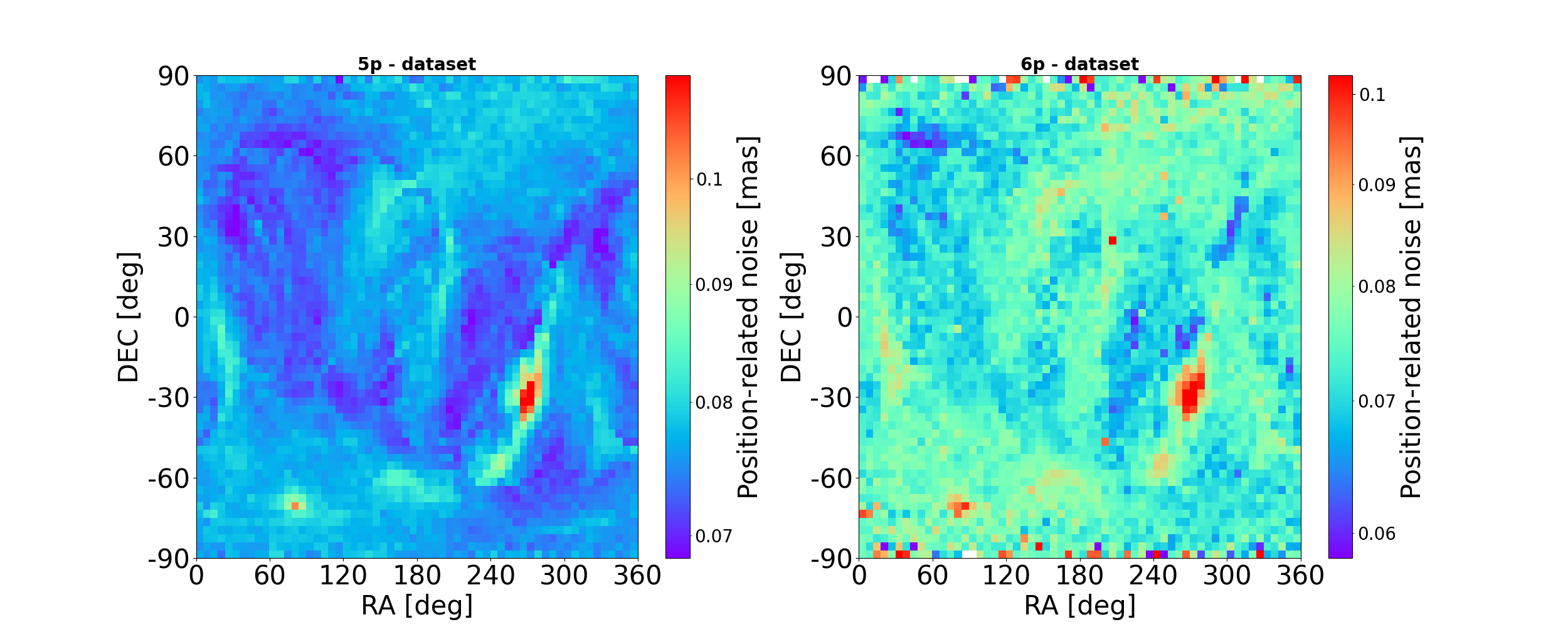}
    \caption{Median attitude excess noise distribution with respect to \glsxtrshort{ra} and \glsxtrshort{dec} in the \glsxtrshort{g3} database of sources brighter than $G$=16.}
    \label{fig:attitude_noise}
\end{figure*}

\noindent
To allow for estimation of $\sigma_{\rm att}$ even if a source's \glsxtrshort{aen} is compatible with 0\,mas, and to smooth out scatter among sources with a similar sky location, we calculated a median attitude excess noise in every \glsxtrshort{ra}-\glsxtrshort{dec} bins described in Table~\ref{tab:bins_params}. Those median attitude excess noises are given in Table~H.1. Figure~\ref{fig:attitude_noise} shows the dependence of the median $\sigma_{\rm att}({\rm \glsxtrshort{ra}},{\rm \glsxtrshort{dec}})$ with the sky direction. It shows a strong dependence on this parameter, with more pronounced error, up to 0.13\,mas, in crowded regions, such as the Magellanic clouds and the center of the Milky Way. For any given source, an estimation of the effective level of attitude excess noise is determined by linearly interpolating the \glsxtrshort{ra}-\glsxtrshort{dec} relationship at the \glsxtrshort{ra} and \glsxtrshort{dec} of the source. In the rest of the article, this interpolated value is called $\sigma_{\rm att}$.

\subsection{Along-scan angle measurement error}
\label{sec:AL_error}

Once $\sigma_{\rm att}$ is estimated for any source given their \glsxtrshort{ra}-\glsxtrshort{dec} direction, it is straightforward to determine the $\sigma_{\rm AL}$ for all the sources with \glsxtrshort{aen}$>$0\,mas. We simply used 
\begin{equation}
    \sigma_{\rm AL} = \sqrt{\sigma_{\rm formal}^2 - \sigma^2_{\rm att}}
\end{equation}

\noindent
As for $\sigma_{\rm att}$ above, to allow for estimation of $\sigma_{\rm AL}$, even if a source's \glsxtrshort{aen} is compatible with 0\,mas and to smooth out scatter among sources with a similar magnitude and color, we calculated a median attitude excess noise in every magnitude--color bins are described in Table~\ref{tab:bins_params}. The median \glsxtrshort{al} measurement errors are given in Table~H.2. It is available online with only an extract shown here at a $Bp-Rp$ close to that of GJ\,832, that is, $Bp-Rp=2.2$. Figure~\ref{fig:AL_error} shows the dependence of the median $\sigma_{\rm AL}$ with the magnitude and color.  
For any given source, an estimation of the effective level of \glsxtrshort{al} measurement error is determined by linearly interpolating the magnitude-color relationship at the $G$-mag and $Bp-Rp$ color of the source. In the rest of the article, this interpolated value is called $\sigma_{\rm AL}$.

\subsection{Calibration noise}
\label{sec:jitter}

The level of the calibration noise truly present in the data depends mainly on the magnitude and the color of the observed sources~\citep{Lindegren2016,Lindegren2018,Lindegren2021}. 
For any source, we calculated a normal model of the $\chi^2_{\rm astro}\sim{\mathcal N}\left(\mu_{\chi^2},\sigma_{\chi^2}\right)$, as thoroughly detailed in Appendix~\ref{sec:details_chi2}. It accounts for the correlations between the co-adjacent astrometric \glsxtrshort{al} angle measurements performed at the same epoch. The mean $\mu_{\chi^2}$ of the distribution of the $\chi^2_{\rm astro}$ is related to $\sigma_{\rm att}$, $\sigma_{\rm AL}$, and $\sigma_{\rm calib}$ by
\begin{equation}\label{eq:meanchi2}
    \mu_{\chi^2} = \frac{N_{\glsxtrshort{al}}}{\sigma^2_{\rm att}+\sigma_{\rm AL}^2}\,\left[(N_{\rm \glsxtrshort{fov}}-5) \, \sigma_{\rm calib}^2 + N_{\rm \glsxtrshort{fov}}\,\sigma^2_{\rm AL}\right]
\end{equation}
where $N_{\rm \glsxtrshort{fov}}$ is the number of \glsxtrlong{fov} (\glsxtrshort{fov}) transits on the detector, and $N_{\glsxtrshort{al}}$ is the average number of \glsxtrshort{al} angles collected per transit, that is, $\approx{\rm int}(N/N_{\rm \glsxtrshort{fov}})$. 
In the \textit{Gaia} archives, the total number of \glsxtrshort{al} angle measurements $N$ is given by \verb+astrometric_n_good_obs_AL+, while $N_{\rm \glsxtrshort{fov}}$ is given by \verb+astrometric_matched_transit+.
This equation leads to an expression of the $\sigma_{\rm calib}$ for any source assumed single, that is,
\begin{equation}
    \sigma_{\rm calib} =  \sqrt{\frac{\chi^2_{\rm astro} \times \left(\sigma_{\rm att}^2 + \sigma_{\rm AL}^2\right) - N_{\glsxtrshort{al}}\, N_{\rm \glsxtrshort{fov}}\,\sigma^2_{\rm AL}}{N_{\glsxtrshort{al}}\,(N_{\rm \glsxtrshort{fov}}-5)}}.
\end{equation}

\noindent
In any of the magnitude and color bins (Table~\ref{tab:bins_params}), the best estimation of $\sigma_{\rm calib}$ is thus that of single sources. The sources are separated into single and multiple stars, whose rate $N({\rm multiple})/N({\rm sources})$ is unfortunately unknown. The distribution of $\sigma_{\rm calib}$ in a given bin is thus the combination of both populations. 
According to \textit{Gaia}'s \glsxtrshort{dr}2 documentation,\footnote{\url{https://gea.esac.esa.int/archive/documentation/GDR2/pdf/GaiaDR2_documentation_1.2.pdf}} 
rather than the median, one can more safely rely on the mode of the \glsxtrlong{uwe} (\glsxtrshort{uwe}) distribution\footnote{\glsxtrshort{uwe}=$\sqrt{\chi^2/(N-5)}$.} to locate the median of single star's distribution. Indeed, the mode is shown to be less affected by multiplicity than the median and is thus a better approximation of single star's median. Conversely to what is adopted in the documentation, we have found that the 41st--percentile is not always a good approximation of the mode. We thus rather localized the mode in the $\sigma_{\rm calib}$ distribution by iteratively excluding sources with a $\sigma_{\rm calib}$ larger than twice 1.483$\times$\glsxtrshort{mad}($\sigma_{\rm calib}$) above the median, where \glsxtrshort{mad} is the \glsxtrlong{mad}. We then defined the mode as the median of this reduced distribution. We found that 3 iterations were necessary and enough to localise the mode. Figure~E.1 shows this mode localization in the cumulative density functions of the $\sigma_{\rm calib}$ distribution at some magnitude-color bins.

\begin{figure*}[hbt]
    \centering
    \includegraphics[width=178.6mm,clip=true,trim=100 0 90 0]{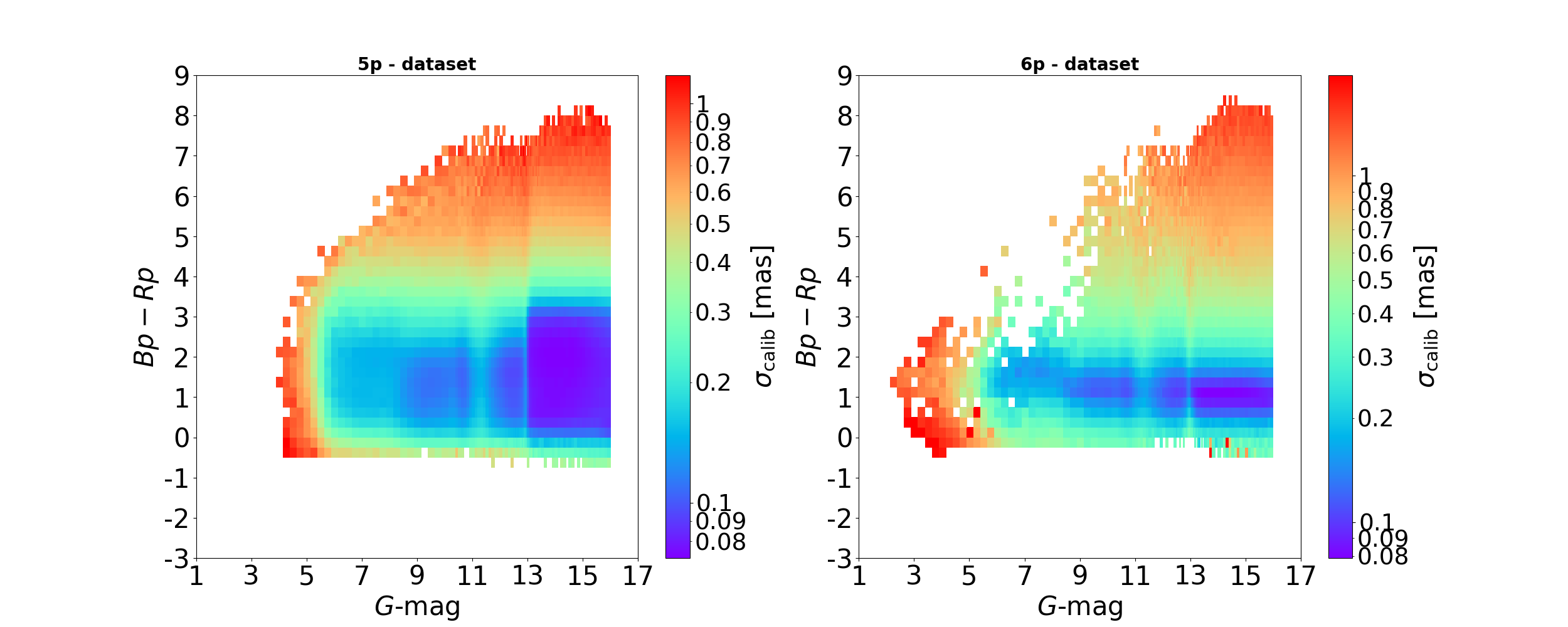}
    \caption{Maps of the calibration noise with respect to both $G$-mag and $Bp-Rp$ color. Left: for the \glsxtrshort{5p} dataset. Right: for the \glsxtrshort{6p} dataset. }
    \label{fig:jitter_noise}
\end{figure*}

Figure~\ref{fig:jitter_noise} shows the distribution of $\sigma_{\rm calib}$ with respect to $G$-magnitude and $Bp-Rp$ color for the \glsxtrshort{5p} and \glsxtrshort{6p} datasets. The $\sigma_{\rm calib}$ of the \glsxtrshort{6p} dataset are systematically higher than the \glsxtrshort{5p} dataset. This is an effect of the poorer-quality of the fit for those stars. For them, a large \glsxtrshort{aen} or \glsxtrshort{ruwe} has to be interpreted with care. In both dataset,  $\sigma_{\rm calib}$ and  magnitude are strongly correlated, especially for bright stars with $G$-mag$<$6, and to a lesser extent for early types with $Bp-Rp<0.5$ and late-types with $Bp-Rp>2.5$. 

\section{Modeling of \textit{Gaia} and \textsc{Hipparcos} astrometry}
\label{sec:simu}
For later use, we define in this section our process for modeling \textit{Gaia} and \textsc{Hipparcos} astrometric data. Our aim is to model the key data, namely \glsxtrshort{pma}, \glsxtrshort{aen}, and \glsxtrshort{ruwe}, which allow us to characterize the presence of companions and determine their main parameters, such as mass and \glsxtrshort{sma}. For any set of fixed companion, star and orbital parameters, we modeled by simulation the system's photocenter orbit as if it was observed by \textit{Gaia} or \textsc{Hipparcos}. In doing so, we accounted for instrumental and measurement noises in \textit{Gaia} and \textsc{Hipparcos} data, and then performed a five-parameter fit of those datasets. We obtained residuals, as well as proper motion and centroid simulated measurements at \glsxtrshort{g3} and \textsc{Hipparcos} epochs, respectively 2016.0~\citep{Gaia2021} and 1991.25~\citep{vanLeeuwen2007}. We explain our method and the technical details of the simulations and the fit procedures in the following sections. 

\subsection{Modeling photocenter orbits}
\label{sec:simu_orbit}

The core of the orbit modeling is the same as the one used in the \say{\glsxtrlong{gaston}} tool or \glsxtrshort{gaston} for short~\citep{Kiefer2019a,Kiefer2021}. We always consider a 2-body system, with a primary A and a secondary B, possibly planetary, brown dwarf or stellar. 
We fix the reference frame of the orbit to be the system's barycenter. With Keplerian parameters fixed for this system, we model the orbit of the photocenter of the system on the plane of the sky. The photocenter semi-major axis, $a_{\rm phot}$ is determined from the total system's semi-major axis, that is, the relative semi-major axis of the companion to the primary star; here written sma, through~\citep{Kiefer2021}: 
\begin{equation}
    a_{\rm phot} =  {\rm \glsxtrshort{sma}} \left( \beta - B \right) \varpi
\end{equation}
with $\varpi$ the parallax, $\beta$=$q$/1+$q$ the mass fraction, and $B$=$L_2$/$L_1$+$L_2$ the luminosity fraction. The relative luminosity of the secondary over the primary is determined from semi-empirical mass-luminosity relation on the main sequence, at a typical age of 5\,Gyr (see~\citealt{Kiefer2021} for more details). By default, we consider that the secondary may contribute to the photocenter's position. 
Depending on the case at hand, one may instead consider a dark companion, whose luminosity is thus not contributing to the photocenter's displacement. An illustration of the result of assuming instead a dark companion is shown in Sect.~\ref{sec:examples} for the case of $\alpha$\,CMa\,B, that is, Sirius\,B, and whose companion Sirius\,A is resolved by \textit{Gaia} and thus not contributing to $\alpha$\,CMa\,B's photocenter's displacement.

The modeled orbits are then sampled at specific epochs, along specific directions, according to the scan law of \textit{Gaia} and \textsc{Hipparcos} during their observation campaign. Noise is finally added to the individual measurements in a way that is specific to each instrument. This is further explained in the next Sects.~\ref{sec:scanlaw} and~\ref{sec:hiplaw}.

\subsection{\textit{Gaia} DR3 sampling, scan-law, and noise}
\label{sec:scanlaw}
\begin{figure}
    \centering
    \includegraphics[width=89.3mm]{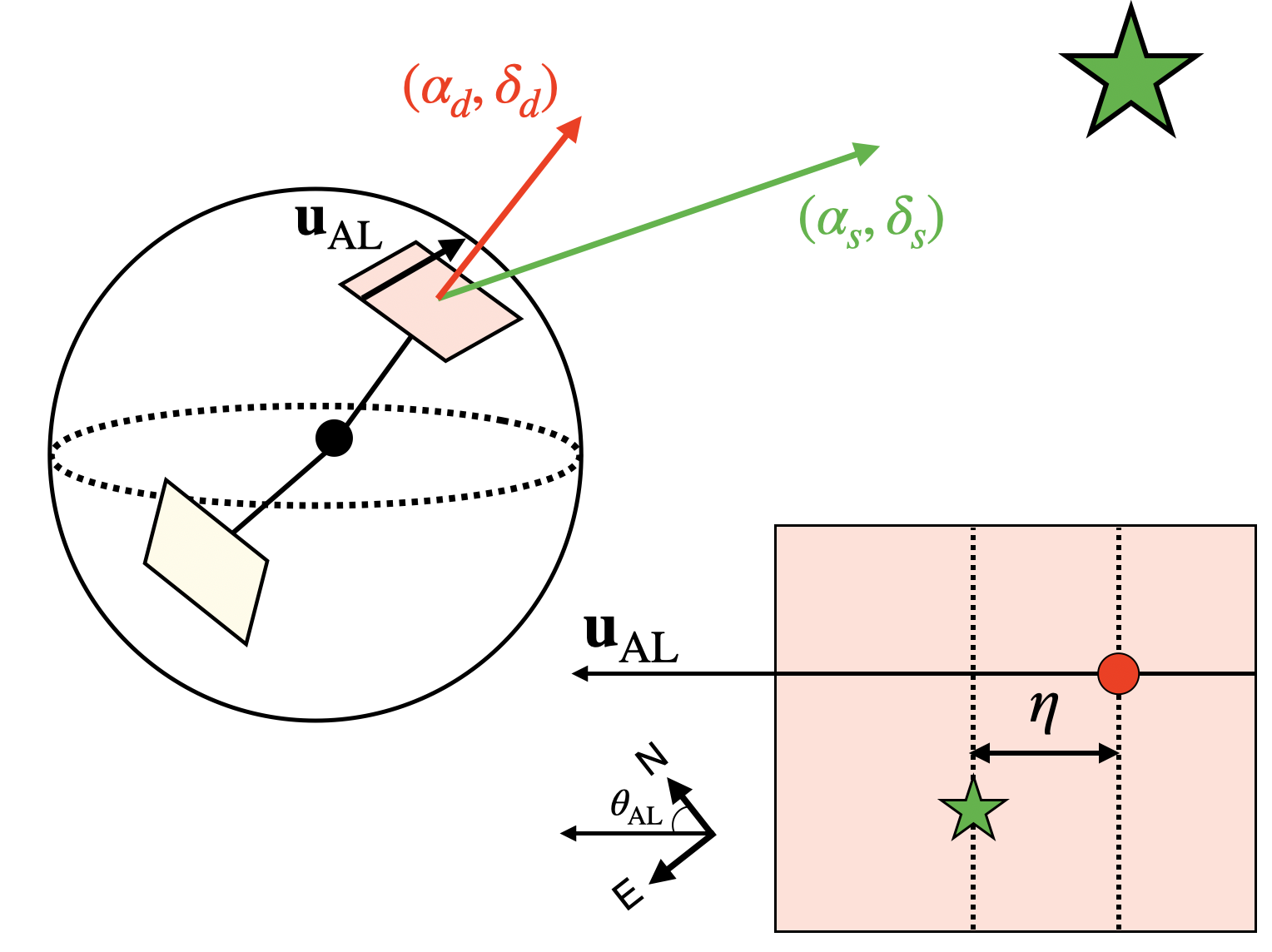}
    \caption{Schematic representation of the orientation of one of \textit{Gaia}'s detectors (red arrow) compared to a star's direction (green arrow). The solid circle represents the celestial sphere as seen from the \textit{Gaia} center of mass, and the dashed-line circle represents the celestial equator. The two quadrilaterals represent \textit{Gaia}'s preceding (light red) and following (light yellow) \glsxtrshort{fov} detectors. On the bottom right, we show the possible location of the star on the detector and the $\eta$ angle that is measured projected along the \glsxtrshort{al} axis ($u_{\rm \glsxtrshort{al}}$). Arbitrary north and east directions are shown with the definition of the \glsxtrshort{pa} of the \glsxtrshort{al} direction. They are not intended to exactly correspond to the top-left drawing but allowed us to define $\theta_{\rm AL}$, the eastward-oriented angle between $u_{\rm \glsxtrshort{al}}$ and the north.} 
    \label{fig:detector_star_diagram}
\end{figure}

We sampled the modeled orbits at the \glsxtrshort{g3} \glsxtrshort{fov}-passage epochs and along the \glsxtrshort{al} direction. The \textit{Gaia} spacecraft is composed of two \glsxtrshort{fov}, the 'preceding' and the 'following'. They are separated around the spin axis of the spacecraft by a basic angle of 106.5$^\circ$~(\citealt{Lindegren2012}; see also Fig.~\ref{fig:detector_star_diagram}). At a given epoch, the spin axis is moreover oriented in a certain direction conferring to the detector a certain orientation of its main axis, the \glsxtrshort{al} axis, $u_{\rm \glsxtrshort{al}}$. The law of the \glsxtrlong{pa} (or \glsxtrshort{pa}) of $u_{\rm \glsxtrshort{al}}$ through time can be found in the \glsxtrshort{gost}. Six to 9 astrometric measurements are performed at the same epoch during the transit of the source across the detector thanks to \textit{Gaia}'s spin, at a speed of 60\arcsec/min~\citep{Lindegren2012}.

The date of passage of a star on the \textit{Gaia} detector and the \glsxtrshort{pa} of $u_{\rm \glsxtrshort{al}}$ can be predicted accurately using the \glsxtrlong{gost} (\glsxtrshort{gost} for short). However, this tool is only accessible online\footnote{\url{https://gaia.esac.esa.int/gost/}}. Instead, we built a code that performs the same predictions using the spacecraft scan law accessible from the \verb+commanded_scan_law+ database. As explained in Fig.~\ref{fig:detector_star_diagram}, we calculate the angle between the direction of each detector $(\alpha_{\rm d},\delta_{\rm d})$ with the direction of the star at the \glsxtrshort{g3} epoch $(\alpha_{\rm s},\delta_{\rm s})$.
We used a gnomonic projection (see, e.g.,~\citealt{Calabretta2002}) to transform this angle into a vector on the plane of the detector.  
More specifically, there is a relationship between $(\eta,\zeta)$ the \glsxtrshort{al} and \glsxtrshort{ac} coordinates on the detector, and the difference of coordinates between the pointing direction of the detector and the direction of the star. Moreover, given that the \glsxtrshort{al}-direction is oriented at a \glsxtrshort{pa}=$\theta_{\rm AL}$, we found this relationship to be
\begin{align}
   \eta =& \frac{\cos \delta_{\rm s} \sin\left(\alpha_{\rm s}-\alpha_{\rm d}\right)}{\sin\delta_{\rm s}\sin\delta_{\rm d} + \cos\delta_{\rm s}\cos\delta_{\rm d}\cos\left(\alpha_{\rm s}-\alpha_{\rm d}\right)}\,\cos\theta_{\rm AL} \nonumber  \\ &- \frac{\sin\delta_{\rm s} \cos\delta_{\rm d} - \cos\delta_{\rm s}\sin\delta_{\rm d}\cos\left(\alpha_{\rm s}-\alpha_{\rm d}\right)}{\sin\delta_{\rm s}\sin\delta_{\rm d} + \cos\delta_{\rm s}\cos\delta_{\rm d}\cos\left(\alpha_{\rm s}-\alpha_{\rm d}\right)}  \,\sin\theta_{\rm AL} \\
\zeta =& \frac{\cos \delta_{\rm s} \sin\left(\alpha_{\rm s}-\alpha_{\rm d}\right)}{\sin\delta_{\rm s}\sin\delta_{\rm d} + \cos\delta_{\rm s}\cos\delta_{\rm d}\cos\left(\alpha_{\rm s}-\alpha_{\rm d}\right)}\,\sin\theta_{\rm AL} \nonumber  \\ &+ \frac{\sin\delta_{\rm s} \cos\delta_{\rm d} - \cos\delta_{\rm s}\sin\delta_{\rm d}\cos\left(\alpha_{\rm s}-\alpha_{\rm d}\right)}{\sin\delta_{\rm s}\sin\delta_{\rm d} + \cos\delta_{\rm s}\cos\delta_{\rm d}\cos\left(\alpha_{\rm s}-\alpha_{\rm d}\right)}  \,\cos\theta_{\rm AL}
\end{align}

\begin{figure}[hbt]
    \centering
    \includegraphics[width=89.3mm]{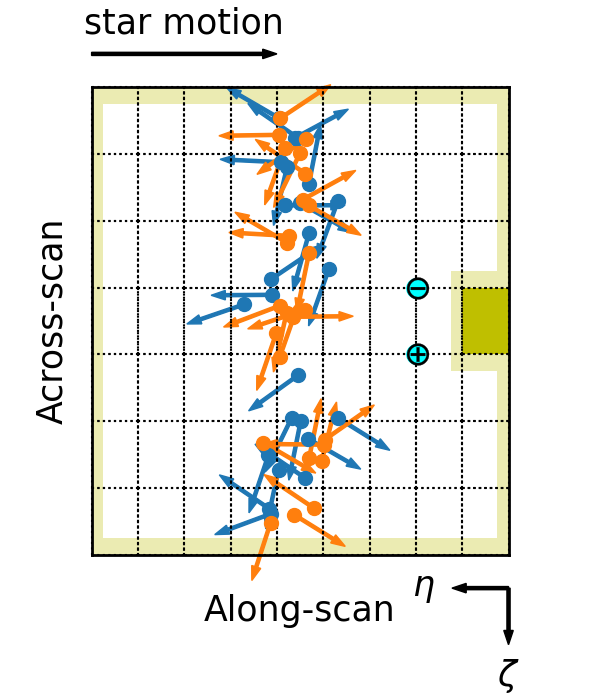}
    \caption{Transits through the detector found for GJ\,832 on the preceding \glsxtrshort{fov} (blue) and the following \glsxtrshort{fov} (orange). Each rectangle is a CCD, and the grid is 9$\times$7. The cyan-filled black symbols represent the \glsxtrshort{fov} origins, with a '+' for the preceding and a '-' for the following \glsxtrshort{fov}. The black arrow at the top shows the direction of the source motion through the \glsxtrshort{fov}. The dots show the average positions of the star on the detector at different epochs. The arrow connected to the dot indicates the average north direction at that epoch. The yellow regions depict the assumed dead zones, with the darker rectangle corresponding to the \glsxtrshort{wfs}2.}
    \label{fig:detector_traces}
\end{figure}

\noindent
Then we imposed that this vector should be contained within the used area of the detector. The zero origin of $\eta$ and $\zeta$ is not located at the center of the detectors and is different in the two \glsxtrshort{fov}s, as explained in~\citep{Lindegren2016}. In terms of CCD (\glsxtrshort{al} $\times$ \glsxtrshort{ac}) unit, compared to the center of the detectors, they are located at (-2.5,+0.5) for the preceding \glsxtrshort{fov} and at (-2.5,-0.5) for the following \glsxtrshort{fov}.
The detectors have a common dimension of 0.66$\times$0.74\,degree$^2$ with a grid of 9$\times$7\,CCDs. Dead zones are the \glsxtrlong{wfs} \glsxtrshort{wfs}2~\citep{Gaia2016} and the area exterior to the detector. We assumed that CCD regions at less than a quarter of a CCD-sized distance to a dead zone is also a dead zone. This led to a better match of the number of predicted transits with the actual number of transits for any given star. We rejected a detection if the star fell on the dead zones. Figure~\ref{fig:detector_traces} shows a representation of the detector and the geometry of the assumed dead zones, with GJ\,832's predicted average positions and \glsxtrshort{al} scan direction orientations. 

The time sampling of the scan law is $\sim$11\,sec. Therefore, during any transit of a source on the detector, several epochs are found, whereas only one epoch is required per transit. The spacecraft rotates at 1\arcsec/s, and the largest of the diagonals of the detector have a dimension of 1$^\circ$. We thus determined, for any transit, the average epoch and average \glsxtrshort{pa} from all the predicted transit epochs found within a 1\,hour window. We thus obtained for a given star all its theoretical epochs of transits through any of the two detectors with their corresponding \glsxtrshort{pa} of the \glsxtrshort{al} direction. 
As a final step, we removed the epochs that fell at known gaps published in the \glsxtrshort{g3} catalog\footnote{\url{https://www.cosmos.esa.int/web/gaia/dr3-data-gaps}}. The position of GJ\,832 on the detector and the \glsxtrshort{pa} of the scan directions during its transits as retrieved from the scan-law is shown in  Fig.~\ref{fig:detector_traces}. 

We verified that the retrieved \glsxtrshort{fov} transits matched those predicted by the \glsxtrshort{gost}. Moreover their number are always close to those given by the \verb+astrometric_matched_transit+ in the \glsxtrshort{g3} catalog.  We noted that our calculation, consistently with the \glsxtrshort{gost}, sometimes overestimated the number of actual \glsxtrshort{fov} transits retained to calculate the astrometric solution in the \glsxtrshort{g3}. This happens most often to bright stars, thus indicating an effect of saturation that led to removal of some of the transits in the solution. In those cases, we randomly selected the correct number of epochs effectively used by \textit{Gaia} in the \glsxtrshort{g3} among all retrieved epochs of \glsxtrshort{fov} passages.

\begin{figure*}[hbt]
    \centering
    \includegraphics[width=178.6mm,clip=true,trim=70 0 150 90]{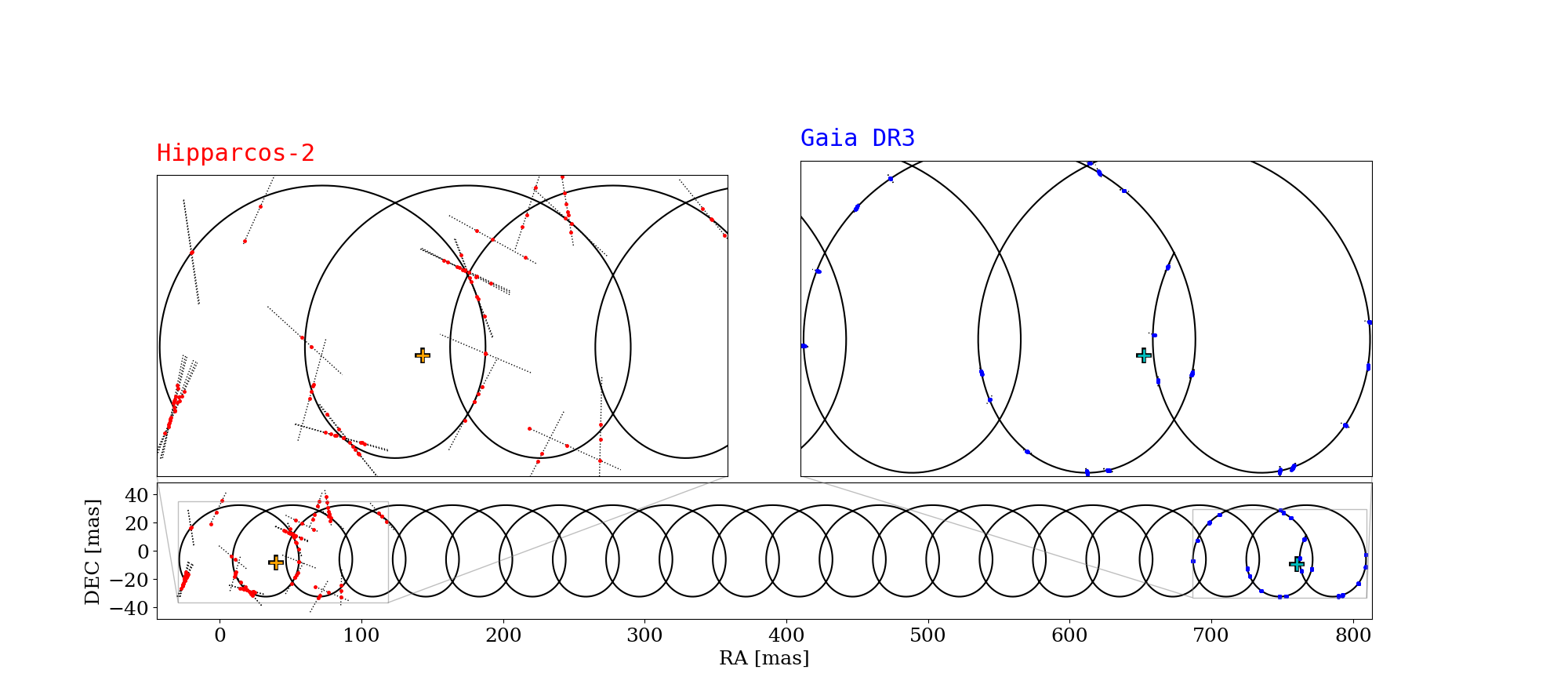}
    \caption{Simulation of an orbital motion as seen by \textsc{Hipparcos} (red dots) and \textit{Gaia} (DR3; blue dots) around GJ\,832 for a companion mass of 100\,M$_{\rm J}$ and sma=1\,au, $e$=0, $I_c$=0$^\circ$. For visualization, we added a virtual proper motion of 30\,mas/yr along the \glsxtrshort{ra} direction. The individual astrometric measurements are scattered along the along-scan directions at each \glsxtrshort{fov} transit epoch with $\sigma_{\rm \glsxtrshort{al}}$=0.095\,mas and $\sigma_{\rm calib}$=0.15\,mas for \textit{Gaia} and an average dispersion of $\sim$4.2\,mas for \textsc{Hipparcos}. The orange and cyan crosses respectively mark the position of the fit centroid on the \textsc{Hipparcos} and \textit{Gaia} datasets.}
    \label{fig:simu_orbit}
\end{figure*}

We then assumed that the \textit{Gaia}'s \glsxtrshort{al} measurements along the \glsxtrshort{al} direction are distributed according to a normal law with as standard deviation, the noise $\sigma_{\rm AL}$ determined by the $G$ and $Bp-Rp$ of the source in consideration (see Sect.~\ref{sec:AL_error}). Adding to this error, we added an epoch-specific offset randomly drawn from a normal distribution with standard deviation $\sigma_{\rm calib}$, determined with respect to the $G$ and $Bp-Rp$ of the source (see Sect.~\ref{sec:jitter}). One such simulation is shown in Fig.~\ref{fig:simu_orbit}. 

\subsection{\textsc{Hipparcos} sampled epochs}
\label{sec:hiplaw}
The \textsc{Hipparcos}-2 \glsxtrlong{iads} (\glsxtrshort{iads};~\citealt{vanLeeuwen2007}) are necessary to model the \glsxtrshort{pma} as determined by~\citet{Kervella2022}. We here only focus on the \glsxtrshort{pma} between \glsxtrshort{g3} and \glsxtrshort{hg} baselines. The location of the centroid of the \textsc{Hipparcos}-2 data in the source barycenter reference frame has to be determined for any modeled orbit to find the \glsxtrshort{hg} average proper motion between epochs 1991.25 and 2016.0. In the \textsc{Hipparcos}-2 database the published source centroid is located along the fitted solution. This is not adequate for us because \textsc{Hipparcos}-2 used, when possible, more elaborate models including acceleration or orbital motion. However, \citet{Kervella2022} only considers the result of a five-parameter fit of the \textsc{Hipparcos}-2 IADs to derive the \textsc{Hipparcos}-2 proper motion and the location of the \textsc{Hipparcos}-2 source centroid at epoch 1991.25. 

Therefore, for any source also observed with \textsc{Hipparcos}, we downloaded the \glsxtrshort{iads} residuals from the \textsc{Hipparcos}-2 Interactive Data Access Tool \footnote{\url{https://www.cosmos.esa.int/web/hipparcos/interactive-data-access};
"the Java tool"}. These data include the orbit number (IORB), the epoch, the cosine and sine of the \textsc{Hipparcos} scan angle (related to the \textit{Gaia} scan angle convention by $\psi$=$\theta_{\rm AL}$-$\pi/2$;~\citealt{Brandt2021}), the residuals of the fitted model (RES) and the formal errors (SRES). 

We removed the data with negative or zero SRES that are rejected observations. To model \textsc{Hipparcos} observations of the astrometric displacement of the photocenter due to an orbital motion, we only need the part of the residuals at each IORB that cannot be due to supplementary non-modeled displacement. We thus calculated corrected residuals (CRES) by removing the local average from common IORB residuals. Then when simulating an orbit, those CRES are added along the \textsc{Hipparcos} scan direction.

Besides, to have an estimation of the typical dispersion associated with a source centroid position, we also calculated the \textsc{Hipparcos}-2 positional error for the considered source from the \glsxtrshort{ra} and \glsxtrshort{dec} positional error published in the \textsc{Hipparcos}-2 catalog: 
\begin{equation}
\sigma_{\rm pos, HIP}=\sqrt{\text{e\_\glsxtrshort{ra}}^2\cos^2(\text{\glsxtrshort{dec}})+\text{e\_\glsxtrshort{dec}}^2}
\end{equation}

\noindent 
An illustration of the \textsc{Hipparcos} data modeled for an arbitrary orbit is shown in Fig.~\ref{fig:simu_orbit}. 

\subsection{\textit{Gaia} and \textsc{Hipparcos} five-parameter model fit}
\label{sec:fit}

For each modeled orbit, we applied a five-parameter\footnote{For \textit{Gaia}, pseudo-color is accounted for in targets with a six-parameter fit in the form of a higher calibration noise (see Sect.~\ref{sec:noise_proxy}).} fit to \textsc{Hipparcos}-2 and \glsxtrshort{g3} simulated data. It included the \glsxtrshort{ra}-\glsxtrshort{dec} centroid of data points, the linear proper motions $\mu_\alpha$ and $\mu_\delta$, and the parallax $\varpi$. Given that we placed ourselves in the barycenter reference frame of the considered system, we thus fit the excesses of (positive or negative) offset, proper motion, and parallax only due to the presence of an orbital motion. Because of the orbital motion, the parallax measured in \glsxtrshort{g3} deviates from the true value. Assuming that the current orbit was the true one, we first estimated the parallax error $\Delta\varpi$ from the fitted parallax excess in a first simulation. We then performed a second simulation, correcting the parallax by $\varpi \rightarrow \varpi - \Delta\varpi$. The 2D-fitted linear model is 

\begin{equation}
    M(t) = (\delta {\rm \glsxtrshort{ra}},\delta {\rm \glsxtrshort{dec}}) + (\delta\mu_\alpha,\delta\mu_\delta) \, (t-t_0) +    \delta\varpi\,\Pi(t) 
\end{equation}

where the $\delta \mu$ are proper motion in $\alpha$=\glsxtrshort{ra}\,$\cos$\,\glsxtrshort{dec} and $\delta$=\glsxtrshort{dec} tangent plane directions, $\Pi(t)$ is the parallax ellipse depending on the coordinate of the star, and $t_0$ the \textsc{Hipparcos}-2 or \glsxtrshort{g3} epochs, respectively 1991.25 and 2016.0. We note that the effect of perspective acceleration~(see, e.g., \citealt{Michalik2014,Halbwachs2023}) that mainly affect high proper motion targets close to Sun, is already corrected in the \glsxtrshort{g3}, so being a second-order effect we can ignore it here~\citep{Lindegren2021}. To compare this linear model to the Hipparocs and \textit{Gaia} measurements, we needed to project this model onto the along-scan directions, with position angle $\theta_{\rm \glsxtrshort{al}}$, determined at the sampled epochs along the orbit: 

\begin{align}
    M_{\rm AL}(t_i) = &\left(\delta {\rm \glsxtrshort{ra}}  + \delta\mu_\alpha \, (t_i-t_0) +  \delta\varpi\,\Pi(t_i)\vert_{\rm \glsxtrshort{ra}} \right)\,\sin\theta_{\rm \glsxtrshort{al}}(t_i) \nonumber \\ & + \left(\delta {\rm \glsxtrshort{dec}} + \delta\mu_\delta\, (t_i-t_0) + \delta\varpi\,\Pi(t_i)\vert_{\rm \glsxtrshort{dec}} \right)\,\cos\theta_{\rm \glsxtrshort{al}}(t_i)
\end{align}

We separated components along \glsxtrshort{dec} (north) and \glsxtrshort{ra} (east) directions. We subtracted this five-parameter model from the simulated data and calculate the residuals. 
For \textit{Gaia}, they are further used in comparison to the tabulated \glsxtrshort{aen} or \glsxtrshort{ruwe} published in the \glsxtrshort{g3} catalog as explained in Sects.~\ref{sec:astro_sig} and~\ref{sec:PMEX}.

To calculate the \glsxtrshort{pma}, the fit \glsxtrshort{g3} proper motion, $\delta\mu_{\rm \glsxtrshort{g3}}$, was combined with the average proper motion between the fit positions of the photocenter at the \textsc{Hipparcos}-2 reference epoch and the \glsxtrshort{g3} reference epoch, $\delta\mu_{\rm \glsxtrshort{hg}}$, through 
\begin{align}
    & {\rm \glsxtrshort{pma}} = \delta\mu_{\rm \glsxtrshort{g3}} - \delta\mu_{\rm \glsxtrshort{hg}} \\ 
    & \text{with   } \delta\mu_{\rm \glsxtrshort{hg}} = \frac{(\alpha,\delta)_{\rm \glsxtrshort{g3}} - (\alpha,\delta)_{\rm HIP} }{24.75\,yr}
\end{align}

\noindent
This modeled \glsxtrshort{pma} is compared to the \glsxtrshort{pma} published in~\citet{Kervella2022}, as explained in Sects.~\ref{sec:astro_sig} and~\ref{sec:PMEX}.

\section{The non-singleness of stars observed with \textit{Gaia}}
\subsection{Astrometric signatures}
\label{sec:astro_sig}

To assess the non-singleness of stars from \glsxtrshort{aen}, \glsxtrshort{ruwe}, and \glsxtrshort{pma}, we defined (and introduce here) the \say{astrometric signatures}, further written as $\alpha$. They properly quantify the deviation of \glsxtrshort{aen}, \glsxtrshort{ruwe} and  \glsxtrshort{pma} beyond the level that they must have had if the sources were single. 

\subsubsection{The residuals astrometric signature}
\label{sec:alpha_mse}

We first introduced the residuals \glsxtrlong{mse} (\glsxtrshort{mse}) related to the $\chi^2_{\rm astro}$ by
\begin{equation}\label{eq:mse_chi2}
    {\rm \glsxtrshort{mse}} = \frac{\chi^2_{\rm astro}}{N-5} \times \left(\sigma_{\rm att}^2 + \sigma_{\rm AL}^2\right)
\end{equation}
This quantity's square root, also known as \glsxtrlong{rse} (\glsxtrshort{rse}),  is an unbiased estimator of the data typical error in the considered sample of measurements. 
The \glsxtrshort{mse} of any source, considering Eqs.~\ref{eq:ruwe_chi2},~\ref{eq:chi2_errors} and~\ref{eq:mse_chi2}, can be estimated in two different ways, either using the \glsxtrshort{aen} or the \glsxtrshort{ruwe}: 

\begin{align}
    {\rm \glsxtrshort{mse}}_{\rm aen} &={\rm AEN}^2 + \sigma^2_{\rm att} + \sigma^2_{\rm AL} \label{eq:aen_mse} \\
    {\rm \glsxtrshort{mse}}_{\rm ruwe}&=\left({\rm \glsxtrshort{ruwe}} \times u_0\right)^2 \, \left(\sigma^2_{\rm att}+\sigma^2_{\rm AL}\right) \label{eq:ruwe_mse}
\end{align}

\noindent
We consider both in the rest of the study. The \glsxtrshort{mse} of a single source (${\rm \glsxtrshort{mse}}_{\rm single}$) only accounts for calibration noise and \glsxtrshort{al} astrometric noise.  The distribution of ${\rm \glsxtrshort{mse}}_{\rm single}$  is obtained from the normal model of the $\chi^2$ distribution of the residuals, determined in Appendix~\ref{sec:details_chi2}, Eqs.~\ref{eq:chi2_mean} and~\ref{eq:chi2_sigma}. It led to a normal distribution ${\mathcal N}(\mu_{\rm \glsxtrshort{mse},single},\sigma_{\rm \glsxtrshort{mse},single})$ of the ${\rm \glsxtrshort{mse}}_{\rm single}$ with
\begin{align}
    \mu_{\rm \glsxtrshort{mse},single} =& \frac{N_{\glsxtrshort{al}}}{N_{\glsxtrshort{al}}\,N_{\rm \glsxtrshort{fov}} - 5}\,\left[(N_{\rm \glsxtrshort{fov}}-5) \, \sigma^2_{\rm calib} + N_{\rm \glsxtrshort{fov}}\,\sigma^2_{\rm AL}\right] \label{eq:mse_single} \\ 
    \sigma_{\rm \glsxtrshort{mse},single}^2=&  \frac{2 N_{\glsxtrshort{al}}}{\left(N_{\glsxtrshort{al}}\,N_{\rm \glsxtrshort{fov}} - 5\right)^2}\,\Bigg[ N_{\glsxtrshort{al}} \left(N_{\rm \glsxtrshort{fov}}-5\right)\,\sigma^4_{\rm calib} \nonumber \\ & \qquad\qquad\qquad +  N_{\rm \glsxtrshort{fov}}\,\sigma^4_{\rm AL}  + 2\,N_{\rm \glsxtrshort{fov}} \,\sigma_{\rm AL}^2\,\sigma_{\rm calib}^2\Bigg] \label{eq:mse_single_sigma}
\end{align}

\begin{figure}[hbt]
    \centering
    \includegraphics[width=89.3mm]{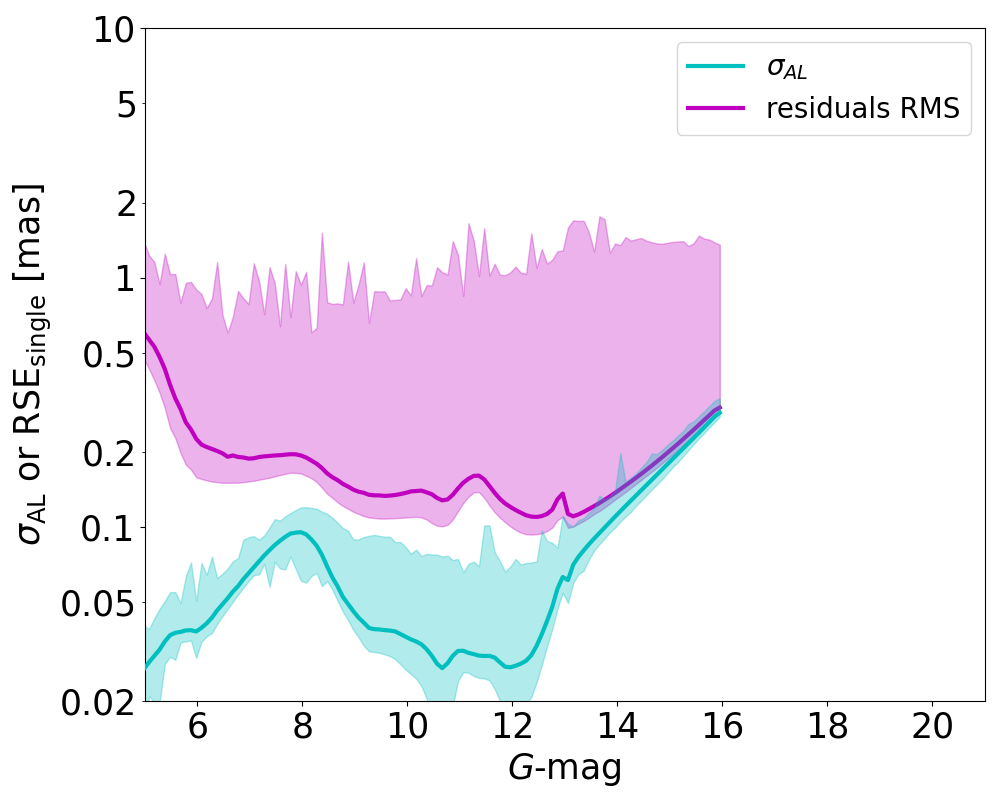}
    \caption{Along-scan astrometric measurement precision variations with respect to $G$-mag for all \glsxtrshort{g3} sources. The magnitude interval is enlarged up to $G$=20 for direct comparison to Fig.A.1 in~\citet{Lindegren2021}. The cyan line shows the median $\sigma_{\rm AL}$, and the magenta line shows the median \glsxtrshort{rse}$_{\rm single}$. Colored regions show the extent of $\sigma_{\rm AL}$ and \glsxtrshort{rse}$_{\rm single}$ with respect to color at each magnitude.}
    \label{fig:aslindegren21}
\end{figure}

\noindent
Ingredients needed to calculate $\mu_{\rm \glsxtrshort{mse},single}$ and $\sigma_{\rm \glsxtrshort{mse},single}$ for any source are thus $G$-mag, $Bp-Rp$, \glsxtrshort{ra} and \glsxtrshort{dec} to estimate the noises; and $N_{\rm \glsxtrshort{fov}}$, as well as $N_{\glsxtrshort{al}}={\rm int}(N/N_{\rm \glsxtrshort{fov}})$. 
The variation of the single-star \glsxtrshort{rse} with respect to magnitude is compared to the level of $\sigma_{\rm AL}$ in Fig.~\ref{fig:aslindegren21}. It compares well and agrees with the same curves plotted in Fig. A.1 in~\citet{Lindegren2021} determined directly from the unpublished time series and images. This shows that our estimation of ${\rm \glsxtrshort{mse}}_{\rm single}$ in Eq.~\ref{eq:mse_single} provides a reliable estimation of the ground level of the \glsxtrshort{mse} for any source.

For a non--single star, when adding the orbital motion, the \glsxtrshort{mse} may positively deviate from ${\rm \glsxtrshort{mse}}_{\rm single}$. We define the residuals astrometric signature as the angular excess that has to be quadratically added to the ground level \glsxtrshort{mse}$_{\rm single}$ to recover the \glsxtrshort{mse} of the residuals of the given source. It thus measures, in units of milli-arcseconds, the strength of non-singleness of the source. It is written $\alpha_{\rm \glsxtrshort{mse}}$ and is formally defined as
\begin{align}\label{eq:astro_sig}
    \alpha_{\rm \glsxtrshort{mse}} = \sqrt{ {\rm \glsxtrshort{mse}} - {\rm \glsxtrshort{mse}}_{\rm single}}
\end{align}

\noindent
Using $\mu_{\rm \glsxtrshort{mse},single}$ as the expectation value of ${\rm \glsxtrshort{mse}}_{\rm single}$ and the \glsxtrshort{aen} (Eq.~\ref{eq:aen_mse}) or the \glsxtrshort{ruwe} (Eq.~\ref{eq:ruwe_mse}) to estimate ${\rm \glsxtrshort{mse}}$, we can determine $\alpha_{\rm \glsxtrshort{mse}}$ for any source. For a single star, as further developed in Sect.~\ref{sec:significance_alpha}, because of the diverse astrometric noises that depends on the star's magnitude, color and sky coordinates, the \glsxtrshort{mse} follows a broadened distribution that extends around ${\rm \glsxtrshort{mse}}_{\rm single}$. It leads to $\alpha_{\rm \glsxtrshort{mse}}$=0 if the \glsxtrshort{mse} is smaller or equal to ${\rm \glsxtrshort{mse}}_{\rm single}$ and positive values otherwise. For non-single sources, $\alpha_{\rm \glsxtrshort{mse}}$ may become strongly positive if the astrometric motion dominates over the astrometric noises. In that sense, $\alpha_{\rm \glsxtrshort{mse}}$ is indeed an astrometric signature.

\begin{figure}[hbt]
    \centering
    \includegraphics[width=89.3mm]{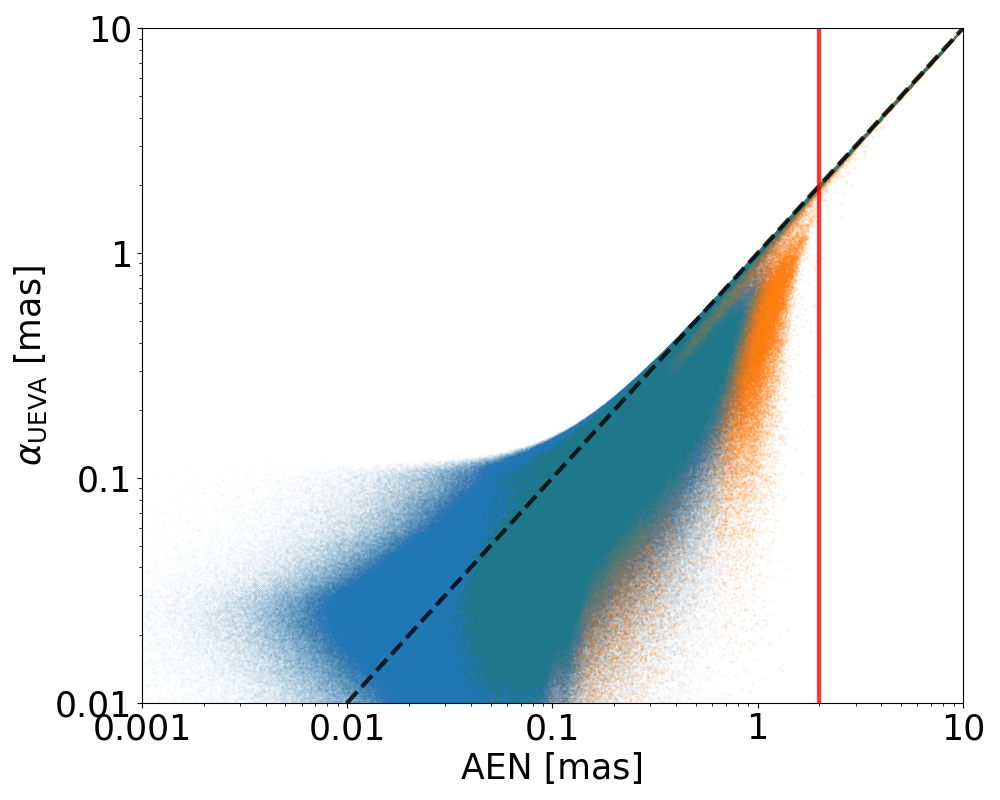}
    \includegraphics[width=89.3mm]{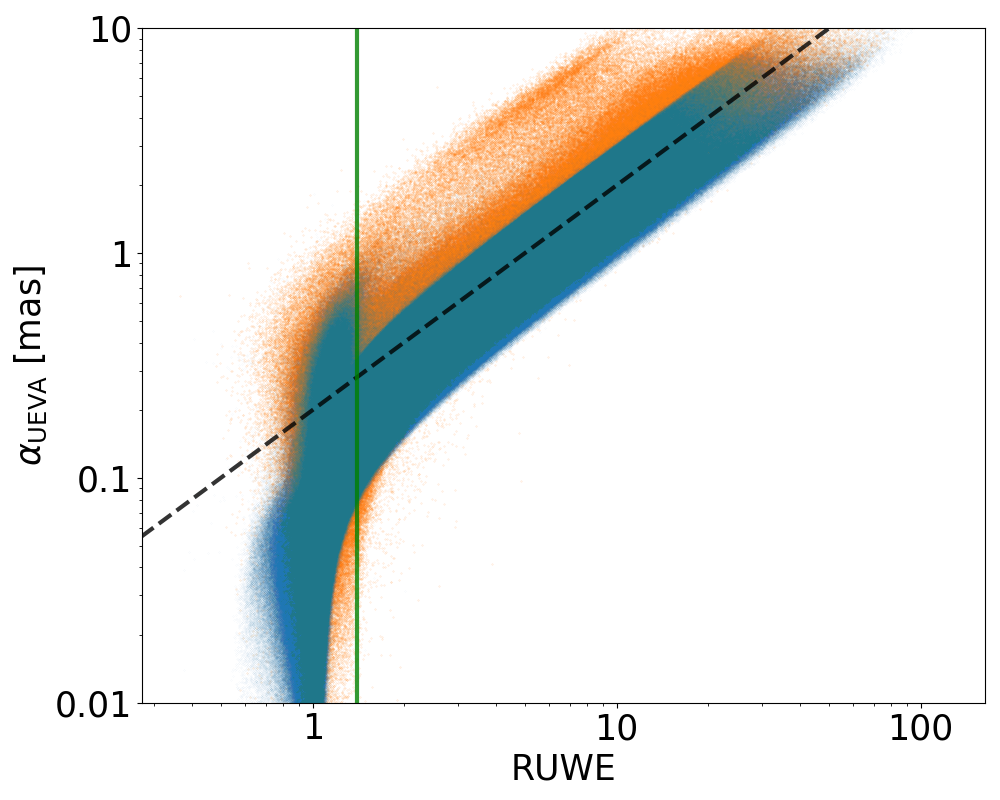}
    \caption{Astrometric signature $\alpha_{\rm \glsxtrshort{mse}}$ calculated from either \glsxtrshort{aen} (top) or \glsxtrshort{ruwe} (bottom) and compared to the quantities in the \glsxtrshort{5p} (blue dots) and \glsxtrshort{6p} (orange dots) datasets. The dashed black line shows, respectively, the $\alpha_{\rm \glsxtrshort{mse}}$=\glsxtrshort{aen} and $\alpha_{\rm \glsxtrshort{mse}}$=0.2\,\glsxtrshort{ruwe} relationships mentioned in the text, through Eqs.~\ref{eq:aen_mse},~\ref{eq:ruwe_mse} and \ref{eq:astro_sig}. The red and green vertical lines show, respectively, the \glsxtrshort{aen}=2\,mas and \glsxtrshort{ruwe}=1.4 thresholds.}
    \label{fig:AEN_RUWE_astrosig}
\end{figure}

Figure~\ref{fig:AEN_RUWE_astrosig} compares the $\alpha_{\rm \glsxtrshort{mse}}$ calculated from  ${\rm \glsxtrshort{mse}}_{\rm aen}$ or ${\rm \glsxtrshort{mse}}_{\rm ruwe}$ to respectively \glsxtrshort{aen} or \glsxtrshort{ruwe} for both \glsxtrshort{5p} and \glsxtrshort{6p} datasets. It shows that the $\alpha_{\rm \glsxtrshort{mse}}$ is almost equal to \glsxtrshort{aen} beyond \glsxtrshort{aen}=2\,mas. The \glsxtrshort{aen} could thus be directly interpreted as an astrometric signature in this regime. While there is a clear linear correspondance between \glsxtrshort{ruwe} and $\alpha_{\rm \glsxtrshort{mse}}$ beyond \glsxtrshort{ruwe}=1.4, with an approximative slope of $\sim$0.2, the thickness of the relation makes a direct astrometric interpretation more difficult. This is worse for the \glsxtrshort{6p} dataset where the range of possible $\alpha_{\rm \glsxtrshort{mse}}$ for a given \glsxtrshort{ruwe} is even more spread out, due to the larger levels of noise compared to the \glsxtrshort{5p} dataset. Irrespective of the dataset, below \glsxtrshort{aen}=2\,mas and \glsxtrshort{ruwe}=1.4 the range of possible $\alpha_{\rm \glsxtrshort{mse}}$ is significantly broader. Some values are as low as 10\,$\mu$as, thus dominated by noise and insignificant.

\subsubsection{The PMa's astrometric signature}
\label{sec:alpha_pma}
Similar to $\alpha_{\rm \glsxtrshort{mse}}$, a \glsxtrshort{pma}'s astrometric signature can determine the excess that has to be quadratically added to \glsxtrshort{pma}$_{\rm single}$ to recover the \glsxtrshort{pma} of the given source measured in~\citet{Kervella2022}. Indeed, for a given orbit, the \glsxtrshort{pma} should be given by a constant vector ($c$), that is, the \say{noiseless} orbital contribution, plus a stochastic vector ($\xi$), that is, the pure noise contribution. The expectation value of the square-norm of the \glsxtrshort{pma} is thus 
\begin{align}
    \left< \Vert{\bf \rm \glsxtrshort{pma}}\Vert^2 \right> &= \left< \Vert {\bf c} + {\bf \xi} \Vert^2 \right>  \nonumber \\
    &= \Vert {\bf c} \Vert^2  + \left< \Vert {\bf \xi} \Vert^2 \right> 
\end{align}
since $ \left< {\bf \xi} \right>$=0 and ${\bf c}$ is constant. The first term is the pure orbital contribution to the \glsxtrshort{pma}, null if there are no orbital motion, that is, the astrometric signature that we seek. The second rightmost term is the squared-norm of \glsxtrshort{pma} for a single star, or ${\rm \glsxtrshort{pma}}_{\rm single}^2$. Therefore, we introduce $\alpha_{\rm \glsxtrshort{pma}}$ the \glsxtrshort{pma}'s astrometric signature, formally defined as
\begin{align}\label{eq:PMa_sig}
    \alpha_{\rm \glsxtrshort{pma}} = \sqrt{ {\rm \glsxtrshort{pma}}^2 - {\rm \glsxtrshort{pma}}_{\rm single}^2}
\end{align}

\noindent
It is theoretically possible to determine the typical level of \glsxtrshort{pma}$_{\rm single}$. However, conversely to the \glsxtrshort{mse}, there are no theoretical formula for estimating this distribution. We thus needed to perform simulations to estimate its mean and standard deviation, depending on the \glsxtrshort{ra}, \glsxtrshort{dec}, $Bp$-$Rp$ and $G$ of the sources. We determined the typical distribution followed by \glsxtrshort{pma}$_{\rm single}$ in Sect.~\ref{sec:significance_PMa}.

\subsection{The significance of the astrometric signatures}
\label{sec:significance_alpha}
\subsubsection{Significance of \texorpdfstring{$\alpha_{\rm UEVA}$}{the residuals' astrometric signature}}

\begin{figure}[hbt]
    \centering
    \includegraphics[width=84mm,clip=true]{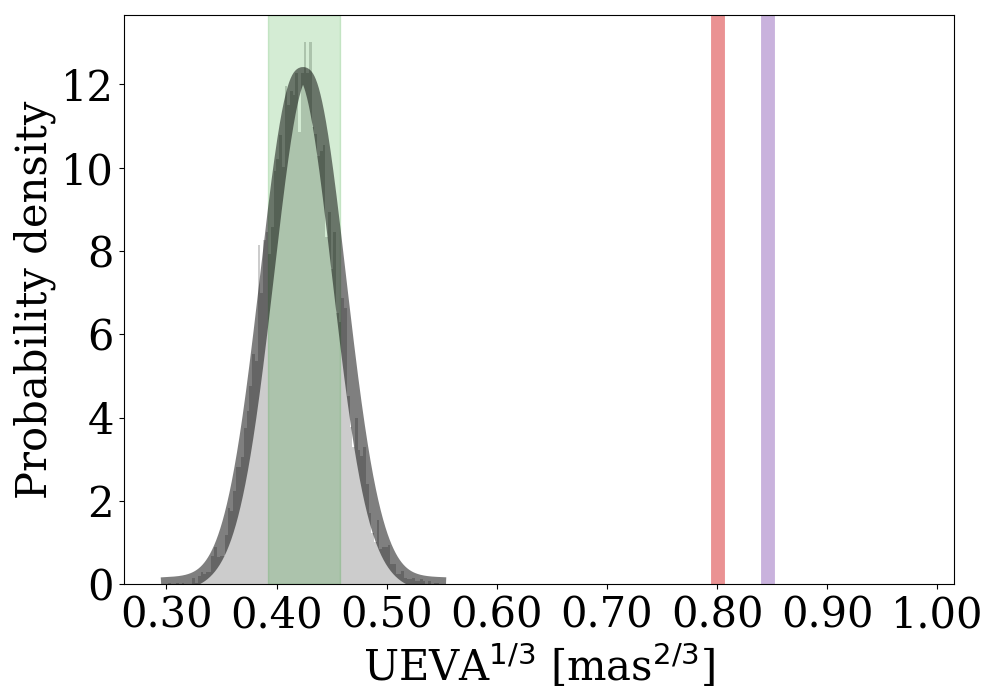}
    \caption{Distribution of \glsxtrshort{mse}$^{1/3}$ expected for a single star applied on the case of HD\,114762. The green area shows the region spanned by the median plus or minus the standard deviation. The \glsxtrshort{mse}$^{1/3}$ estimates from the \glsxtrshort{aen} and \glsxtrshort{ruwe} published in the \glsxtrshort{g3} archive are shown in red and purple, respectively. The thick black line shows the normal model derived from Eqs.~\ref{eq:mse_single} and~\ref{eq:mse_single_sigma}. All values of noises used in the models are given in Table~\ref{tab:examples_param}.}
    \label{fig:intrinsic_MSE}
\end{figure}

As introduced in Sect.~\ref{sec:alpha_mse}, the \glsxtrshort{mse}$_{\rm single}$ could be modeled by a normal distribution ${\mathcal N}(\mu_{\rm \glsxtrshort{mse},single},\sigma_{\rm \glsxtrshort{mse},single})$ whose parameters are written in Eqs.~\ref{eq:mse_single} and~\ref{eq:mse_single_sigma}. Rigorously speaking, as explained in Appendix~\ref{sec:details_chi2}, the \glsxtrshort{mse}$_{\rm single}$ is a linear combination of $\chi^2$ and normal distributions, with the main terms distributed according to the $\chi^2$ law. Under the prescription of \citet{Wilson1931} (see also \citealt{canal2005}), the cubic-root of the \glsxtrshort{mse} more closely resembles a normal distribution. We thus used \glsxtrshort{mse}$^{1/3}$ and assumed that it followed a normal distribution ${\mathcal N} \left(\mu_{1/3},\sigma_{1/3}\right)$. We approximated its parameters by $\mu_{1/3}$=$\mu^{1/3}$ and $\sigma_{1/3}$=$ \sigma\,\mu^{-2/3} /3$ by applying error propagation. 
Figure~\ref{fig:intrinsic_MSE} shows the resulting distribution of 100,000 simulations of \glsxtrshort{mse}$^{1/3}$ obtained when assuming that a source -- here, for example, HD\,114762 -- is a single star, and compares it to the normal model. 

The significance of $\alpha_{\rm \glsxtrshort{mse}}$ naturally corresponds to the $p$-value of the \glsxtrshort{mse}$^{1/3}$ as calculated from either the \glsxtrshort{aen} or the \glsxtrshort{ruwe} from the \glsxtrshort{g3}, within the ${\mathcal N} \left(\mu_{1/3},\sigma_{1/3}\right)$ single-star distribution. This $p$-value is converted to an $N$--$\sigma$ significance, following the \say{normal law} relationship between the 1-2-3-$\sigma$ levels and the 31.6-4.6-0.27\% $p$-values. We defined that $\alpha_{\rm \glsxtrshort{mse}}$ is significant at $N_\sigma$--$\sigma$ if $\chi^2_{\rm \glsxtrshort{fov}}$ $>$ \verb+chi2.ppf+($x_{N_\sigma}/100$,$N_{\rm \glsxtrshort{fov}}-5$) where \verb+chi2.ppf+ is the function of the python's \verb+scipy.stats+-module, and $x_{N_\sigma}$ corresponds to the usual percentage at $N_\sigma$--$\sigma$ ($N$=1: $x_{N_\sigma}$=68.3\%; $N$=2: $x_{N_\sigma}$=95.4\%; $N$=3: $x_{N_\sigma}$=99.73\%). An $\alpha_{\rm \glsxtrshort{mse}}$ significant at 2--$\sigma$ would thus imply that the \glsxtrshort{aen} or \glsxtrshort{ruwe} would have a less than 4.6\% chance to occur if the star was single.

\begin{figure}
    \centering
    \includegraphics[width=89.3mm]{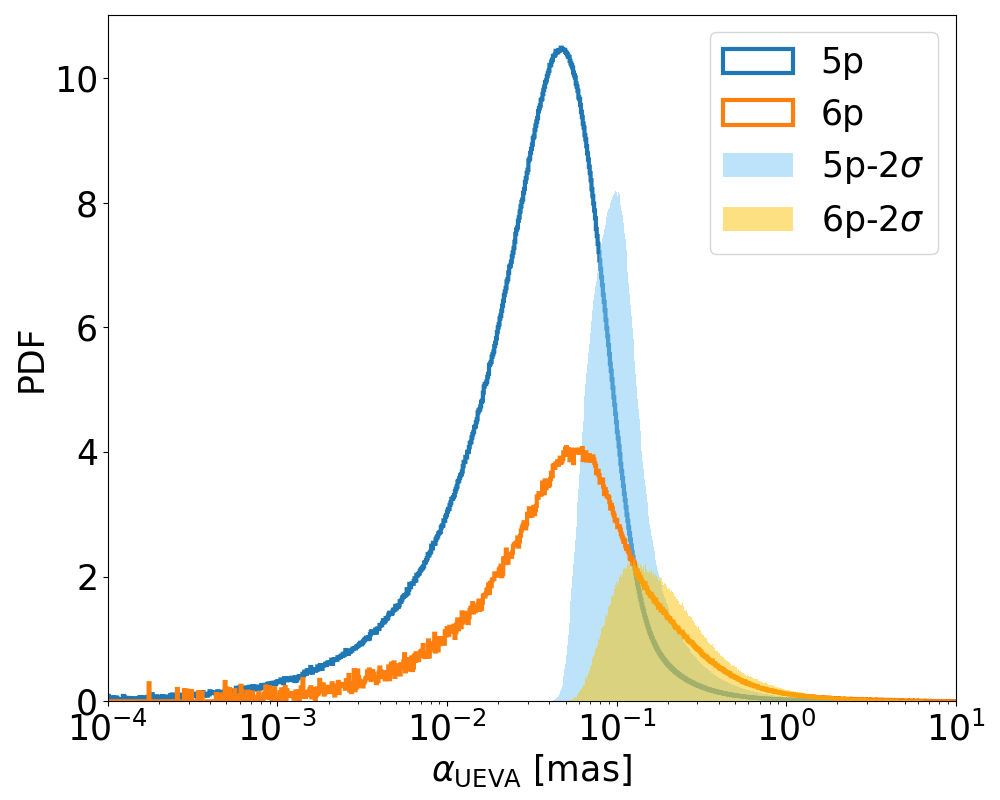}
    \caption{Probability density functions of the astrometric signature in the \glsxtrshort{5p} (blue line) and \glsxtrshort{6p} (orange line) datasets. The colored histograms show the distribution of $\alpha_{\rm \glsxtrshort{mse}}$ whose significance is greater than 2$\sigma$, respectively blue and golden for the \glsxtrshort{5p} and \glsxtrshort{6p} datasets.}
    \label{fig:astrosig_distribution}
\end{figure}

We showed in Fig.~\ref{fig:astrosig_distribution} the distribution of $\alpha_{\rm \glsxtrshort{mse}}$ among all datasets. Its full range goes from about 0.1$\mu$as to 10\,mas, but the $\alpha_{\rm \glsxtrshort{mse}}$ with a significance $>$2--$\sigma$ rather span the range that is beyond 10\,$\mu$as. 

\begin{figure}[hbt]
    \includegraphics[width=89.3mm,clip=true,trim=10 0 10 0]{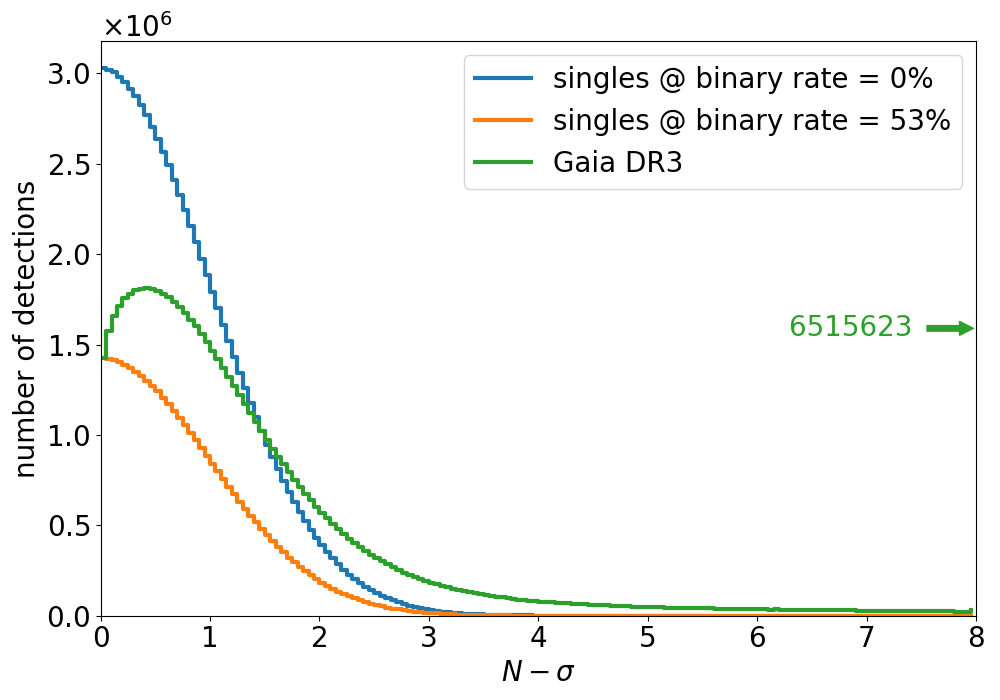}
    \caption{Number of detections per bin of significance among the 76 million sources (green line) compared to the expected numbers for single stars if the global binary rate among all \glsxtrshort{g3} sources is 0\% (blue line) and 53\% (orange line). The rightward arrow shows the number of sources with significance levels greater than greater than $8\sigma$.}
    \label{fig:detection_sigmas}
\end{figure}

The sample with an $\alpha_{\rm \glsxtrshort{mse}}$ significance larger than 2--$\sigma$ contains 19.1\,$\times$\,10$^6$ sources, that is, about 25\% of the sample of 75.9\,$\times$\,10$^6$ \textit{Gaia} sources with $G$$<$16 and $\varpi$$>$0. Close to 9\% of the sources (6.7\,$\times$\,10$^6$) have a significance larger than\footnote{The numerical calculation of the $N_\sigma$ significance for a given $\alpha_{\rm \glsxtrshort{mse}}$ involves calculating a p-value from a normal law, which is numerically limited to $N_\sigma$=8. Beyond this level, the $p$-value is thus 0 and $N_\sigma$=$\infty$.} 8--$\sigma$.
Figure~\ref{fig:detection_sigmas} shows ${\rm d}f_{\rm detec}(N_\sigma)$ the number of detections per bin of $\left(N_\sigma,N_\sigma+{\rm d}N_\sigma\right)$--$\sigma$ significance. They are compared to the theoretical distribution for only single stars, that is, ${\rm d}f_{\rm single}$=$N_{\rm single}\times \exp(-N_\sigma^2/2)\,{\rm d}N_\sigma/\sqrt{2\pi}$. The number of single stars, $N_{\rm single}$, is determined with respect to an assumed binary (and multiple) rate in the sample, $\Gamma_b$=$N_{\rm binary}/N_{\rm sample}$, using $N_{\rm single}$=$(1-\Gamma_b)\,\times\,N_{\rm sample}$. Assuming a $\Gamma_b$ of 0\% (blue curve in Fig.~\ref{fig:detection_sigmas}), the single star distribution cannot explain the rate of detections beyond 1.5--$\sigma$. Moreover, since many systems must be non-single, considering 100\% of single-stars in the sample obviously overestimates the number of detection below 1.5--$\sigma$. A more realistic value for $\Gamma_b$ can be found by assuming that all the sources at 0--$\sigma$ significance must be single. We then fixed $N_{\rm single}$ in such a way that ${\rm d}f_{\rm single}(N_\sigma$=$0)$ matches ${\rm d}f_{\rm detec}(N_\sigma$=$0)$. As illustrated with an orange line in Fig.~\ref{fig:detection_sigmas}, we found a single-star rate of 47\% and thus a $\Gamma_b$ of 53\% in this sample of sources brighter than $G$=16. The pollution from \glsxtrlong{fp} (\glsxtrshort{fp}) clearly equates/dominates over true positives below 2--$\sigma$. Beyond 2--$\sigma$, with a single-star rate of conservatively 47--100\%, we estimate that about 9--19\% of the selected binary or planetary systems could be single-star false positives. Thus, more than 80\% of the selected sample beyond 2--$\sigma$ significance are bona-fide binary, multiple or planetary systems. Beyond 3--$\sigma$ the sample reaches 16\% of the 76 millions sources, leading to single-star false positives rate of 0.9--2\%. Thus, virtually all $>$3--$\sigma$ sources are true binary, multiple or planetary systems, but about 14--24\% of the non-single star sample is lost compared to using the 2--$\sigma$ threshold. In the perspective of identifying binary, multiple or planetary systems candidates, we thus recommend, and adopt, that an $\alpha_{\rm \glsxtrshort{mse}}$, whose significance is above the 2$\sigma$ level, has to be considered as a strong candidate, with a 9--19\% chance that it actually is a single-star \glsxtrshort{fp}.

We use HD\,114762 as an illustration for the significance of astrometric signature. This system was known for a long time for hosting a candidate planet HD\,114762\,b with an $m\sin i$ of 11\,M$_{\rm J}$~\citep{Latham1989,Kane2011}. Confirming suspicion~\citep{Cochran1991,Hale1995,Halbwachs2000}, it was further shown using the astrometry from \textit{Gaia} DR1, then \glsxtrshort{g3}, that HD\,114762\,b was actually an M--dwarf~\citep{Kiefer2019a,Winn2022,Arenou2023}. This was first shown using the significant value of \glsxtrshort{aen} of 1.09\,mas~\citep{Kiefer2019a}; then a proper astrometric orbit solution was obtained using the private \glsxtrshort{g3} timeseries~\citep{Arenou2023}. In the light of the new framework introduced above, we found indeed that its \glsxtrshort{g3} residuals astrometric signature rejected the single star hypothesis.

If HD\,114762 was a single star -- or orbited by an undetectable companion -- the \textit{Gaia} measurements would lead to \glsxtrshort{aen}$_{\rm single}$=0.251$\pm$0.035\,mas, \glsxtrshort{ruwe}$_{\rm single}$=1.11$\pm$0.13, and \glsxtrshort{mse}$_{\rm single}$=0.076$\pm$0.018\,mas$^{2}$. 
In comparison, using the values of \glsxtrshort{aen}=0.708\,mas and \glsxtrshort{ruwe}=3.16 found in the \glsxtrshort{g3} for HD\,114762, we determined that \glsxtrshort{mse}$_{\rm aen}$=0.514\,mas$^2$ and \glsxtrshort{mse}$_{\rm ruwe}$=0.607\,mas$^2$. 
The astrometric signature deduced in both cases by applying Eq.~\ref{eq:astro_sig} are $\alpha_{\rm astro, aen}$=0.662\,mas and $\alpha_{\rm astro, ruwe}$=0.729\,mas. The $p$-value of \glsxtrshort{mse}$_{\rm ruwe}^{1/3}$ and \glsxtrshort{mse}$_{\rm aen}^{1/3}$ in the distribution of \glsxtrshort{mse}$_{\rm single}^{1/3}$ corresponds to a significance $>$9--$\sigma$ (see Fig.~\ref{fig:intrinsic_MSE}). A single star may explain the \glsxtrshort{aen} and the \glsxtrshort{ruwe} in close to 0\% of the simulations. Both the \glsxtrshort{aen} and the \glsxtrshort{ruwe} thus indicate the presence of a companion around HD\,114762. 

Assuming that the 84-day companion is responsible for this astrometric signature and that the star's semi-major axis $a_\star$$\sim$$\alpha_{\rm \glsxtrshort{mse}}/\varpi$ with the parallax $\varpi$=26\,mas, and crudely applying $M_c= a_\star \left(M_\star/{\rm M}_\odot\right)^{2/3} \left(P/{\rm year}\right)^{-2/3}$ with $M_\star$=1.05\,M$_\odot$~\citep{Winn2022}, we find a possible mass of the companion of 86\,M$_{\rm J}$. The mass obtained is indeed much larger than 11\,M$_{\rm J}$ but is less than its most recent estimation $\sim$0.3\,M$_\odot$~\citep{Winn2022}. This just shows that $\alpha_{\rm \glsxtrshort{mse}}$ is not a measure of $a_\star$, leading us to develop a more sophisticated framework for interpreting $\alpha_{\rm \glsxtrshort{mse}}$ and infer main parameters of companions, as explained in Sect.~\ref{sec:PMEX}.

\subsubsection{Significance of \texorpdfstring{$\alpha_{\rm PMa}$}{the PMa's astrometric signature} }
\label{sec:significance_PMa}

The significance of $\alpha_{\rm \glsxtrshort{pma}}$ corresponds to the $p$-value of the \glsxtrshort{pma}$^{2/3}$ within the distribution of \glsxtrshort{pma}$^{2/3}_{\rm single}$. This $p$-value is converted to an $N$--$\sigma$ significance (see Sect.~\ref{sec:significance_alpha} for more details). As mentioned in Sect.~\ref{sec:alpha_pma}, the distribution of \glsxtrshort{pma}$^{2/3}_{\rm single}$ is not known a priori because it strongly depends on the time sampling of the astrometric observations further fitted by a five-parameter model to measure the \glsxtrshort{g3} proper motion, as well as the \glsxtrshort{hg} relative positions. Conversely to $\alpha_{\rm \glsxtrshort{mse}}$, determining the distributions of $\alpha_{\rm \glsxtrshort{pma}}$ thus requires performing many simulations of a single star observations given the main parameters, scan law, and noises of the given source.

We used the system of GJ\,832 to illustrate the significance of \glsxtrshort{pma} beyond the single star hypothesis. GJ\,832 is an M0V star at a distance of 5\,pc and with a mass of 0.48$\pm$0.05\,M$_\odot$. Its planetary system was discovered by~\citet{Bailey2009}, reporting one Jupiter-like planet with a period of 3416$\pm$131\,days and minimum mass of 0.64$\pm$0.06\,M$_{\rm J}$. A second Earth-like planet was proposed for detection~\citep{Wittenmyer2014} but finally identified as a stellar activity artifact~\citep{Gorrini2022}. The \glsxtrshort{sma} and minimum mass of GJ\,832\,b were further updated to 3.6$\pm$0.4\,au and minimum mass of 0.74$\pm$0.06\,M$_{\rm J}$~\citep{Gorrini2022}. This system is illustrative for us because it harbors one of the smallest mass planets leading to significant astrometric acceleration detected by combining \textsc{Hipparcos} and \textit{Gaia}, and astrometric excess noise in the \glsxtrshort{g3}.

\begin{figure}[hbt]
    \centering
    \includegraphics[width=80mm,clip=true]{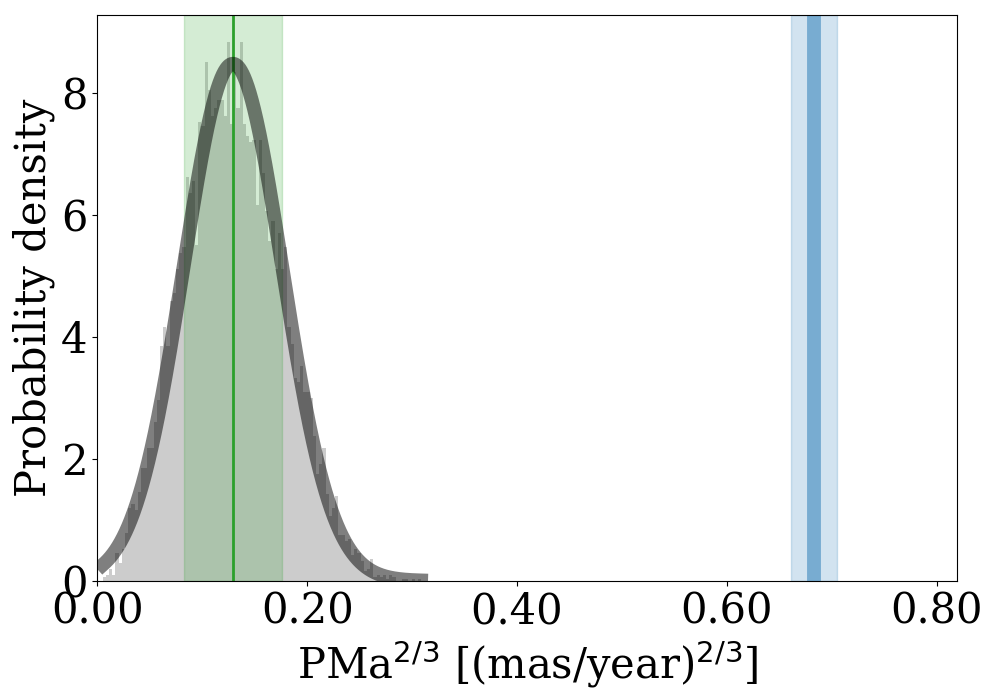}
    \caption{Distribution of \glsxtrshort{pma}$^{2/3}$ generated from noise only for the case of GJ\,832. The caption is the same as in Fig.~\ref{fig:intrinsic_PMa}. Here, the blue line and blue region show the PMa (at power 2/3) and its uncertainty taken from~\citet{Kervella2022}. The thick black line shows the normal model derived from the simulations themselves (see text for explanation).}
    \label{fig:intrinsic_PMa}
\end{figure}

Figure~\ref{fig:intrinsic_PMa} shows the \glsxtrshort{pma}$^{2/3}$ distribution obtained for GJ\,832 in the hypothesis that it is a single star. We used the values of noises, including the \textsc{Hipparcos} position error, that are given in Table~\ref{tab:examples_param}. 
It leads to \glsxtrshort{pma}$_{\rm single}$=0.060$\pm$0.032\,mas\,yr$^{-1}$
and \glsxtrshort{pma}$^{2/3}_{\rm single}$=0.153$\pm$0.054\,(mas\,yr$^{-1}$)$^{2/3}$, while \citet{Kervella2022} measures \glsxtrshort{pma}=0.565$\pm$0.027\,mas\,yr$^{-1}$, and equivalently \glsxtrshort{pma}$^{2/3}$=0.683$\pm$0.022\,(mas\,yr$^{-1}$)$^{2/3}$. If comparing to a zero-point \glsxtrshort{pma} offset of zero, that is, when neglecting noise, the \glsxtrshort{pma}$^{2/3}$ would have an apparent \glsxtrlong{snr} (\glsxtrshort{snr}) of 30. But the distribution of \glsxtrshort{pma}$^{2/3}_{\rm single}$, because of noise, strongly departs from zero as shown in Fig.~\ref{fig:intrinsic_PMa}. It leads to rather consider a positive zero-point offset and a larger error, that imply a more modest \glsxtrshort{snr} of 10 for the \glsxtrshort{pma} of GJ\,832. We thus stress that the noise brings a major contribution to the zero-point offsets and errors that are used to determinate the \glsxtrshort{snr}. For GJ\,832, the $p$-value of \glsxtrshort{pma}$^{2/3}$ in the distribution of \glsxtrshort{pma}$^{2/3}_{\rm single}$, implies a significance of the \glsxtrshort{pma} $>$9--$\sigma$ (see Fig.~\ref{fig:intrinsic_PMa}). Thus, it firmly indicates the presence of a companion in this system.  

Since the orbital period of GJ\,832\,b is $\sim$6\,yrs, that is, smaller than the \textit{Gaia}--\textsc{Hipparcos} baseline of 24.5\,years, we would tend to crudely interpret the \glsxtrshort{pma} here as the average orbital speed during \glsxtrshort{g3} observations. Assuming thus that ${\rm \glsxtrshort{pma}}/\varpi\propto\sqrt{G\,M_c^2/({\rm \glsxtrshort{sma}}\,M_\star})$, it leads to estimate the mass of GJ\,832\,b at 3.6\,au to $M_c\sim 0.1$\,M$_{\rm J}$. This mass is on the order of magnitude of the expected mass, though underestimated. The tool that we developed in Sect.~\ref{sec:PMEX} allowed us also to properly infer the main parameters of companions from the knowledge of \glsxtrshort{pma}.

\section{The GaiaPMEX method}
\label{sec:PMEX}
We aim at calculating the confidence regions of possible mass and sma of a companion, for a \textit{Gaia} source with a given $\alpha_{\rm \glsxtrshort{mse}}$ -- either determined from \glsxtrshort{aen} or \glsxtrshort{ruwe} -- and/or $\alpha_{\rm \glsxtrshort{pma}}$. To do so, rather than using a \glsxtrlong{mcmc} approach, which is time consuming, we perform a mass-\glsxtrshort{sma} grid search within a Bayesian framework. At each node of the mass-\glsxtrshort{sma} grid, as summarized in Fig.~\ref{fig:onenodescheme}, the values determined in the \glsxtrshort{g3} and in \textsc{Hipparcos}-\textit{Gaia}-(E)DR3 studies~\citep{Brandt2021,Kervella2022} are compared to modeled \glsxtrshort{pma} and \glsxtrshort{mse} (see Sect.~\ref{sec:simu}). We defined a likelihood function in Sect.~\ref{sec:likelihood}. The adopted Bayesian framework is explained in Sect.~\ref{sec:bayesian_inversion} and summarized in Fig.~\ref{fig:allnodescheme}. 

\subsection{A uniform grid of \texorpdfstring{$\log M_c$}{log(Mc)} and \texorpdfstring{$\log$\,sma}{log(sma)}}
\label{sec:grid}

\begin{figure*}[hbt]
    \centering
    \includegraphics[width=178.6mm,clip=true,trim=0 0 0 0]{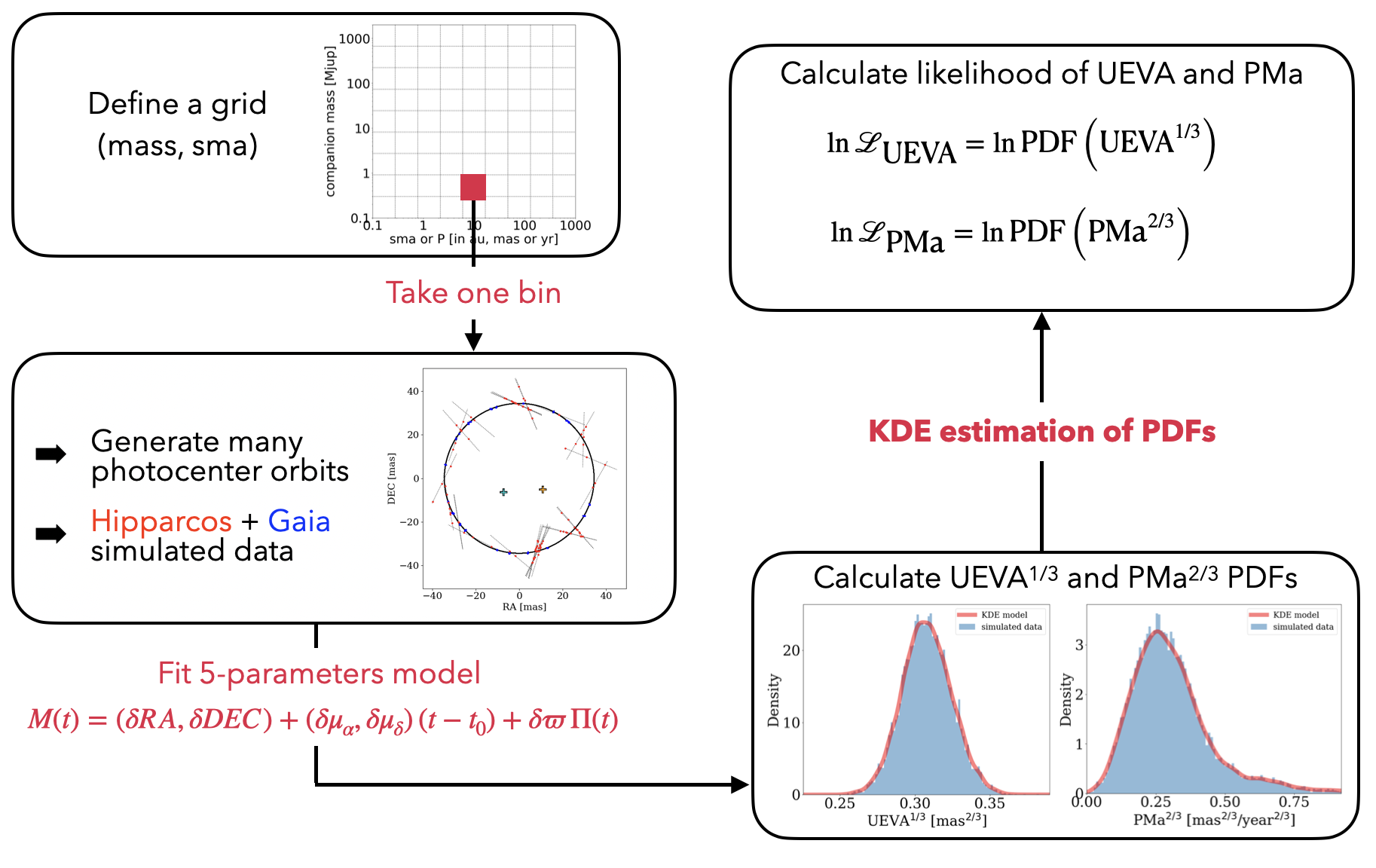}
    \caption{Summary sketch of the analysis performed on one single bin of the grid.}
    \label{fig:onenodescheme}
\end{figure*}

To probe different orders of magnitudes for the mass and \glsxtrshort{sma}, we define a uniform 2D-grid on $\log$$M_c$ and $\log$\,\glsxtrshort{sma} with log-scaled $\Delta$mass and $\Delta$\glsxtrshort{sma}, as sketched in Fig.~\ref{fig:onenodescheme}. In each bin, we draw $\log$$M_c$ and $\log$\,\glsxtrshort{sma} within a uniform distribution bounded by the bin edges. We also draw Keplerian parameters of the possible companion orbits, namely $e$ the eccentricity, $\omega$ the periastron longitude, $\Omega$ the longitude of ascending node, $\phi$ the phase, and $I_c$ the orbit inclination, following the distributions summarized in Table~\ref{tab:distro}. Parallax, $\varpi$, and stellar mass, $M_\star$, are drawn from normal distribution defined by prior knowledge on those parameters. The parallax is taken from the \glsxtrshort{g3} catalog. 
By default, if $M_\star$ is not given as input, it is first searched in the \glsxtrshort{g3} \glsxtrlong{cu8} (\glsxtrshort{cu8}) catalog~\citep{Creevey2023}. We specifically looked for the \glsxtrlong{flame} (or \glsxtrshort{flame}) available for more than 140 millions sources with $M_\star$$>$$0.5$\,M$_\odot$, the \verb+mass-Flame+ that is determined from combining photometry, parallax and stellar models. Complying with studies of stellar masses reported in~\citet{Arenou2023} and~\citet{Babusiaux2023}, as well as the recommendations in the \textit{Gaia} documentation\footnote{\url{https://gea.esac.esa.int/archive/documentation/GDR3/pdf/GaiaDR3_documentation_1.3.pdf}} and more specifically the \say{astrophysical parameters} section\footnote{\url{https://gea.esac.esa.int/archive/documentation/GDR3/Gaia_archive/chap_datamodel/sec_dm_astrophysical_parameter_tables/ssec_dm_astrophysical_parameters.htm\#astrophysical_parameters-flags_flame}}, the uncertainty on the mass is assumed $\sim$10\% if $M_\star$$>$0.7\,M$_\odot$ and 0.1\,M$_\odot$ if the mass is $<$0.7\,M$_\odot$. If the star is a giant we only use the \glsxtrshort{cu8} stellar mass if it is within 1-2\,M$_\odot$, and we assume a 30\%-uncertainty. If missing in this catalog, the stellar mass is instead estimated from the two first letter and digit of the spectral type given in the SIMBAD database\footnote{\url{https://simbad.u-strasbg.fr/simbad/}} assuming that the star is on the main sequence. 

\begin{table}[hbt]
    \centering
    \caption{Distribution of parameters sampled at each tested bin of the mass-\glsxtrshort{sma} grid.}
    \label{tab:distro}
    \begin{tabular}{lcc}
        Parameter &  type & bounds or law \\
        \hline 
        $\log M_c$ & uniform &  $\log M_c$$\pm$$\Delta\log M_c$  \\
        $\log \text{\glsxtrshort{sma}}$ & uniform &  $\log \text{\glsxtrshort{sma}}$$\pm$$\Delta\log \text{\glsxtrshort{sma}}$ \\ 
        $e$ & uniform &  $0$--$0.9$ \\
        $\omega$ & uniform &  $0$--$\pi$ \\
        $\Omega$ & uniform &  $0$--$2\pi$ \\
        $\phi$ & uniform &  $0$--$1$ \\
        $I_c$ & uniform or $\sin i$ &  $0$--$\pi/2$ \\
        $\varpi$ & normal & $\mathcal N$(PLX,$\sigma_{\rm PLX}^2$)  \\
        $M_\star$ & normal & $\mathcal N$($M_\star$,$\sigma_{M_\star}^2$)  \\
        \hline
    \end{tabular}
\end{table}

\subsection{Modeling of PMa and UEVA}
\label{sec:simu_PMa_UEVA}
In each bin of the (mass, \glsxtrshort{sma})-grid, we modeled by simulation many \textit{Gaia} observations of photocenter orbits, due to the reflex motion of the source due to a companion at given mass -- or $\log({\rm mass})$ -- and \glsxtrshort{sma} -- or $\log({\rm \glsxtrshort{sma}})$.\footnote{\glsxtrshort{sma} can be replaced by $P$ the orbital period, as one can choose to probe either \glsxtrshort{sma} or $P$. Moreover, the \glsxtrshort{sma} can be expressed in au (by default) or in mas (in which case \glsxtrshort{sma} stands for angular separation).} For each modeled observation, we added noise in \textit{Gaia} data (see Sects.~\ref{sec:AEN} and~\ref{sec:formal_error}), and then performed a five-parameter fit. Technical details on this modeling by simulation are explained in Sect.~\ref{sec:simu_orbit}. Each simulation leaded to a value of \glsxtrshort{pma} and a value of \glsxtrshort{mse}.

For each bin of the mass-\glsxtrshort{sma} grid drawn in Sect.~\ref{sec:grid}, the full set of simulations obtained at a given ($\log$$M_c$,$\log$\,\glsxtrshort{sma})-bin leaded to distributions of possible \glsxtrshort{mse} and \glsxtrshort{pma} that would be measured in the \glsxtrshort{g3} if the companion had such a mass and \glsxtrshort{sma}. We found that at least 100 simulations per bin are necessary to lead to a fine quality map. Running 300 simulations per bin performed better, leading to cleaner noiseless maps, with a computation time that was still tractable, though 3$\times$ longer.\footnote{Beyond 300 simulations per bin, run times increased exponentially.} The maps that are shown here were obtained with 100 simulations per bin.

\subsection{A Bayesian scheme for the 2D posterior distribution}
\label{sec:bayesian_scheme}

From the modeled distributions, we determined a likelihood of the actual values of \glsxtrshort{mse} -- either determined from \glsxtrshort{aen} or \glsxtrshort{ruwe} -- and \glsxtrshort{pma} given the (mass,\glsxtrshort{sma})$_n$-model ${\mathcal L} = p({\rm \glsxtrshort{pma},\glsxtrshort{mse}} | {\rm mass}_n, {\rm \glsxtrshort{sma}}_n)$. It is derived in Sect.~\ref{sec:likelihood} below. This likelihood was further used for a Bayesian inversion to determine the posterior distributions on mass and \glsxtrshort{sma} of the hypothetical companion, as explained in Sect.~\ref{sec:bayesian_inversion}.

\subsubsection{A log-likelihood of PMa and UEVA}
\label{sec:likelihood}
\begin{figure}[hbt]
    \centering
    \includegraphics[height=34mm]{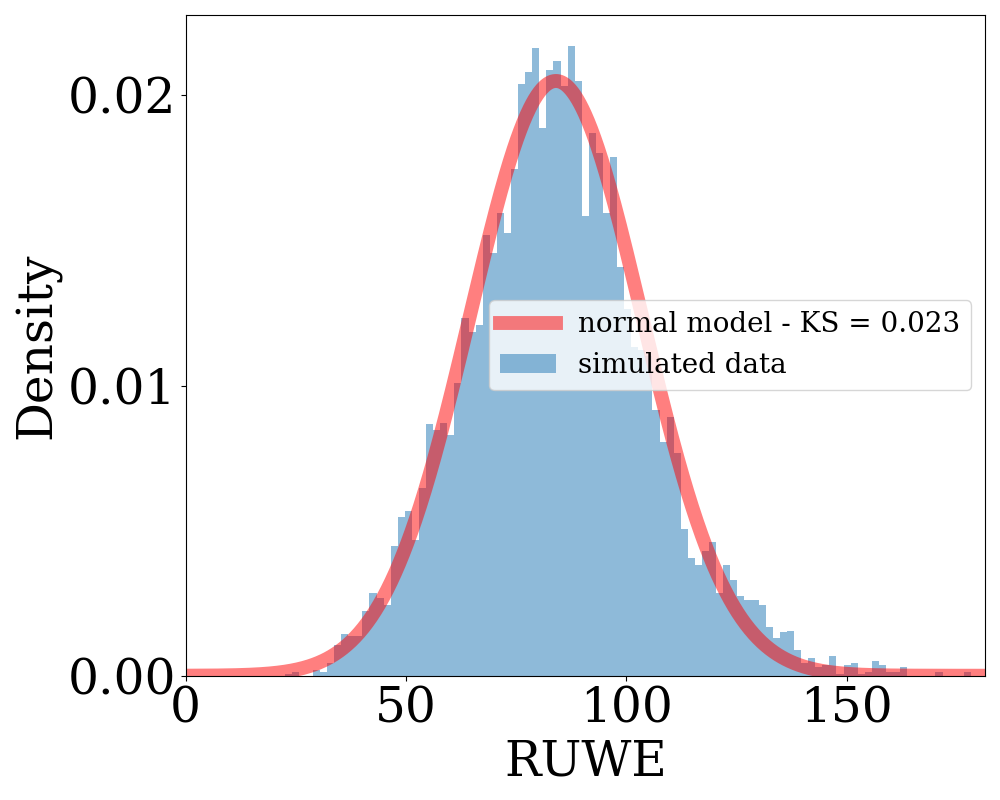}
    \includegraphics[height=34mm]{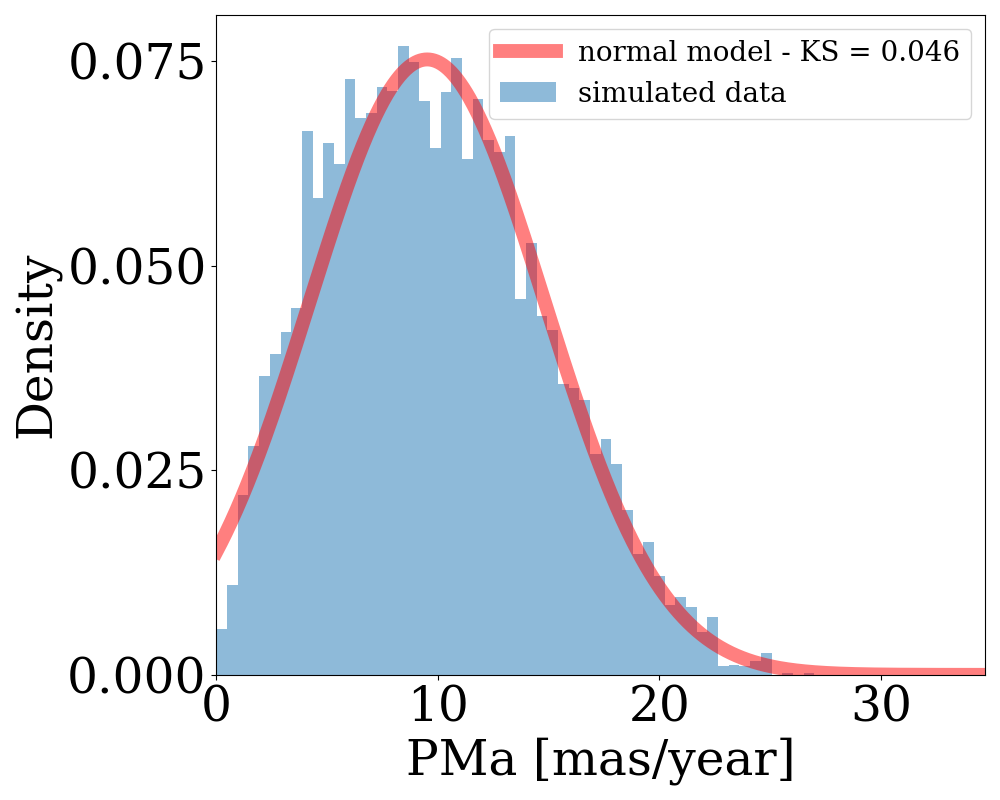}
    \includegraphics[height=34mm]{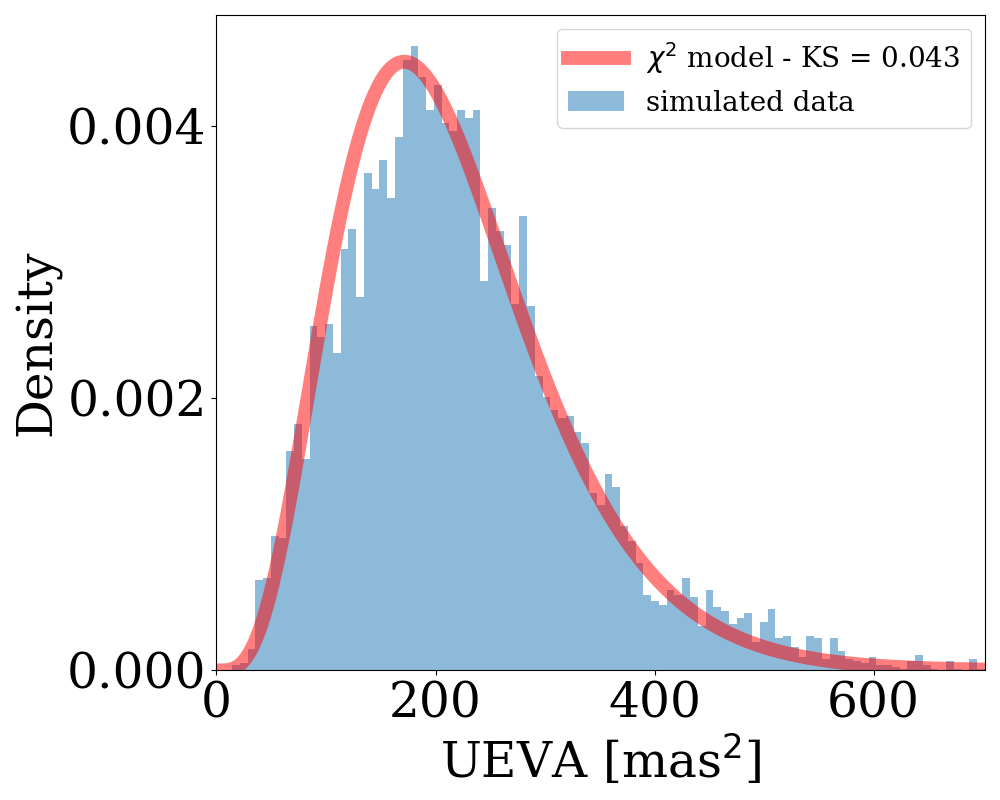}
    \includegraphics[height=34mm]{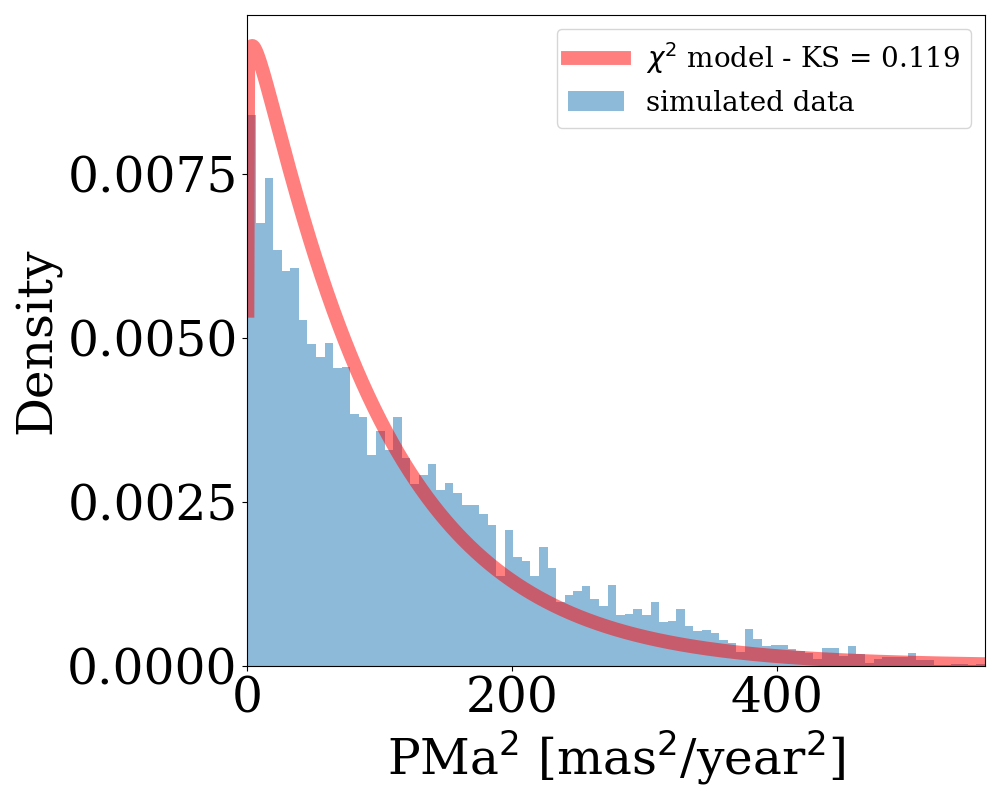}
    \includegraphics[height=34mm]{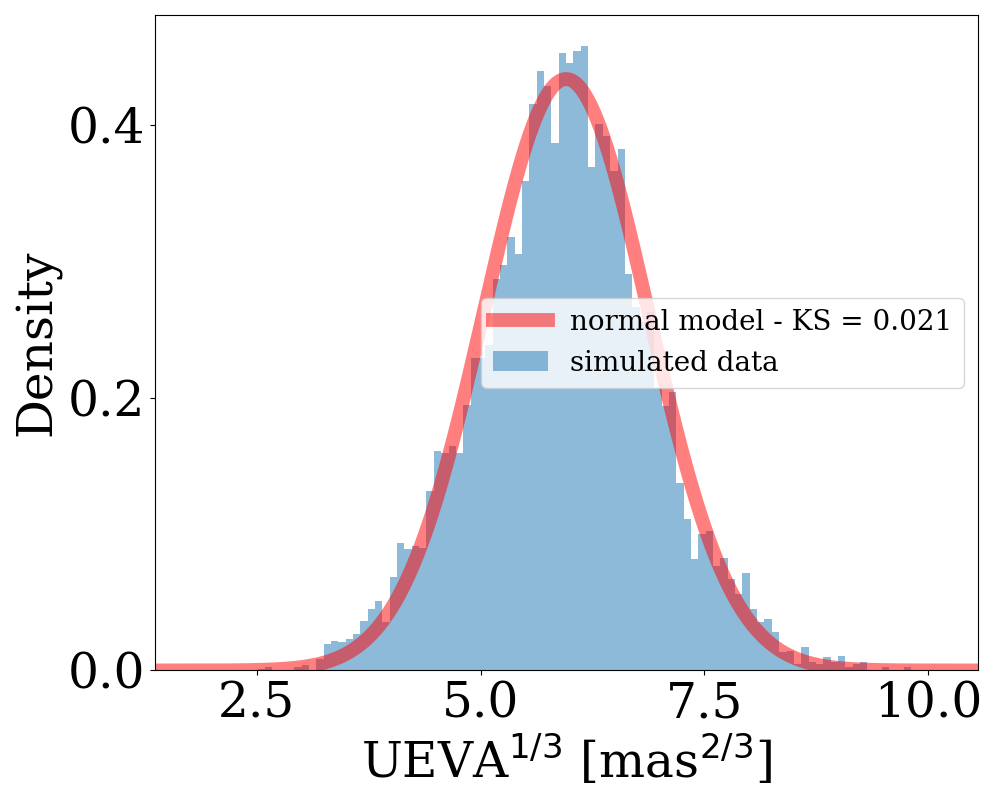}
    \includegraphics[height=34mm]{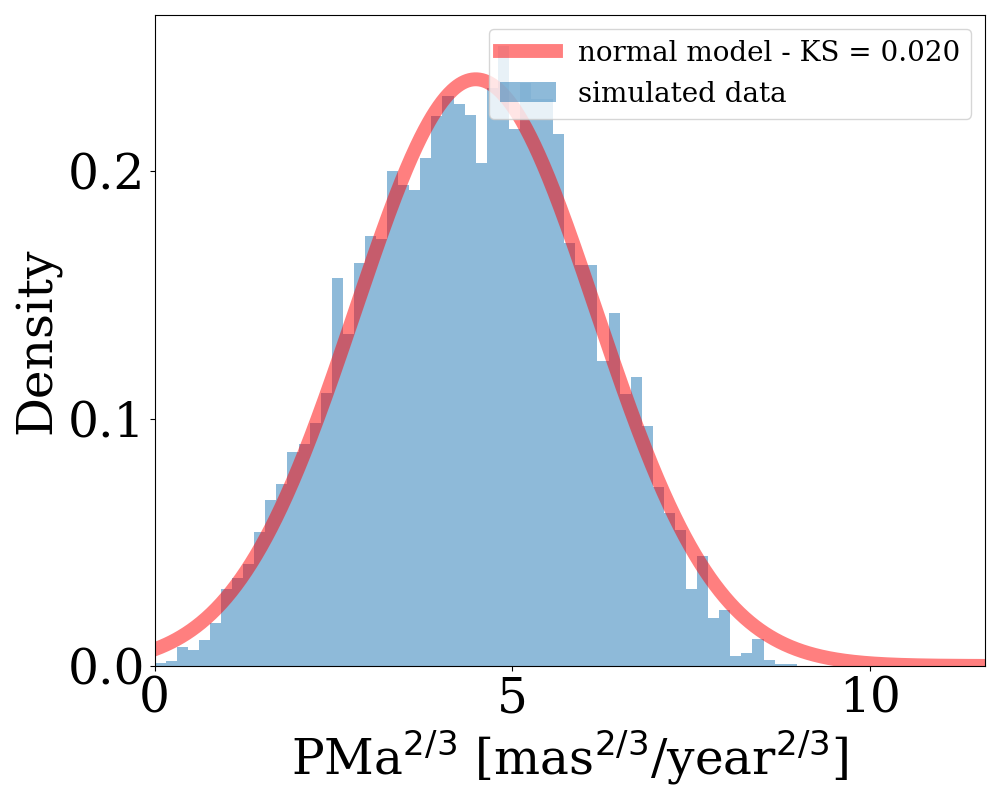}
    \caption{Probability density functions of \glsxtrshort{ruwe}, \glsxtrshort{mse}, \glsxtrshort{mse}$^{1/3}$, and \glsxtrshort{pma}, \glsxtrshort{pma}$^2$ and \glsxtrshort{pma}$^{2/3}$ are modeled for a companion with mass 100\,M$_{\rm J}$ and \glsxtrshort{sma}=1\,au around a system similar to GJ\,832 (i.e., an  M-type star at parallax of 200\,mas). Modeled data are shown in the blue histograms, and normal or $\chi^2$ laws are shown as red curves. The results of a Kolmogorov-Smirnov test statistics are shown in the legend.}
    \label{fig:normal_transform}
\end{figure}

One of the important issues met when defining a log-likelihood for \glsxtrshort{pma} and \glsxtrshort{mse}, was to determine their \glsxtrlong{pdf} (\glsxtrshort{pdf}) with $\ln{\mathcal L}$=$\ln\text{\glsxtrshort{pdf}}(\text{data})$. They typically do not follow a normal-law, since both quantities are always positive. We show in Sect.~\ref{sec:astro_sig} that, because the \glsxtrshort{pdf} of the squares \glsxtrshort{pma}$^2$ and \glsxtrshort{mse} are similar to $\chi^2$-distributions, their transformations to \glsxtrshort{pma}$^{2/3}$ and \glsxtrshort{mse}$^{1/3}$ follow close-to-normal laws~\citep{Wilson1931,canal2005}. This ensured that our data distributions had \glsxtrshort{pdf} that were more compact and symmetrical as shown in Fig.~\ref{fig:normal_transform}.

\begin{figure}[hbt]
    \centering
    \includegraphics[width=89.3mm]{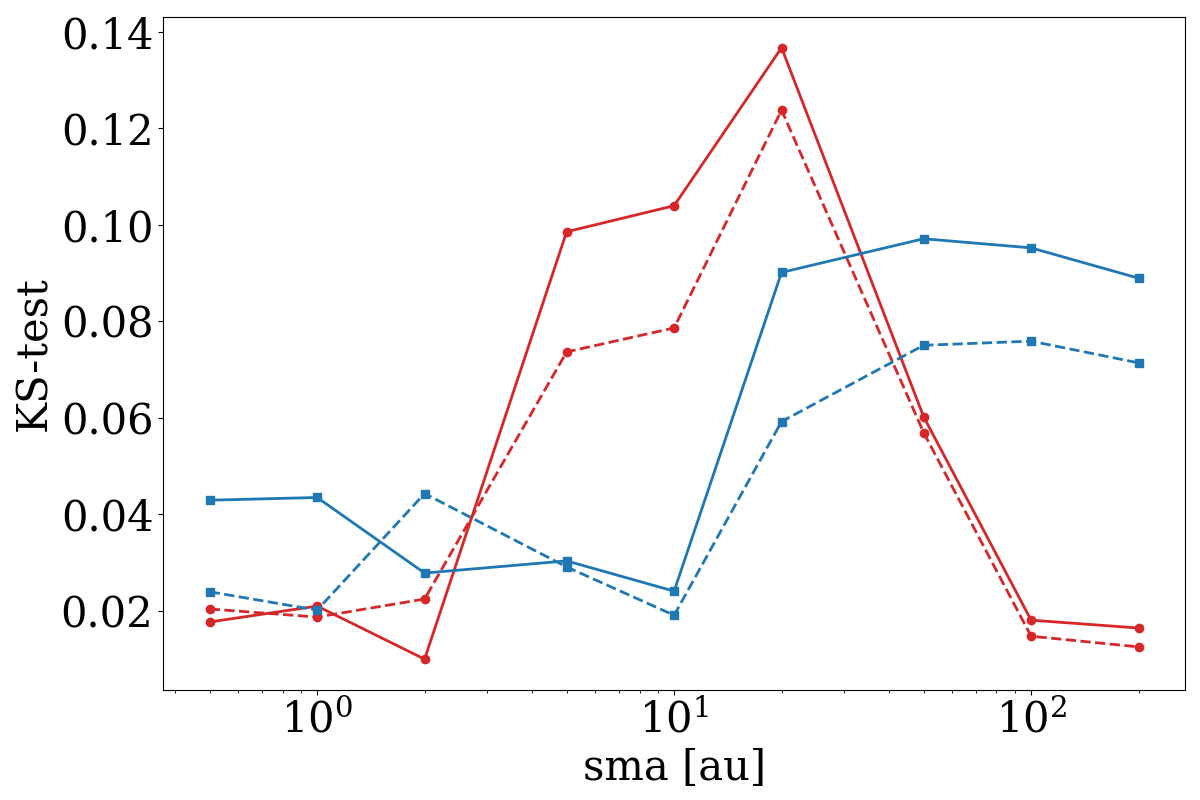}
    \caption{Kolmogorov-Smirnov test statistics of the normal law with the distributions of \glsxtrshort{mse} and \glsxtrshort{pma} (respectively red and blue solid lines) and \glsxtrshort{mse}$^{1/3}$ and \glsxtrshort{pma}$^{2/3}$ (respectively red and blue dashed lines). }
    \label{fig:KStest}
\end{figure}

Figure~\ref{fig:normal_transform} illustrates the typical differences in the distributions of \glsxtrshort{mse} and \glsxtrshort{mse}$^{1/3}$ as well as \glsxtrshort{pma}, \glsxtrshort{pma}$^2$ and \glsxtrshort{pma}$^{2/3}$, for the case of a hypothetical 100-M$_{\rm J}$ companion at 1\,au around GJ\,832. Figure~\ref{fig:KStest} also shows the difference in the KS-test of the normal law with the distributions of \glsxtrshort{mse} or \glsxtrshort{pma} and respectively those of \glsxtrshort{mse}$^{1/3}$ or \glsxtrshort{pma}$^{2/3}$ modeled at different values of \glsxtrshort{sma}. Generally for most \glsxtrshort{sma} from 0.5 to 100\,au, the $X^{1/3}$--transformation is a closer match to the normal law than simply \glsxtrshort{mse} or \glsxtrshort{pma}. We thus adopt using \glsxtrshort{mse}$^{1/3}$ and \glsxtrshort{pma}$^{2/3}$ for calculating the likelihoods.

For any (mass,\glsxtrshort{sma})-bin of the grid, a Gaussian kernel density estimation (with the \verb+gaussian_KDE+ library from \verb+scipy+) was performed on the distributions of the modeled \glsxtrshort{mse}$^{1/3}$ and \glsxtrshort{pma}$^{2/3}$. This gives a good approximation of the true \glsxtrshort{pdf} of those quantities, as long as the sampling is dense enough, which is why we preferred using distributions that are not too extended and long-tailed. They can be used to derive the log-likelihoods as:
\begin{align}
\ln {\mathcal L}_{\rm \glsxtrshort{mse}} &= \ln \left[ {\rm \glsxtrshort{pdf}}_{{\rm \glsxtrshort{mse}}^{1/3}}\left({\rm \glsxtrshort{mse}}_{\rm \glsxtrshort{ruwe}}^{1/3}\right) \right]\nonumber \\
\ln {\mathcal L}_{\rm \glsxtrshort{pma}} &=\ln  \left[ {\rm \glsxtrshort{pdf}}_{{\rm \glsxtrshort{pma}}^{2/3}}\left({\rm \glsxtrshort{pma}}_{\rm \glsxtrshort{hg}DR3}^{2/3}\right) \right]
\end{align}

\noindent
Those log-likelihood give the probability of the data given a certain (mass,\glsxtrshort{sma})-companion model and distributions of other Keplerian parameters. They are then used in \glsxtrshort{pmex} in the inversion of the Bayesian formula to obtain the probability of the models given the data, and confidence regions on mass and \glsxtrshort{sma}, as explained in Sect.~\ref{sec:bayesian_inversion}.


\subsubsection{Bayesian inversion}
\label{sec:bayesian_inversion}

Every bin $n$ of the grid has thus a likelihood  ${\mathcal L}_n$ and a corresponding log-likelihood $\ln\mathcal L_n$. The bin $n_{\rm max}$ at which $\log\mathcal L_n$ is maximized, reaching $\ln\mathcal L_{\rm max}$, is the (mass,\glsxtrshort{sma})-model for which the data are best matching the \glsxtrshort{mse} and \glsxtrshort{pma} distribution. At any other bin, we measured a \glsxtrlong{lr} (\glsxtrshort{lr}) through $\Delta\ln{\mathcal L_n}=\ln{\mathcal L_{\rm max}}-\ln{\mathcal L_n}$. 

\begin{figure*}
    \centering
    \includegraphics[width=178.6mm,clip=true]{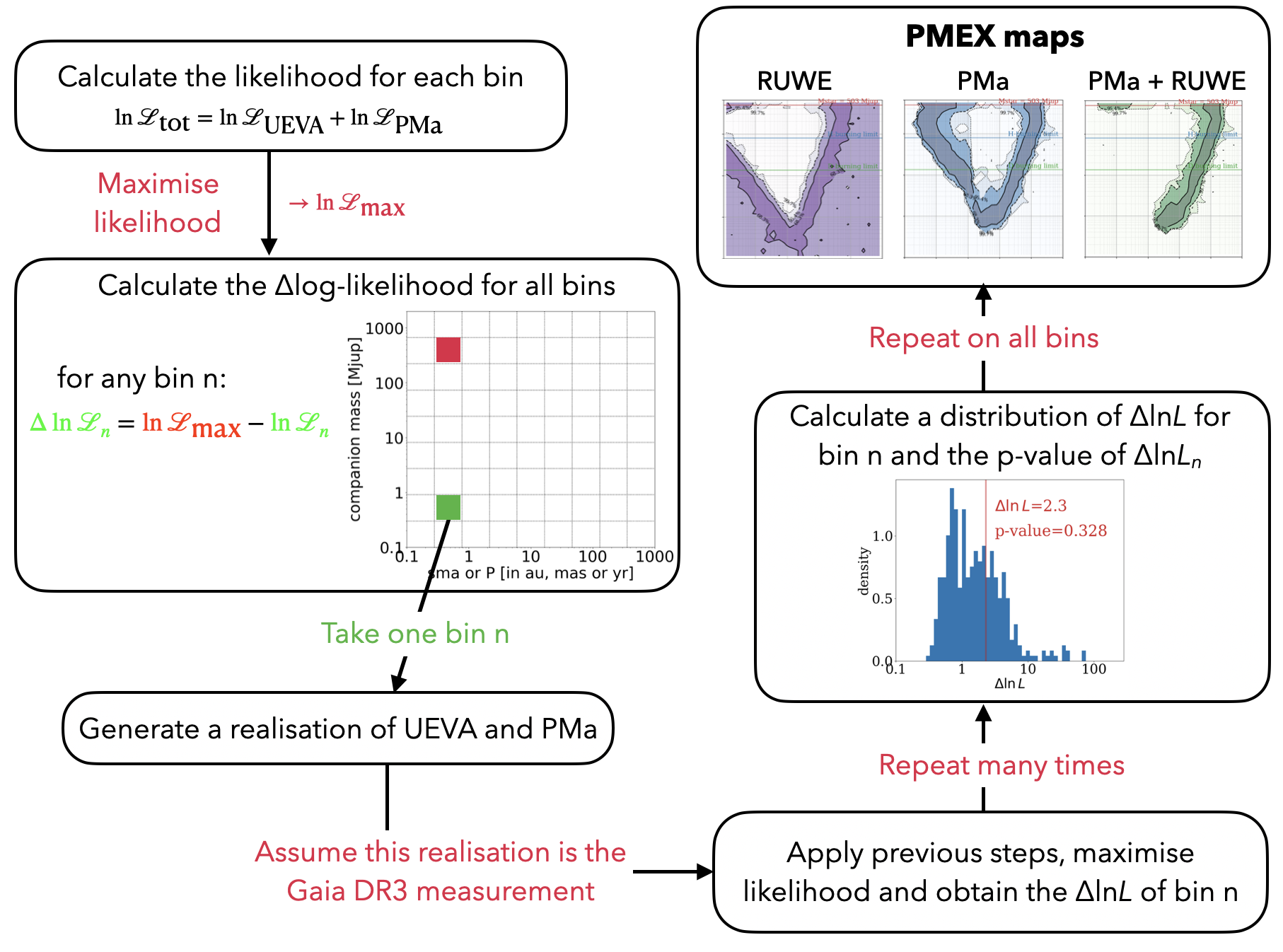}
    \caption{Summary sketch of the Bayesian analysis performed on all bins of the grid to recover the posterior probability function on mass and \glsxtrshort{sma}.}
    \label{fig:allnodescheme}
\end{figure*}

To derive a probability function for mass and \glsxtrshort{sma}, we needed to determine at each bin $n$, what is the p-value of $\Delta\ln{\mathcal L_n}$. This is summarized in Fig.~\ref{fig:allnodescheme}. In the ideal case of the likelihood-ratio test of some null hypothesis, the Wilks theorem states that with a large number of data, $2\,\Delta\ln{\mathcal L}=2\,\left(\text{max}\ln{\mathcal L}-\ln{\mathcal L}_{\rm null}\right)$ should follow a $\chi^2$ distribution with $k$ degrees of freedom~\citep{Wilks1938,Silvey1970}; $k$ being the difference between the maximum number of degree of freedom ($\text{\glsxtrshort{dof}}_{\rm max}=N_{\rm data}-N_{\rm param}$), and the number of degree of freedom in the region constrained by the null-hypothesis where some of the parameters are fixed ($\text{\glsxtrshort{dof}}_{\rm null}=N_{\rm data}-N_{\rm param,unfixed}$). Here, fixing the only 2 parameters to vary, the mass and the \glsxtrshort{sma}, we have  $k$=$\text{\glsxtrshort{dof}}_{\rm max}$-$\text{\glsxtrshort{dof}}_{\rm null}=2$. However, since the number of data is small ($N_{\rm data}$=2) the Wilks theorem does not apply. 

To convert the \glsxtrshort{lr} at bin $n$ into a p-value, we must find the empirical distribution of the \glsxtrshort{lr} at that bin.
This was done by assuming that the models (mass$\pm$$\Delta$mass, \glsxtrshort{sma}$\pm$$\Delta$\glsxtrshort{sma})$_n$ is the true one and draw many possible \glsxtrshort{mse} and \glsxtrshort{pma} from these models, as if they were those measured by \glsxtrshort{g3}. For each drawn \glsxtrshort{mse} \& \glsxtrshort{pma}, we apply the same grid search as explained in the above paragraph, finding the likelihood optimum. At the considered bin $n$, this led to a distribution of $\Delta$$\ln{\mathcal L}$. The corresponding percentile $p_n$ of $\Delta$$\ln{\mathcal L_n}$ within this distribution has a frequentist interpretation, as for $\Delta \chi^2$ inference on confidence limits~\citep{NumericalRecipes}. It is the confidence level with which a confidence region $r_n$ may contain the true (mass, \glsxtrshort{sma}) and whose boundary passes through model $n$\footnote{Said differently, if we were to retry multiple times the same observations at the same epochs of the same source with \textit{Gaia}, such confidence region $r_n$ would contain the true mass and \glsxtrshort{sma} in $100\, p_n\%$ of the cases.}. Details on how the $\ln{\mathcal L}$ is calculated for a given bin are explained in Sect.~\ref{sec:likelihood}. 

We thus produced the \glsxtrshort{pmex} constraints map that shows the shape of the posterior probability density functions on the Log-mass and Log-\glsxtrshort{sma} (or Log-period) with respect to the $\alpha_{\rm \glsxtrshort{mse}}$, \glsxtrshort{hg} \glsxtrshort{pma} (\citealt{Kervella2022}), or both combined. The combination of both observables is most important in enabling the exclusion of either small period or long period companions, as discussed in Sect.~\ref{sec:maps}.

\section{The GaiaPMEX maps}
\label{sec:maps}

\subsection{Constraints from AEN, RUWE, and PMa on the mass and sma of a companion}

\begin{figure*}[hbt]
    \centering    
    \setlength{\unitlength}{1mm}
    \begin{picture}(178.6,160)
    \put(0,75){\includegraphics[width=89.mm,clip=true]{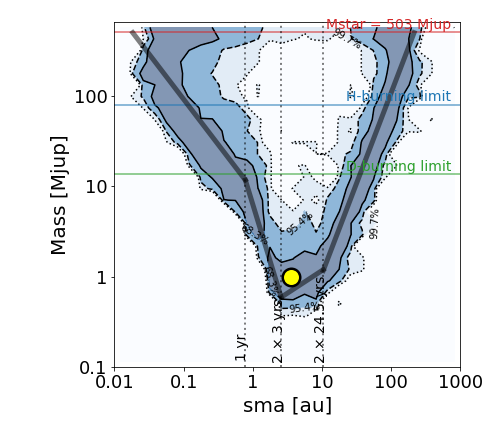}}
    \put(90,75){\includegraphics[width=89.mm,clip=true]{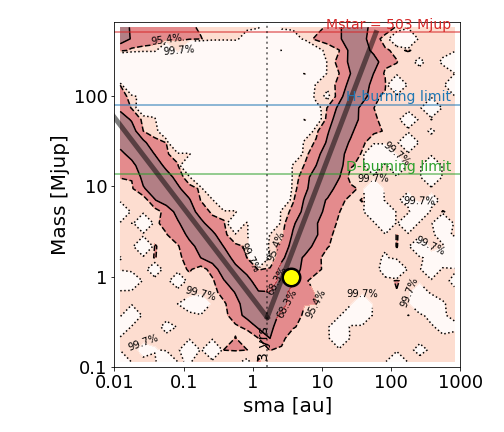}}
    \put(0,0){\includegraphics[width=89.mm,clip=true]{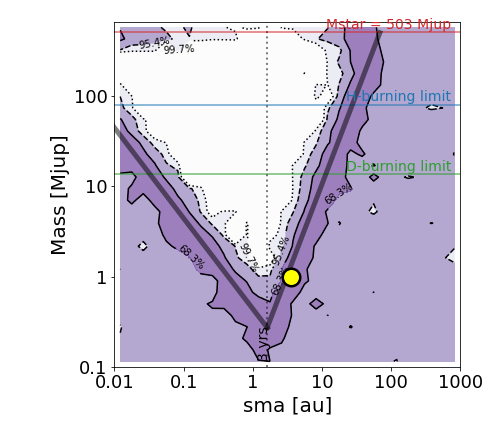}}
    \put(90,0){\includegraphics[width=89.mm,clip=true]{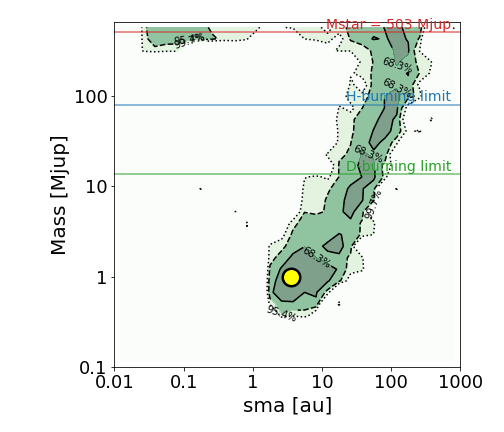}}
    \put(45,149){\glsxtrshort{pma}}
    \put(135,149){\glsxtrshort{aen}}
    \put(45,74){\glsxtrshort{ruwe}}
    \put(131,74){\glsxtrshort{pma}+\glsxtrshort{ruwe}}
    \end{picture}
    \caption{\glsxtrshort{pmex} constraints on mass and \glsxtrshort{sma} of a candidate companion around GJ\,832. Top-left: Using \glsxtrshort{pma}. Top-right: Using $\alpha_{\rm \glsxtrshort{mse}}$ from \glsxtrshort{aen}. Bottom-left: Using $\alpha_{\rm \glsxtrshort{mse}}$ from \glsxtrshort{ruwe}. Bottom-right: Combining \glsxtrshort{pma} and $\alpha_{\rm \glsxtrshort{mse}}$. The colored regions inside the contours show the 68.3\%, 95.4\%, and 99.73\% confidence intervals. The thick dark lines show the  model relationships developed in Sects.~\ref{sec:PMa_curve} and~\ref{sec:AEN_curve}. The yellow dot shows the properties of the known exoplanet companion in this system derived from \glsxtrshort{rv}, ${\rm \glsxtrshort{sma}}_b$=3.6\,au and $M_b$=0.99\,M$_{\rm J}$~\citep{Philipot2023b}.}
    \label{fig:GJ832_PMEX}
\end{figure*}

\begin{figure*}[hbt]
    \centering
    \setlength{\unitlength}{1mm}
    \begin{picture}(178.6,160)
    \put(0,75){\includegraphics[width=89.mm,clip=true]{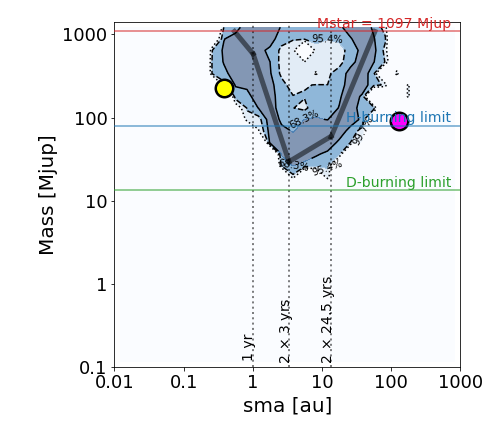}}
    \put(90,75){\includegraphics[width=89.mm,clip=true]{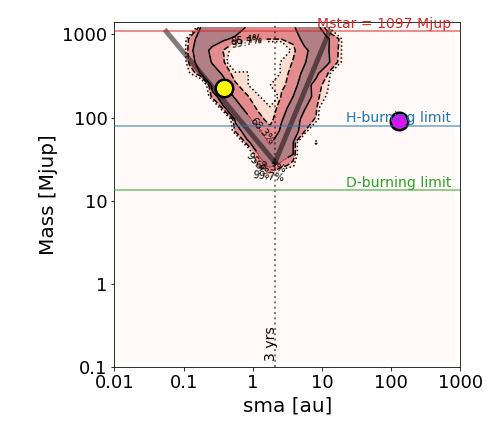}}
    \put(0,0){\includegraphics[width=89.mm,clip=true]{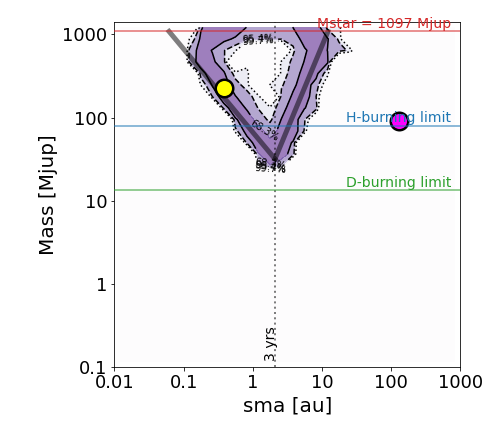}}
    \put(90,0){\includegraphics[width=89.mm,clip=true]{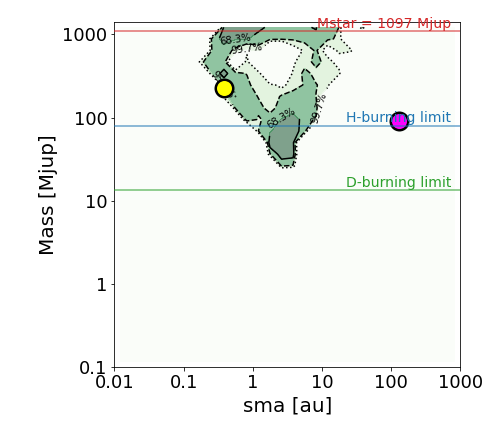}}
    \put(45,149){\glsxtrshort{pma}}
    \put(135,149){\glsxtrshort{aen}}
    \put(45,74){\glsxtrshort{ruwe}}
    \put(131,74){\glsxtrshort{pma}+\glsxtrshort{ruwe}}
    \end{picture}
    \caption{Same as Fig.~\ref{fig:GJ832_PMEX} for HD\,114762. The yellow dot shows the properties of the known massive companion HD\,114762\,Ab in this system, with ${\rm \glsxtrshort{sma}}_b$=0.35\,au and $M_b$=0.2\,M$_\odot$. The pink dot shows the wide binary companion HD\,114762\,B with ${\rm \glsxtrshort{sma}}_B$=130\,au and $M_B$=0.088\,M$_\odot$.}
    \label{fig:HD114762_PMEX}
\end{figure*}

We use the systems of GJ\,832 and HD\,114762 to illustrate the outcomes of the \glsxtrshort{pmex} approach. Details on HD\,114762 and GJ\,832 are given respectively in Sects.~\ref{sec:significance_alpha} and~\ref{sec:significance_PMa} and their main parameters are summarized in Table~\ref{tab:examples_param}. 
Figs.~\ref{fig:GJ832_PMEX} and~\ref{fig:HD114762_PMEX} show the confidence regions, or maps, at confidence levels 68.3 (1--$\sigma$ region), 95.4 (2--$\sigma$ region), 99.73 \% (3--$\sigma$ region), of the mass and \glsxtrshort{sma} of a candidate companion around GJ\,832 and HD\,114762 as calculated with \glsxtrshort{pmex}, given either only \glsxtrshort{ruwe} or \glsxtrshort{aen} -- through calculating \glsxtrshort{mse} as explained in Sect.~\ref{sec:alpha_mse} -- or only \glsxtrshort{pma}, or combining the constraints from \glsxtrshort{pma} and \glsxtrshort{ruwe}. 

The maps give the possible \glsxtrshort{sma} and mass of a companion assuming it is responsible of the measured \glsxtrshort{pma}, \glsxtrshort{aen} or \glsxtrshort{ruwe}. The mass--\glsxtrshort{sma} degeneracy drawn by the confidence regions follows typical curves, essentially U-shaped for the \glsxtrshort{pma}, and V-shaped for the \glsxtrshort{aen} or the \glsxtrshort{ruwe}. They present mainly three features: a short-period (SP) branch, a long-period (LP) branch, and the near equal-mass binary solutions forming an horizontal branch around the mass of the primary (B-branch hereafter). The SP and LP semi-linear branches curve up into the B-branch because of the contribution of the secondary in the position of the photocenter. The \glsxtrshort{pmex} combination of the constraints from \glsxtrshort{pma} and \glsxtrshort{ruwe}, as shown on the green map in Figs.~\ref{fig:GJ832_PMEX} and~\ref{fig:HD114762_PMEX}, combines the shape of the \glsxtrshort{pma} and \glsxtrshort{ruwe} maps. Most importantly it lifts up a fraction of the degeneracy, leaving confidence regions of smaller extent. 

In the case of GJ\,832, as shown in Fig.~\ref{fig:GJ832_PMEX}, the $\alpha_{\rm \glsxtrshort{mse}}({\rm AEN})$ maps form a confidence region at 2--$\sigma$ leading to a mass$>$0.1\,M$_{\rm J}$ with 95.4\% confidence and a mass upper-limits at any \glsxtrshort{sma}$<$10\,au. However, the \glsxtrshort{ruwe} is only leading to the 2--$\sigma$ upper-limit on mass at any \glsxtrshort{sma}. This is explained by the difference in the significance of the $\alpha_{\rm \glsxtrshort{mse}}$ from either \glsxtrshort{ruwe} (1.7--$\sigma$) or \glsxtrshort{aen} (2.5--$\sigma$). The \glsxtrshort{pma} maps lead to a well-defined 3--$\sigma$ confidence region with mass$>$0.5\,M$_{\rm J}$ at 99.7\% confidence and an upper-limit on mass within 1--40\,au. Both 1--$\sigma$ confidence regions of the $\alpha_{\rm \glsxtrshort{mse}}$ and \glsxtrshort{pma} maps encircles the minimum mass 0.74$\pm$0.06\,M$_{\rm J}$ at \glsxtrshort{sma}=3.6$\pm$0.4\,au for GJ\,832\,b~\citep{Gorrini2022}, as well as the mass of $0.99^{+0.09}_{-0.08}$\,M$_{\rm J}$ found by combining the \glsxtrshort{rv} and the \textsc{Hipparcos}--\textit{Gaia} proper motions of GJ\,832 in~\citet{Philipot2023b}. Nonetheless, the confidence regions still leave a lot of degeneracy in the \glsxtrshort{sma} and mass solutions. Combining the constraints from \glsxtrshort{pma} and \glsxtrshort{ruwe} leads to reject most of the SP-branch. The remaining LP-branch spans $\sim$2--500\,au with mass as low as 0.5\,M$_{\rm J}$. The 1--$\sigma$, or 68\%, confidence region is restricted to within the LP-branch with \glsxtrshort{sma}=2--8\,au and true mass within 0.5-2\,M$_{\rm J}$. Again, the 1--$\sigma$ confidence region encircles the known possible mass and \glsxtrshort{sma} of GJ\,832\,b. 

In the case of HD\,114762, as shown in Fig.~\ref{fig:HD114762_PMEX}, the confidence regions drawn from \glsxtrshort{aen}, \glsxtrshort{ruwe} and \glsxtrshort{pma} agreed well on a companion mass $>$20\,M$_{\rm J}$ for \glsxtrshort{sma} within 0.1-100\,au. The \glsxtrshort{ruwe} and \glsxtrshort{aen} even constrained \glsxtrshort{sma} to be lower than 20\,au. Combining \glsxtrshort{pma} and \glsxtrshort{ruwe} together led to still largely degenerated 2 and 3--$\sigma$ confidence regions. The 1--$\sigma$ region is now centered about 3\,au and mass=50\,M$_{\rm J}$, and the 2--$\sigma$ regions spans the stellar domain, embedding the known mass=225$\pm$14\,M$_{\rm J}$ at 0.38\,au of HD\,114762\,b~\citep{Winn2022}. The wide-orbit companion at 130\,au HD\,114762\,B cannot be compatible with neither \textit{Gaia} nor \textit{Gaia}--\textsc{Hipparcos} astrometry and thus did not contribute significantly to the photocenter's motion as captured by \textit{Gaia}. 

The broken linear relationships drawn upon the U and V-shaped confidence regions in the \glsxtrshort{pma}, \glsxtrshort{aen} and \glsxtrshort{ruwe} maps are related to their physical interpretation. The semi-empirical laws behind those curves draw approximate constraints on the mass and \glsxtrshort{sma} for any system. We further elaborated on these insights in Sects.~\ref{sec:PMa_curve} and~\ref{sec:AEN_curve}.

\subsection{The mass-sma relationship constrained by the \texorpdfstring{$\alpha_{\rm PMa}$}{PMa astrometric signature}}
\label{sec:PMa_curve}

The typical broken U-shape of the mass-\glsxtrshort{sma} relationships that we found in the \glsxtrshort{pma} map is shown in Fig.~\ref{fig:GJ832_PMEX}. It can be divided into four sections.

First, if $P$$\gg$24.5\,yr, then the \glsxtrshort{hg} baseline covers a part of the orbit. The contribution of the orbital motion in the \glsxtrshort{hg} PM vector and in the \glsxtrshort{g3} PM vector are the mean orbital speed during respectively 24.5 and 3\,yr. In this regime, the \glsxtrshort{pma} is indeed an approximation of the orbital acceleration fixed by the relative \glsxtrshort{sma} of the companion, with \glsxtrshort{pma}\,$\propto 1/{\rm \glsxtrshort{sma}}^2$. This approximation is increasingly accurate with increasing orbital period, as the orbital motion becomes more linear within 24.5\,yr. If $P$$<$24.5\,yr, for an orbital period sufficiently small, the contribution of the orbital motion in the 24.5-yr averaged PM becomes negligible.

In fact, the central point between those two main regimes is close to a period of 49\,yr, that is, about $P/2$=24.5\,yr. Indeed, for a star on a pole-on circular orbit with $P/2$=24.5\,yr, the orbital contribution to the \glsxtrshort{hg} PM (PM$_{\rm \glsxtrshort{hg},orbit}$ hereafter) that is, the average orbital speed along half the period, is $\sim$$4\,a_\star/P$; the orbital contribution to the \glsxtrshort{g3} PM (PM$_{\rm \glsxtrshort{g3},orbit}$ hereafter) is the instantaneous speed $\sim$$2\pi\,a_\star/P$; and they lead to a ratio PM$_{\rm \glsxtrshort{hg},orbit}$/PM$_{\rm \glsxtrshort{g3},orbit}$$\sim$0.6. For $P/2$$<$24.5\,yr, the orbital motion contribution to the PM$_{\rm \glsxtrshort{hg},orbit}$  becomes even smaller than half the orbital motion contribution to PM$_{\rm \glsxtrshort{g3},orbit}$. Conversely, for $P/2$$>$24.5\,yr, the PM$_{\rm \glsxtrshort{hg},orbit}$ and PM$_{\rm \glsxtrshort{g3},orbit}$ strengths gradually reach equality with increasing $P$.

Below 49\,yr, assuming that PM$_{\rm \glsxtrshort{hg},orbit}$ becomes rapidly negligible, we find three different regimes that depend on \glsxtrshort{g3} sampling and baselines at $P$$<$1\,yr, 1$<$$P$$<$6\,yr and 6$<$$P$$<$49\,yr:

\begin{itemize}
    \item $P$$<$$1$\,yr: The \textit{Gaia} data undersample the orbits, leading to strong aliasing of the orbital signal. The fit proper motion varies a lot from one value of \glsxtrshort{sma} to another, and almost nulling at resonances between the main sampling frequency and the orbital frequency (see also ~\citealt{Kervella2019}). There is nonetheless an average trend of the $\alpha_{\rm \glsxtrshort{pma}}$ that is linearly increasing with \glsxtrshort{sma}, and thus $\alpha_{\rm \glsxtrshort{pma}}$\,$\propto {\rm \glsxtrshort{sma}}$;
    \item $1$$<$$P$$<$$6$\,yr: Single orbits are better phase-covered, and there is more than half an orbit cycle monitored during the 3 yr of the \glsxtrshort{g3}. In this regime, the smaller the period, the more cycles are monitored and the less sensitive the $\alpha_{\rm \glsxtrshort{pma}}$ is to the amplitude of the motion. Thus, $\alpha_{\rm \glsxtrshort{pma}}$\,$\propto {\rm \glsxtrshort{sma}}\,P \propto {\rm \glsxtrshort{sma}}^{5/2}$;
    \item $6$$<$$P$$<$$49$\,yr: Less than half a phase of the orbital motion is covered by the \textit{Gaia} observations and the fit proper motion is oriented along, and in intensity proportional to, the average orbital velocity of the star along this orbit segment. Thus, the $\alpha_{\rm \glsxtrshort{pma}}$ is approximately proportional to the instantaneous orbital speed, $\alpha_{\rm \glsxtrshort{pma}}$\,$\propto 1/\sqrt{{\rm \glsxtrshort{sma}}}$.
\end{itemize}

Adding the correct dependence on the star mass $M_\star$, the mass of the companion $M_c$, and the parallax $\varpi$, these relations lead to the four mass--\glsxtrshort{sma} log-linear relations observed in the \glsxtrshort{pma} maps: 
\begin{align}
  \text{$\alpha_{\rm \glsxtrshort{pma}}$} \propto  \begin{cases}
        M_c \, M_\star^{-1}\,\varpi \, {\rm \glsxtrshort{sma}} &\text{in (1): $<$1\,yr}~~\Rightarrow M_c \propto {\rm \glsxtrshort{sma}}^{-1} \\
        M_c \, M_\star^{-3/2} \,\varpi \, {\rm \glsxtrshort{sma}}^{5/2}  & \text{in (2): 1--6\,yr}~~\Rightarrow M_c \propto {\rm \glsxtrshort{sma}}^{-5/2} \\
        M_c\,M_\star^{-1/2} \,\varpi \,{\rm \glsxtrshort{sma}}^{-1/2} & \text{in (3): 6--49\,yr}~~\Rightarrow M_c \propto {\rm \glsxtrshort{sma}}^{1/2} \\
        M_c \, \varpi \,{\rm \glsxtrshort{sma}}^{-2} & \text{in (4): $>$49\,yr}~~\Rightarrow M_c \propto {\rm \glsxtrshort{sma}}^2
    \end{cases} 
\end{align}

\noindent
Those four regimes are all linear in $\log$-$\log$ space but with different slopes. We note that the constraints stop following these linear models as the luminosity of the companion becomes comparable with the luminosity of the primary. When this occurs, as the luminosity of the companion increases, the photocenter semi-major axis shrinks more and more until reaching almost zero at a mass ratio\footnote{In our analysis, the photocenter semi-major axis never reaches exactly zero because in any bin, we explored an interval of mass and \glsxtrshort{sma}, and thus we almost never had exactly $q=1$.} $q\sim1$.

\subsection{The mass-sma relationship constrained by the \texorpdfstring{$\alpha_{\rm \glsxtrshort{mse}}$}{residuals astrometric signature}}
\label{sec:AEN_curve}

We now explain the typical V-shape of the mass-\glsxtrshort{sma} relationship that we observed in the \glsxtrshort{mse} maps, that is, those of \glsxtrshort{aen} and \glsxtrshort{ruwe}, shown in Fig.~\ref{fig:GJ832_PMEX}. Two regimes need to be distinguished. The regimes with $P$ either shorter than 3\,yr, when more than one orbital cycle is monitored, or longer than 3\,yr, when only a part of a single orbit is covered by \textit{Gaia}. 

\begin{itemize}
    \item If $P$<3\,yr, several orbital cycles happen during the \glsxtrshort{g3} observation campaign. The residuals amplitude, and thus $\alpha_{\rm \glsxtrshort{mse}}$, varies in proportion to the extent of the motion. The star's semi-major axis $a_\star$ itself is proportional to \glsxtrshort{sma}, $\alpha_{\rm \glsxtrshort{mse}}$\,$\propto {\rm \glsxtrshort{sma}}$.
    \item If $P$$>$3\,yr, only a part of a single orbit is covered by \glsxtrshort{g3}'s observations and is locally fit by a linear motion. The $\alpha_{\rm \glsxtrshort{mse}}$ measures the non-linearity of the motion, that is, the acceleration. Therefore, in this regime, $\alpha_{\rm \glsxtrshort{mse}}$\,$\propto 1/{\rm \glsxtrshort{sma}}^2$. 
\end{itemize}     
Adding the correct dependence on the star mass $M_\star$, the mass of the companion $M_c$ and the parallax $\varpi$, these relations lead to the mass--\glsxtrshort{sma} log-linear relations observed in the \glsxtrshort{mse} maps: 
\begin{align}
   \alpha_{\rm \glsxtrshort{mse}} \propto  \begin{cases}
         M_c \, M_\star^{-1} \, \varpi \, {\rm \glsxtrshort{sma}} &\text{in (1): $<$3\,yr}~~\Rightarrow M_c \propto {\rm \glsxtrshort{sma}}^{-1} \\
         M_c \, \varpi \, {\rm \glsxtrshort{sma}}^{-2}  &\text{in (2): $>$3\,yr}~~\Rightarrow M_c \propto {\rm \glsxtrshort{sma}}^2
    \end{cases} 
\end{align}

\noindent
The $\log$(mass) decreases linearly with the $\log$(\glsxtrshort{sma}) down to about $P$=3\,yr that corresponds to a range of \glsxtrshort{sma} within 1--3\,au depending on the stellar mass. Then the $\log$(mass) increases linearly with the $\log$(\glsxtrshort{sma}) up to very long period, until it reaches stellar masses. We note that, as for the \glsxtrshort{pma}, the constraints stop following these linear models as the luminosity of the companion becomes comparable with the luminosity of the primary, that is, beyond $M_c$$\sim$0.5\,$M_\star$.

\subsection{The mass-sma relationships and the minimum mass of companion}
The relations expressed above between $M_c$ and \glsxtrshort{sma}, at given $\varpi$ and $M_\star$, can be summarized as
\begin{equation}
    M_c = C_{\ell} \, \frac{\left(\alpha_{\rm PMa}\right){\rm~or~} \left(\alpha_{\rm \glsxtrshort{mse}}\right) }{\varpi} \,M_\star^{\frac{2-n_\ell}{3}} \, {\rm sma}^{n_\ell} \label{eq:relation_summarized}
\end{equation}
 with $\ell$ a number indexing the considered period regime, $C_\ell$ a multiplicative constant, and $n_\ell$ an exponent. To determine the $C_\ell$'s, we considered an arbitrary source and modeled by simulation 1000 values of $\alpha_{\rm \glsxtrshort{mse}}$ and $\alpha_{\rm \glsxtrshort{pma}}$ at a $P$ of the corresponding regime -- either 0.1, 2, 30 or 100\,yrs -- and at fixed $M_c$=10\,M$_{\rm J}$, $M_\star$=1\,M$_\odot$ and $\varpi$=1000\,mas. The inclination, eccentricity and $\omega$ were set randomly following the distributions of Table~\ref{tab:distro}. This led, for given \glsxtrshort{sma}=$P^{2/3}\,M_\star^{1/3}$, $M_c$, $M_\star$ and $\varpi$, to distributions of $\alpha_{\rm \glsxtrshort{mse}}$ and $\alpha_{\rm \glsxtrshort{pma}}$, and thus, inverting Eq.~\ref{eq:relation_summarized}, to measurements of the constants, $C_{\ell,{\rm simu}} \pm \sigma_{C,\ell}$. Because this might lead to disconnected line segments at the transition between regimes, that is, at $P$=1, 3, 6, or 49\,yr, we also imposed a continuity condition by noticing that from one regime $\ell$ to another contiguous $\ell+1$, there is a simple condition to fulfill: 
 \begin{equation}\label{eq:continuity}
     \log C_{\ell+1} - \log C_\ell = \frac{2}{3} \, \left(n_\ell - n_{\ell+1}\right)\,\log P
 \end{equation}
To do so, we minimized a least-squares problem by varying for either the case of $\alpha_{\rm \glsxtrshort{pma}}$ or $\alpha_{\rm \glsxtrshort{mse}}$ only $C_1$ and minimizing the following objective function $f$: 
\begin{equation}
f = \sum_\ell\frac{\left(C_\ell - C_{\ell, {\rm simu}}\right)^2}{ \sigma_{C,\ell}^2} 
\end{equation}
where any $C_{\ell\neq 1}$ is determined by using Eq. \ref{eq:continuity} and, for example, $\log C_{3} - \log C_1 =   \left(\log C_{3} - \log C_2\right) + \left(\log C_{2} - \log C_1\right)$.
We expected that the $C_\ell$ determined by this method would depend on the source considered due to the source-dependent noise levels, the shape of the unit parallax ellipse, and the \textit{Gaia} and \textsc{Hipparcos} scan law. We thus used 20 arbitrary sources with diverse magnitudes, colors, and RA-DEC and determined their global average $C_\ell$ rounded at the significant digit, given standard deviations of about 10--20\%:
\begin{align}
\alpha_{\rm \glsxtrshort{mse}} & \Longrightarrow
    \begin{cases}
        \text{$\ell$=1: $<$3\,yr,} & n_1=-1\text{,}~~C_1=2300 \\
        \text{$\ell$=2: $>$3\,yr,} & n_2=+2\text{,}~~C_2=260
    \end{cases} \label{eq:mass_sma_AEN}  \\ \nonumber \\
\alpha_{\rm \glsxtrshort{pma}} & \Longrightarrow
  \begin{cases}
        \text{$\ell$=1: $<$1\,yr,} & n_1=-1\text{,}~~C_1=6800\\
        \text{$\ell$=2: 1--6\,yr,} & n_2=-5/2\text{,}~~C_2=6800\\
        \text{$\ell$=3: 6--49\,yr,} & n_3=+1/2\text{,}~~C_3=190 \\
        \text{$\ell$=4: $>$49\,yr,} & n_4=+2\text{,}~~C_4=3.8
    \end{cases}  \label{eq:mass_sma_PMa}
\end{align}

\noindent
Reporting these numbers in Eq.~\ref{eq:relation_summarized}, the curve segments in these different regimes are added to the \glsxtrshort{pmex} maps in Figs.~\ref{fig:GJ832_PMEX} and~\ref{fig:HD114762_PMEX}. Given the actual mass uncertainty associated with the maps at any \glsxtrshort{sma}, the $C_\ell$ given here should only be considered as indicative, which is why we did not quote their uncertainties. These curves give an immediate approximate idea of what \glsxtrshort{pmex} maps should look like without doing the full computation. They are not intended as a substitute for \glsxtrshort{pmex} calculations. They do not produce, conversely to \glsxtrshort{pmex}, exact constraints on the \glsxtrshort{sma} and mass of the companion given the observed \glsxtrshort{aen}, \glsxtrshort{ruwe} and \glsxtrshort{pma}. Nevertheless, these curves can allow for rapid estimation of the mass of the companion as a function of its possible \glsxtrshort{sma} given either \glsxtrshort{aen}, \glsxtrshort{ruwe}, or \glsxtrshort{pma}. 

If the $\alpha_{\rm \glsxtrshort{mse}}$ or the $\alpha_{\rm \glsxtrshort{pma}}$ are sufficiently significant, typically more than 2--$\sigma$, these relationships enable measuring of the minimum mass of the companion, located at the minimum of all curves. This is reached at an \glsxtrshort{sma}$\sim$2-3\,au: 
\begin{align}
\alpha_{\rm \glsxtrshort{mse}} & \Longrightarrow
    \begin{cases}
        M_{c,\text{min}} = 1150 \,M_\star^{2/3} \,\left(\alpha_{\rm \glsxtrshort{mse}}/\varpi\right) \quad (\text{M}_{\rm J})\\
        \text{\glsxtrshort{sma}}_{\rm min} = 2.1\,M_\star^{1/3} \quad (\text{au})
    \end{cases} \label{eq:mass_minimum_AEN} \\ \nonumber  \\
\alpha_{\rm \glsxtrshort{pma}} & \Longrightarrow
    \begin{cases}
        M_{c,\text{min}} = 340 \,M_\star^{2/3} \,\left(\alpha_{\rm \glsxtrshort{pma}}/\varpi\right) \quad (\text{M}_{\rm J}) \\
        \text{\glsxtrshort{sma}}_{\rm min} = 2.9 \,M_\star^{1/3}  \quad (\text{au})
    \end{cases}  \label{eq:mass_minimum_PMa}
\end{align}

\noindent
These values can be used to identify planet candidates among catalogs of targets observed with \textit{Gaia}.

\section{Illustrative cases of GaiaPMEX results}
\label{sec:examples}

\subsection{HD 81040: Evidence for a short-period companion}

HD\,81040 is well-known as being the system of the first discovery from \glsxtrshort{g3} data of an exoplanet companion using only an orbital fit of the astrometric data~\citep{Arenou2023}.\footnote{See also the dedicated \textit{Gaia} ESA webpage at \url{https://www.cosmos.esa.int/web/gaia/iow_20220131}} HD\,81040 is a solar-type G2/3V star of magnitudes $V$=7.72 and $G$=7.57, at a distance of 32.56$\pm$1.31\,pc from Earth. The planetary candidate HD\,81040\,b was first discovered by~\citet{Sozzetti2006} with an orbital period of 1001\,days, eccentricity of 0.53 and an m$\sin i$ of 6.86\,M$_{\rm J}$. Using the \glsxtrshort{pma} calculated from the \textit{Gaia} EDR3 and \textsc{Hipparcos} data, there was further indications that this companion was indeed planetary with a mass of 7.24$^{+1.00}_{-0.37}$\,M$_{\rm J}$~\citep{Li2021,Winn2022}. Ultimately, a joint fit of the astrometric \glsxtrshort{g3} data and the \glsxtrshort{rv} data for this star fully confirmed a mass of 8.04$_{-0.54}^{+0.66}$\,M$_{\rm J}$. At the orbital period, best constrained with \glsxtrshort{rv}, of 1001\,days, the planet is predicted to have an \glsxtrshort{sma}=1.94$\pm$0.02\,au (ref. \textit{Gaia} ESA webpage). 

\begin{figure}[hbt]
    \centering
    \setlength{\unitlength}{1mm}
    \begin{picture}(89.3,80)
    \put(0,0){
    \includegraphics[width=89.3mm,clip=true]{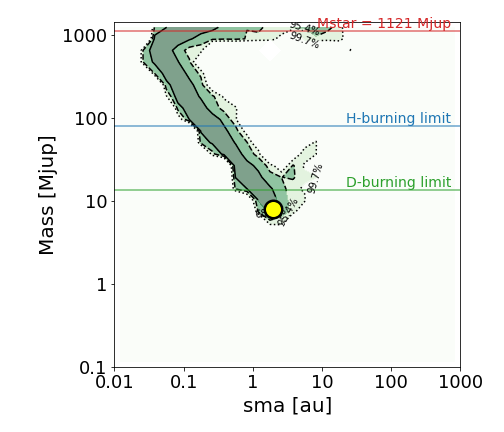}}
    \put(40,77){PMa + RUWE}
    \end{picture}
    \caption{Same as Fig.~\ref{fig:GJ832_PMEX} but only showing the combined \glsxtrshort{pma}+\glsxtrshort{ruwe} map for HD\,81040. The known exoplanet companion is indicated as a yellow circle (see text). Individual maps from~\glsxtrshort{ruwe}, \glsxtrshort{aen}, and \glsxtrshort{pma} constraints are shown in Appendix~F, Fig. F.1.}
    \label{fig:HD81040_PMEX}
\end{figure}

Figure~\ref{fig:HD81040_PMEX} shows the map constraining the mass and \glsxtrshort{sma} of a candidate companions around HD\,81040 calculated with \glsxtrshort{pmex} from combining \glsxtrshort{ruwe} and \glsxtrshort{pma} constraints. Other individual maps using only \glsxtrshort{aen}, \glsxtrshort{ruwe} and \glsxtrshort{pma} are shown in the Appendix F, Fig.~F.1. The parameters and results found for HD\,81040 are summarized in Table~\ref{tab:examples_param}. The astrometric signatures of \glsxtrshort{aen} and \glsxtrshort{ruwe} are both significant at $>$6.7--$\sigma$, with $p$-value$\sim$0.00, while the $\alpha_{\rm \glsxtrshort{pma}}$ cannot reject the single star hypothesis at a 1.5--$\sigma$ significance and a $p$-value of 0.13. Nevertheless, the constraints on mass--sma calculated by \glsxtrshort{pmex} from the \glsxtrshort{pma} stay compatible with the known planet in this system, but do not strongly reject an edge-on inclination. 

Most interestingly, when combining the constraints from the \glsxtrshort{ruwe} and the \glsxtrshort{pma}, the LP-branch disappeared from the \glsxtrshort{pmex} map, only leaving an SP-branch. The case of HD\,81040 is thus illustrative of the identification by \glsxtrshort{pmex} of a short-period companion, typically detectable using \glsxtrshort{rv}. Conversely, in such a case, it may allow for putting constraints on the mass of a known \glsxtrshort{rv} companion, as done with \glsxtrshort{gaston} in~\citet{Kiefer2019a,Kiefer2019b,Kiefer2021}. Here, the \glsxtrshort{pmex} maps agree well with the  mass and \glsxtrshort{sma} of HD\,81040\,b found by fitting the astrometric time series and the \glsxtrshort{rv}. 

\subsection{AF Lep: Evidence for a long-period companion}

AF Lep is an F8V star, initially characterized as an RS CVn. It has been shown that its variability is actually due to the presence of a companion, a super-Jupiter with mass $\sim$3\,M$_{\rm J}$ at 8\,au, detected through the combination of astrometric acceleration and direct imaging~\citep{Franson2023,DeRosa2023,Mesa2023}. The main parameters of AF Lep are summarized in Table~\ref{tab:examples_param}. This system was selected for follow-up with direct imaging because it shows a significant astrometric acceleration, or \glsxtrshort{pma}, compatible with a planet mass. This was a long-shot, because using \glsxtrshort{pma} only, one cannot reject that this system is actually a short period binary that would remain unresolved by direct imaging, while lower mass or lower \glsxtrshort{sma} could remain plausible, also leading to a non-detection with direct imaging. 

\begin{figure}[hbt]
    \centering
    \setlength{\unitlength}{1mm}
    \begin{picture}(89.3,80)
    \put(0,0){
    \includegraphics[width=89.3mm,clip=true]{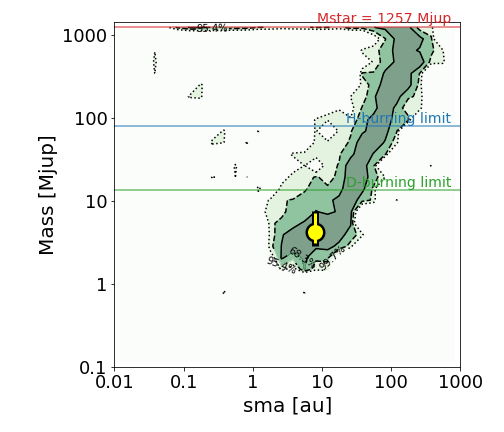}}
    \put(40,77){PMa + RUWE}
    \end{picture}
    \caption{Same as Fig.~\ref{fig:HD81040_PMEX} but for AF\,Lep. There are supplementary figures in Appendix F, Fig.~F.2. The known exoplanet companion is indicated as a yellow circle~\citep{Mesa2023}.}
    \label{fig:AFlep_PMEX}
\end{figure}

The \glsxtrshort{pmex} maps obtained for AF\,Lep are shown in Fig.~\ref{fig:AFlep_PMEX} and~F.2. In agreement with previous work on this source, the astrometric signature of the \glsxtrshort{pma} is significant at 4-$\sigma$, that is, compatible with the single star hypothesis with a p-value of 6.334\,$\times$\,$10^{-5}$. Both astrometric signatures found for \glsxtrshort{aen} and \glsxtrshort{ruwe} cannot reject the single star hypothesis for AF\,Lep with a $p$-value of 0.96 and a significance $<$0.1--$\sigma$. It leads to strict upper-bounds on the companion mass at any \glsxtrshort{sma}, since otherwise the \glsxtrshort{aen} and \glsxtrshort{ruwe} would have been more significant. 
 
 Interestingly, the mass--sma constraints from \glsxtrshort{pma} are largely degenerate, but combining them with those from \glsxtrshort{ruwe} leads to rejection of most low-\glsxtrshort{sma} solutions, that is, the SP-branch. This shows that combining the \glsxtrshort{pma} and the \glsxtrshort{ruwe} would have led to much tighter constraints on the possible mass and \glsxtrshort{sma} of a companion around AF\,Lep. The detected companion with 4.3$_{-1.2}^{+2.9}$\,M$_{\rm J}$ at \glsxtrshort{sma}=7.99$^{+0.85}_{-0.92}$\,au using direct imaging~\citep{Mesa2023,Franson2023} falls indeed within the 1--$\sigma$ bounds of the LP-branch.

\subsection{HD 23596: An almost detection with astrometry}

HD\,23596 is a V=7.2-mag F8 star at 52\,pc from the Sun that is known to host a companion planet. It was first discovered as an 8.2-M$_{\rm J}$ super-Jupiter with the ELODIE spectrograph~\citep{Perrier2003} and further characterized with a similar m$\sin i$ by~\citet{Wittenmyer2009} (7.71$\pm$0.39\,M$_{\rm j}$) and~\citet{Stassun2017} (9.03$\pm$0.74\,M$_{\rm J}$) using supplementary High Resolution Spectrograph (HRS) data and new $M_\star$ estimations. It was then re-established as a 14-M$_{\rm J}$ low-mass brown dwarf combining RVs and \textsc{Hipparcos}--\textit{Gaia} \glsxtrshort{pma}~\citep{Feng2022,Xiao2023}. The orbital period of this companion is 4.31$^{+0.069}_{-0.055}$\,years, with an \glsxtrshort{sma} of 2.90$\pm$0.08\,au, a mass of 14.6$^{+1.5}_{-1.3}$\,M$_{\rm J}$ and an inclination being of either 34.0$^{+3.6}_{-2.9}$$^\circ$ (prograde) or 146.0$^{+2.9}_{-3.6}$$^\circ$ (retrograde)~\citep{Xiao2023}. The inclination significantly non edge-on explains the lower 8.2-M$_{\rm J}$ $m\,\sin(i)$ initially found for this companion. 

\begin{figure}[hbt]
    \centering
    \setlength{\unitlength}{1mm}
    \begin{picture}(89.3,80)
    \put(0,0){
    \includegraphics[width=89.3mm,clip=true]{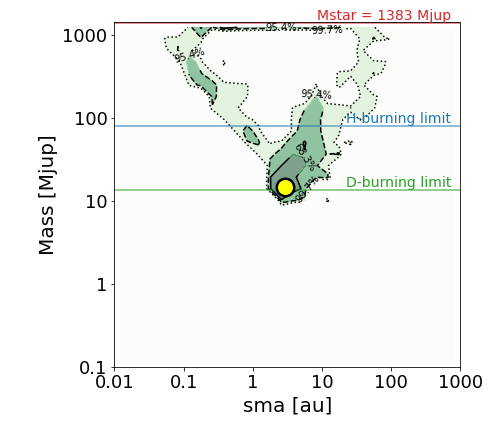}}
    \put(40,77){PMa + RUWE}
    \end{picture}
    \caption{Same as Fig.~\ref{fig:HD81040_PMEX} but for HD\,23596. There are supplementary figures in Appendix F, Fig.~F.3.}
    \label{fig:HD23596_PMEX}
\end{figure}

The \glsxtrshort{pmex} analysis combining the constraints from both \glsxtrshort{ruwe} and \glsxtrshort{pma} of HD\,23596 led to the confidence regions on companion mass and \glsxtrshort{sma} shown in Fig.~\ref{fig:HD23596_PMEX}. The individual maps from either \glsxtrshort{ruwe}, \glsxtrshort{aen} or \glsxtrshort{pma} contraints are shown in the Appendix in Fig.~F.3. The combination of \glsxtrshort{pma} and \glsxtrshort{ruwe} leads to infer a companion in the brown dwarf domain, with a narrow constraint on mass within 10--30\,M$_{\rm J}$ as well as on \glsxtrshort{sma} within 2--5\,au at 68.3\% confidence. This is in perfect agreement with the known companion of HD\,23596. Surprisingly, this source was not identified as a non-single star and does not appear in the non-single star catalog~\citep{vizierI357}\footnote{\url{https://vizier.cds.unistra.fr/viz-bin/VizieR?-source=I/357}}. This shows that \textit{Gaia} in combination with \textsc{Hipparcos}, with the help of the \glsxtrshort{pmex} tool, can detect and characterise a companion even without including \glsxtrshort{rv}s. 
More such planet candidates with strong constraints on the mass and the sma from only \textit{Gaia}+\textsc{Hipparcos} astrometry will be presented in other forthcoming papers~(paper II,~\citealt{Kiefer2024c}; paper III,~\citealt{Lagrange2024a}).

\subsection{\texorpdfstring{$\alpha$}{alpha} CMa B: The effect of considering a dark companion}

The case of $\alpha$\,CMa\,B is an instructive opportunity to show the effect of considering either a dark or a non-dark companion in the \glsxtrshort{pmex} analysis. Indeed, $\alpha$\,CMa\,B is nothing else than Sirius\,B, the white dwarf companion of the massive 240\,Myr old Sirius. It was first hypothesized by~\citet{Bessel1844}, then officially discovered by A. G. Clark in 1862, and further characterized by~\citet{Flammarion1877}. Even though its average separation with Sirius\,A is only 19.6\,au, the close distance to the Sun of this system lead to an angular separation of 7.6\arcsec~\citep{vandenbos1960}, and thus Sirius B was resolved in the \glsxtrshort{g3}. The respective mass of Sirius A \& B are 2.14\,M$_\odot$ and 1.05\,M$_\odot$~\citep{Gatewood1978}. Based on the orbital analysis of the Sirius AB system using astrometry, an additional companion Sirius C around either A or B was suspected to exist in this system, with a putative orbital period of 6\,years~\citep{Volet1932,Benest1995}. Around Sirius\,B's, this companion may have a mass $<$20\,M$_{\rm J}$ and an orbit \glsxtrshort{sma} of 1--2.5\,au~\citep{Bonnet2008}. Using Keck/NIRC2 observations,~\citet{Lucas2022} further excluded any companion of mass $>$10\,M$_{\rm J}$ down to 0.2\,au, $>$2.4\,M$_{\rm J}$ down to 0.5\,au and $>$0.7--1.2\,M$_{\rm J}$ beyond 1\,au around Sirius\,B.

\begin{figure}[hbt]
    \centering
    \setlength{\unitlength}{1mm}
    \begin{picture}(89.3,140)
    \put(0,70){\includegraphics[width=80.3mm,clip=true]{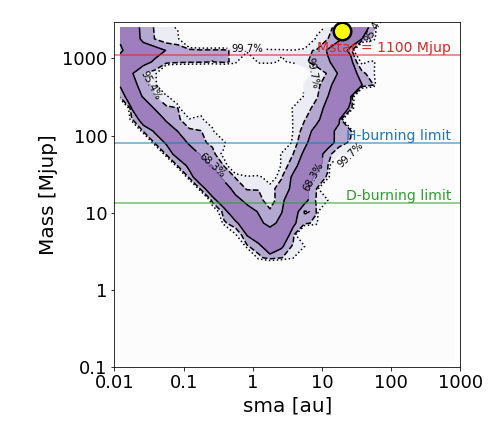}}
    \put(0,0){\includegraphics[width=80.3mm,clip=true]{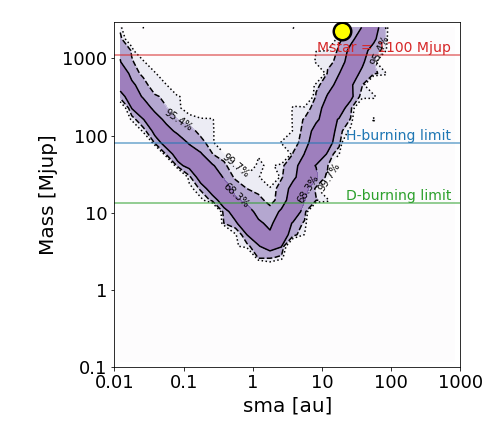}}
    \put(25,137){RUWE + luminous secondary}
    \put(28,67){RUWE + dark secondary}
    \end{picture}
    \caption{Same as Fig.~\ref{fig:GJ832_PMEX} for $\alpha$\,CMa\,B but only showing the constraints from \glsxtrshort{ruwe} since it was never observed with \textsc{Hipparcos}. Top: Considering a luminous companion with both components at age $\sim$5\,Gyr. Bottom: Considering a dark companion.}
    \label{fig:alpCMaB_PMEX}
\end{figure}

In \glsxtrshort{g3}, as summarized in Table~\ref{tab:examples_param}, the $\alpha_{\rm \glsxtrshort{mse}}$, from both \glsxtrshort{aen} and \glsxtrshort{ruwe}, of Sirius B is significant, $>$9--$\sigma$. Sirius\,B was not observed with \textsc{Hipparcos}, implying no \glsxtrshort{pma} for this star. The map derived from \glsxtrshort{ruwe} is shown in Fig.~\ref{fig:alpCMaB_PMEX}. We considered, by default in the top figure, a luminous companion. Within ~\citet{Lucas2022} constraints, the \glsxtrshort{ruwe} allows at 2--$\sigma$ for a companion with a mass as high as 10\,M$_{\rm J}$ at less than 0.5 au. However, with a mass $\sim$2.1\,M$_\odot$, the known companion of Sirius\,B, that is, Sirius\,A, at an average \glsxtrshort{sma}$\sim$20\,au, seems to explain the observed \glsxtrshort{ruwe} without the need to invoke a supplementary companion.

Nonetheless, for Sirius\,A, this approach is problematic for two reasons. First, the age of the system is overestimated since in the current version of \glsxtrshort{pmex}, the mass--luminosity relation is determined only for an age of 5\,Gyr. Second, and most importantly, even though it is much more luminous than B, Sirius\,A is well resolved by \textit{Gaia} and cannot act in the position of the photocenter of Sirius\,B. The main effect of considering dark or luminous companion is to change the shape of the confidence region at companion mass on the order of the magnitude of the mass of the source, here 1.05\,M$_\odot$. A luminous companion narrows down the photocenter's orbit. For a given constant photocenter semi-major axis, at short periods it requires the companion's \glsxtrshort{sma} to increase, while at long periods it requires the acceleration to increase and thus the \glsxtrshort{sma} to decrease. Considering instead a companion whose light does not contribute to the photocenter (dark or resolved), the confidence region rather more closely follows the curves derived in Sects.~\ref{sec:AEN_curve} and~\ref{sec:PMa_curve} that only describe the reflex astrometric motion of the main source under the gravitational pull of a companion. 

The \glsxtrshort{pmex} map derived from \glsxtrshort{ruwe} for Sirius\,B and considering a dark companion is shown in Fig.~\ref{fig:alpCMaB_PMEX}, bottom. We note that Sirius A is within the 3--$\sigma$ region and upon the edge of the 2--$\sigma$ region. Thus, we confirm that Sirius\,A may indeed explain the \glsxtrshort{ruwe}, although it would tend to generate, on average, a \glsxtrshort{ruwe} larger than the one published in the archives. 

Finally, the \glsxtrshort{pmex} maps of Sirius\,B cannot exclude any companion with a mass located below the confidence regions, since the perturbation from this other companion would be subdominant. It thus remains possible that another hidden companion exist around Sirius\,B. The \verb+IPD_frac_multi_peak+ and \verb+IPG_gof_harmonic_ampl+ of $\alpha$\,CMa\,B are moderate, respectively 18\% and 0.05, but according to our analysis performed in Appendix~\ref{sec:IPD}, they imply that a source within 200--500\,mas, that is, 0.5--1.3\,au, with a flux ratio within 10$^{-4}$--10$^{-3}$ is acting on the shape of the \glsxtrshort{psf}. Interestingly, the \citet{Lucas2022} analysis allows for a $<$1--2\,M$_{\rm J}$ companion at 0.5--1.3\,au below a limiting contrast of $\sim$10$^{-3}$. This strongly suggest digging further within the 1--au surroundings of the white dwarf Sirius\,B in the quest of planets.

\subsection{\texorpdfstring{$\beta$}{Beta} Pictoris: An issue with the RUWE}
\label{sec:betapic}
We focus here on the system of $\beta$\,Pictoris to bring forward again that the \glsxtrshort{ruwe} and \glsxtrshort{aen} cannot be interpreted directly as binarity indicators, but only through determining the astrometric signature within residuals and its significance.  The main parameters of $\beta$\,Pictoris are summarized in Table~\ref{tab:examples_param}. $\beta$\,Pictoris is a south-hemisphere 20-Myr old A5V star~\citep{Mamajek2014} located at 19.6\,pc from the Sun with a $G$-mag of 3.82. It is known for being the most furbished planetary system, after the Solar system, in the wealth of body types that planet formation can produce. It hosts a widely extended dusty and gaseous debris disk, exocomets, asteroids, and giant exoplanets with masses of $\sim$8 and $\sim$10\,M$_{\rm J}$. It is a system of great importance and a focus of attention from the community for understanding the running processes during the first hundred Myr of the Solar system and of planetary systems in general. In the \glsxtrshort{g3} database, $\beta$\,Pictoris is announced with a \glsxtrshort{ruwe} of 3.07 and an \glsxtrshort{aen} of 1.39\,mas. It looks therefore at first sight that significant deviation to the five-parameter model has been detected in this system, if we follow the guidelines that \glsxtrshort{ruwe}$>$1.4 indicates a non well-behaved or non-single system (as recommended in the DR3 documentation). 
However, here it is not the case. 

At a $\chi^2_{\rm astro}$=66,641.58 for 231 good \glsxtrshort{al} measurements, the \glsxtrshort{uwe} for $\beta$\,Pic is 17.2 and seems anomalously large, apparently corroborating the large \glsxtrshort{ruwe}. Considering the level of noises for this bright blue source of the \glsxtrshort{6p}-dataset that we found in Sect.~\ref{sec:noise_proxy}, that are $\sigma_{\rm calib}$=1.548\,mas, $\sigma_{\rm \glsxtrshort{al}}$=0.012\,mas, and $\sigma_{\rm \glsxtrshort{al}}$=0.074\,mas, the $\chi^2$ expected for a single source according to Eqs.~\ref{eq:chi2_mean} and~\ref{eq:chi2_sigma} is 80,000$\pm$24,000. And it corresponds to \glsxtrshort{uwe}=18.8$\pm$2.8, in good agreement with the value found above, indicating that 17.2 is, in fact, not an anomalously large \glsxtrshort{uwe}. It is therefore surprising that the \glsxtrshort{ruwe} of $\beta$\,Pic is 3 times larger than 1.0, while the \glsxtrshort{ruwe} was introduced as a renormalized version of the \glsxtrshort{uwe} to recenter this goodness-of-fit indicator around unity for well-behaved sources. The renormalizing factor $u_0$ in Eq.~\ref{eq:ruwe_chi2} is thus ill-defined for sources in the \glsxtrshort{6p}-dataset at the ($G$,$Bp-Rp$) corresponding to $\beta$\,Pic, that is, at $G$=3.82 and $Bp-Rp$=0.261. The \glsxtrshort{aen}=1.39\,mas is conversely well defined for $\beta$\,Pic. According to the relation of Eq.~\ref{eq:chi2_errors}, the \glsxtrshort{aen} implies $\chi^2_{\rm astro}$=77,922. This is in good agreement with the value expected for a single star determined above. The astrometric signature derived from \glsxtrshort{aen} and \glsxtrshort{ruwe} is undefined because their corresponding \glsxtrshort{mse} are smaller than the single star's expected \glsxtrshort{mse}$_{\rm single}$ in Eq.~\ref{eq:astro_sig}. 

\begin{figure}[hbt]
    \centering
    \includegraphics[width=89.3mm]{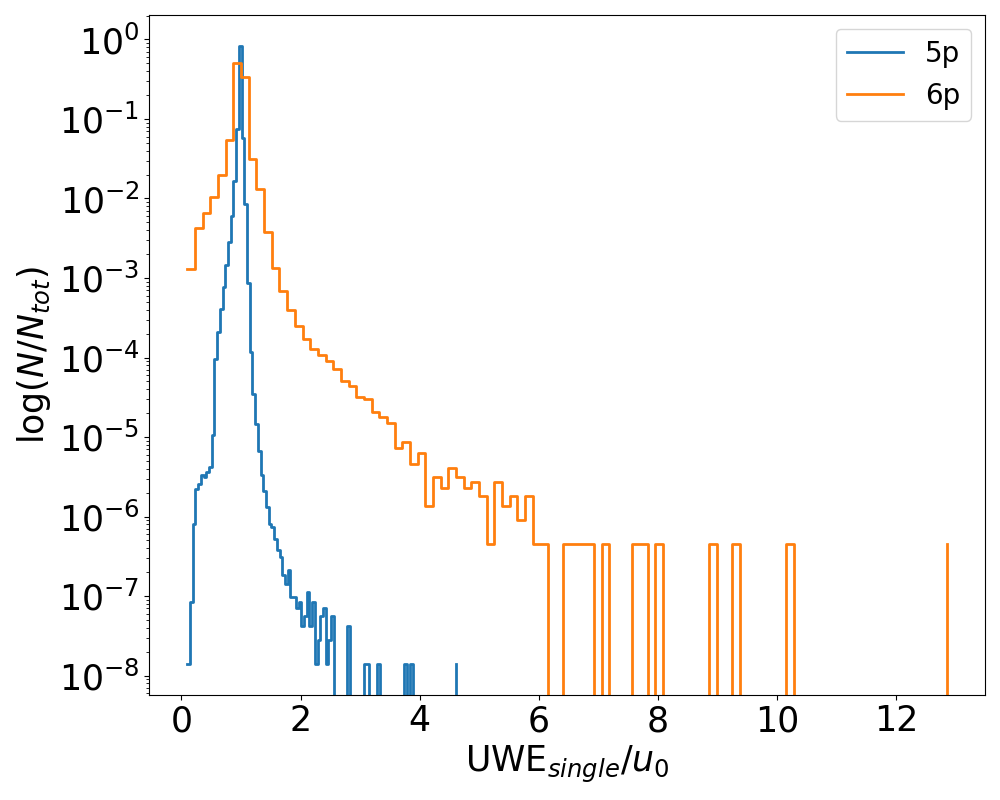}
    \caption{Statistics of the ratio of the single star's \glsxtrshort{uwe} over the $u_0$ published for all sources with $G$$<$16. Sources are separated into \glsxtrshort{5p} and \glsxtrshort{6p} datasets.}
    \label{fig:uwe_u0}
\end{figure}

Figure~\ref{fig:uwe_u0} compares the \glsxtrshort{uwe} obtained for single stars from Eq.~\ref{eq:chi2_mean} to the $u_0$ found in the \glsxtrshort{g3} auxiliary data\footnote{\url{https://www.cosmos.esa.int/web/gaia/auxiliary-data}} over the whole sources database with $G$$<$16. The ratio of both quantities is \glsxtrshort{ruwe}$_{\rm single}$, that should ideally be equal to 1. The distribution of this ratio peaks indeed at 1, but in the \glsxtrshort{6p}-dataset, there are strong tails on both sides toward lower and higher values. In the \glsxtrshort{5p}-dataset, some \glsxtrshort{ruwe}$_{\rm single}$ deviate from $1$, although much tightly than in the \glsxtrshort{6p}-dataset. In the \glsxtrshort{5p} dataset, 462 over 71,042,992 sources (6.5\,10$^{-4}$\%) have a \glsxtrshort{ruwe}$>$1.4, while in the \glsxtrshort{6p} dataset there are 14,890 over 2,203,807 sources (0.68\%) with a \glsxtrshort{ruwe}$>$1.4.
In conclusion, regardless of the dataset, it is safer to interpret the values of \glsxtrshort{aen} and \glsxtrshort{ruwe} only through calculating the astrometric signature of the \glsxtrshort{mse}, as defined in Sect.~\ref{sec:astro_sig}. 

\begin{figure*}[hbt]
    \centering    
    \setlength{\unitlength}{1mm}
    \begin{picture}(178.6,160)
    \put(0,75){\includegraphics[width=89.mm,clip=true]{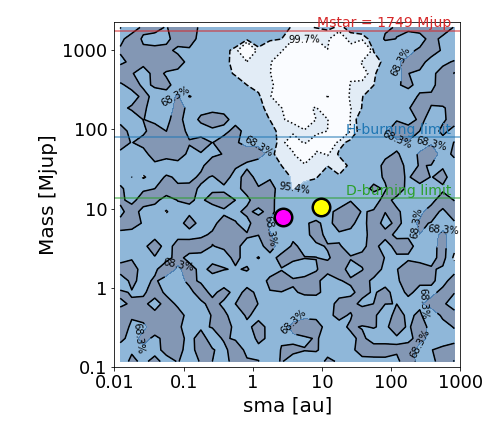}}
    \put(90,75){\includegraphics[width=89.mm,clip=true]{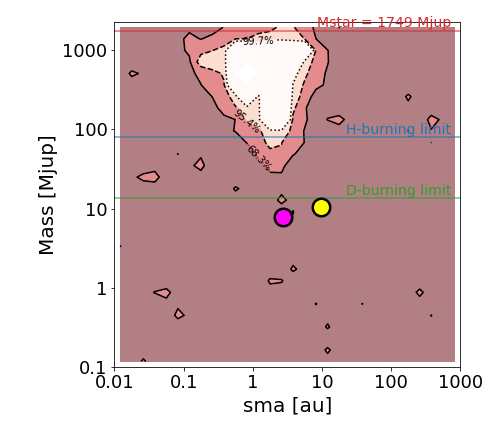}}
    \put(0,0){\includegraphics[width=89.mm,clip=true]{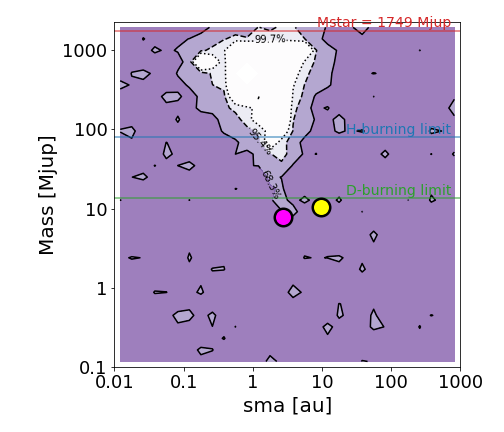}}
    \put(90,0){\includegraphics[width=89.mm,clip=true]{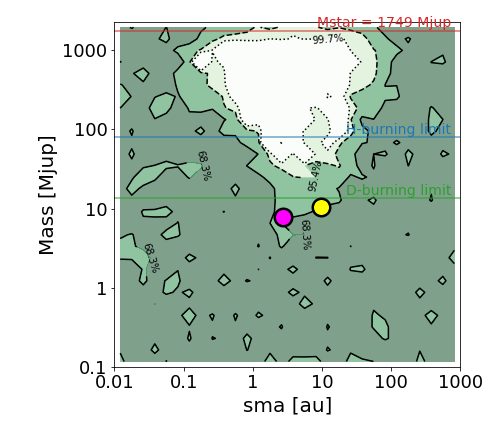}}
    \put(45,149){\glsxtrshort{pma}}
    \put(135,149){\glsxtrshort{aen}}
    \put(45,74){\glsxtrshort{ruwe}}
    \put(131,74){\glsxtrshort{pma}+\glsxtrshort{ruwe}}
    \end{picture}
    \caption{\glsxtrshort{pmex} constraints on mass and \glsxtrshort{sma} of a candidate companion around $\beta$\,Pictoris. Same caption as Fig.~\ref{fig:GJ832_PMEX}. The yellow and pink dots show the properties of the known exoplanet companions b \& c in this system, with ${\rm \glsxtrshort{sma}}_b$=9.8$\pm$0.4\,au and $m_b$=10--11\,M$_{\rm J}$, and ${\rm \glsxtrshort{sma}}_c$=2.7$\pm$0.02\,au and $M_c$=7.8$\pm$0.4\,M$_{\rm J}$~\citep{Lagrange2020}.}
    \label{fig:betapic_PMEX}
\end{figure*}

The \glsxtrshort{pmex} maps of $\beta$\,Pictoris are shown in Fig.~\ref{fig:betapic_PMEX}. Consistently with the above analysis, the \glsxtrshort{aen} and \glsxtrshort{ruwe} both lead to sma and mass of a hypothetical companion to $\beta$\,Pic compatible with a single star, that is, mass=0\,M$_{\rm J}$ or sma=0\,au. The two planets (b: 10--11\,M$_{\rm J}$ at 9.8$\pm$0.4\,au; c: 7.8$\pm$0.4\,M$_{\rm J}$ at 2.7$\pm$0.02\,au;~\citealt{Lagrange2020}) are compatible with the \glsxtrshort{aen} and \glsxtrshort{ruwe}, as well as the \glsxtrshort{pma}. There are no evidence in the \glsxtrshort{g3} for any other yet unknown companion around $\beta$\,Pictoris.

\section{The current sensitivity of \textit{Gaia} to the presence of companions}
\label{sec:discussion}
 
We determined the detection rates of exoplanets of different mass and \glsxtrshort{sma} that one can expect from using either \glsxtrshort{pma}, \glsxtrshort{mse} (determined from \glsxtrshort{aen} or \glsxtrshort{ruwe}), or \glsxtrshort{pma} $\bigcup$ \glsxtrshort{mse}\footnote{$\bigcup$ standing for the logical OR.}. They strongly depend on the stellar mass $M_\star$ and parallax $\varpi$. We call \say{detection} any value of $\alpha_{\rm \glsxtrshort{mse}}$ or $\alpha_{\rm \glsxtrshort{pma}}$ more significant than 2--$\sigma$, that is, for which the \glsxtrshort{mse}$^{1/3}$ or the \glsxtrshort{pma}$^{2/3}$ are above the 95.4th percentile of the respective single star's distributions.

Considering pre-main to \glsxtrlong{ms} (\glsxtrshort{ms}) stars, we explored a 2D-grid, with 30 bins per dimension uniformly spaced in log-scale, of $M_\star$ from 0.08 to 2.5\,M$_\odot$ and $\varpi$ from 1 to 1000\,mas. At each bin with given ($M_\star\pm\Delta M_\star$,$\varpi \pm \Delta \varpi$), 
we modeled \glsxtrshort{mse} and \glsxtrshort{pma}  by simulation (see Sect.~\ref{sec:simu}) of photocentric orbits as observed by \textit{Gaia} and \textsc{Hipparcos}, due to companions with a mass $M_c$ within bins delimited by 0.1, 0.2, 0.5 1, 2, 5, 10 and 20\,M$_{\rm J}$ and with an \glsxtrshort{sma} within different orbital regimes: Mercury-Earth type (0.1--1\,au), Earth--Mars type  (1--3\,au), Jupiter--Saturn type (3--10\,au) and Uranus--Neptune type (10--30\,au). The $G$-mag and the $Bp-Rp$ color corresponding to a given $M_\star$ and $\varpi$ are calculated from~\citet{Pecaut2012} and~\citet{Pecaut2013}'s spectral type to flux conversion tables for pre-\glsxtrshort{ms} to \glsxtrshort{ms} stars\footnote{\url{https://www.pas.rochester.edu/~emamajek/EEM_dwarf_UBVIJHK_colors_Teff.dat}}. The absolute magnitude $M_G$ is converted to apparent magnitude $G$ using the distance modulus calculated from the parallax. For simplicity, and to draw the general picture, we assumed zero extinction. The results are thus susceptible to be only informative, especially beyond 100\,pc. We fixed the other properties (such as \glsxtrshort{ra}, \glsxtrshort{dec}, parallax unit ellipse, \textsc{Hipparcos} IADs) and the \glsxtrshort{g3} epochs sampling, to those of GJ\,832. At each bin, the noises $\sigma_{\rm calib}$ and $\sigma_{\rm AL}$ levels are fixed with respect to the median $G$ and $Bp-Rp$. Since our initial sample covers 3 to 16 $G$-mags, our study of sensitivity in \glsxtrshort{g3} is limited to this range.

For each bin, we simulated $N$=1000 orbits given the range of companion \glsxtrshort{sma} and $M_c$ and host star's $M_\star$ and $\varpi$, randomizing other orbital parameters according to the distributions defined in Table~\ref{tab:bins_params}. We counted the percentage $r$ -- or detection rate -- of \glsxtrshort{mse}$^{1/3}$ and \glsxtrshort{pma}$^{2/3}$ exceeding the 95.4th--percentile in their respective single stars' distributions. We determined when $r$$>$20, 50, 90 and 99\%. We were peculiarly interested in the possibility to detect a planet by considering \glsxtrshort{pma}, \glsxtrshort{mse} or both, that is, \glsxtrshort{pma} $\bigcup$ \glsxtrshort{mse}. For single stars, the frequency of \glsxtrshort{fp} beyond the 2--$\sigma$ threshold is 4.6\%, when \glsxtrshort{pma} or \glsxtrshort{mse} are considered separately, and 9.2\%, when considering \glsxtrshort{pma} $\bigcup$ \glsxtrshort{mse}. This is the worst case scenario (largest \glsxtrshort{fp} frequency) in which noise rather than orbital motion causes a significance larger than 2--$\sigma$. In the general case, some true companion detections might be  serendipitous, that is, due to noise rather than orbital motion, and it is not possible to determine exactly the fraction of \glsxtrshort{fp} in this case. At best, a percentage $r$$>$4.6\% (respectively 9.2\%) indicates an increased sensitivity of \textit{Gaia} to the detection of companions in the given range of mass and \glsxtrshort{sma}. At worst, the fraction of \glsxtrshort{fp} is $4.6/r$ (respectively $9.2/r$). In particular, if $r$ is close to 4.6\% (respectively 9.2\%), then the fraction of \glsxtrshort{fp} is close to 100\%. 

\begin{figure*}[hbt]
    \centering
    \includegraphics[width=180mm,clip=true]{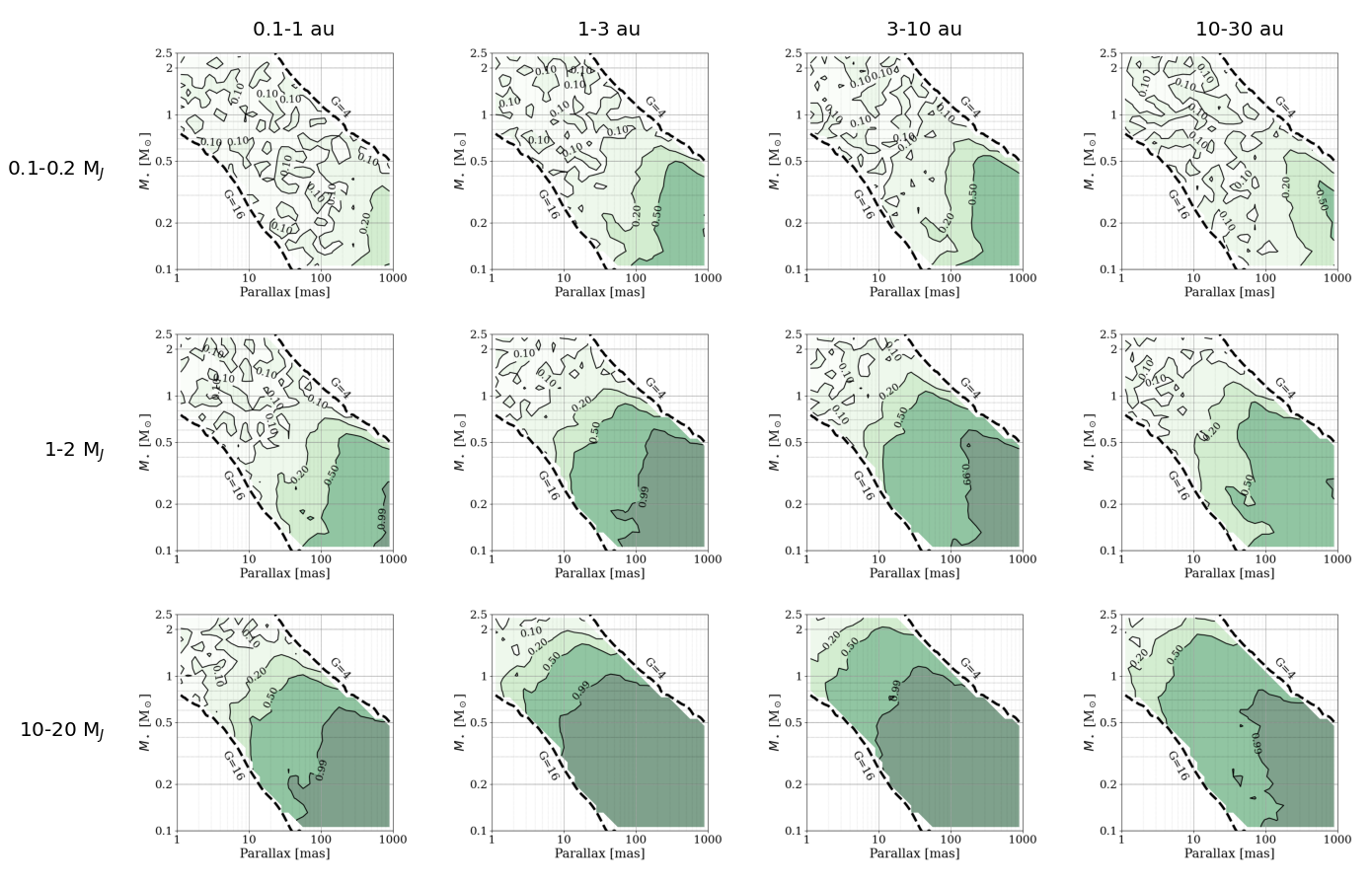}
    \caption{Theoretical detection rates with respect to $M_\star$ and the parallax. We varied the mass and the \glsxtrshort{sma} of the companion according to the column and row labels. White regions show the $M_\star$ and the parallax for which the probability of companion detection is less than 10\%; darker green regions show a probability of detection of $>$20\%, $>$50\% and $>$99\%. The dashed lines show the $M_\star$-parallax relations at the limiting magnitudes of $G$=3 and $G$=16 derived from~\citet{Pecaut2013} tables. Similar figures are obtained considering only \glsxtrshort{ruwe} or \glsxtrshort{pma} in Appendix G, in respectively Fig.~G.1 and G.2.}
    \label{fig:detec_rate_mstar_plx}
\end{figure*}

A map of the detection rates -- or equivalently, \textit{Gaia}'s sensitivity to companion detection -- with respect to star mass and parallax from using \glsxtrshort{pma}, $\alpha_{\rm \glsxtrshort{mse}}$ or both is shown in Fig.~\ref{fig:detec_rate_mstar_plx}. 
Planets with mass $<$1\,M$_{\rm J}$ around stars located farther than 100\,pc ($\varpi$$<$10\,mas) from the Sun lead to significant astrometric signal in less than 20\% of the simulations. But Jupiter-mass planets (1--2\,M$_{\rm J}$) might be detected with a $>$20\% chance around stars less massive than 1\,M$_\odot$ and up to 100\,pc distance. Similarly, planets of mass$<$0.1\,M$_{\rm J}$ around solar-like stars (0.5--2\,M$_{\rm J}$) have a less than 20\% chance of detection with \textit{Gaia} whatever their \glsxtrshort{sma} and whatever the distance to the Sun. But, if their host is an M-type star closer than 10\,pc and if their \glsxtrshort{sma} is within 1--10\,au, planets of any mass $>$0.1\,M$_{\rm J}$ have a more than 20\% chance of being detected at 2--$\sigma$. Moreover, within 1--10\,au and if their host star is an M-type star closer than 5\,pc, Jupiter-mass planets (1--2\,M$_{\rm J}$) were detected in $>$99\% of simulations, and Neptune/Saturn-mass planets (0.1--0.2\,M$_{\rm J}$) in $>$50\% of the simulations. Finally, super-Jupiter and brown dwarfs are easily detected at a rate $>$99\% up to large distance, even beyond 100\,pc and around A-type stars with M$_\star$$>$2\,M$_\odot$. 
 
\begin{figure}
    \centering
    \setlength{\unitlength}{1mm}
    \begin{picture}(89.7,70)
    \put(0,0){
    \includegraphics[width=80.3mm,clip=true]{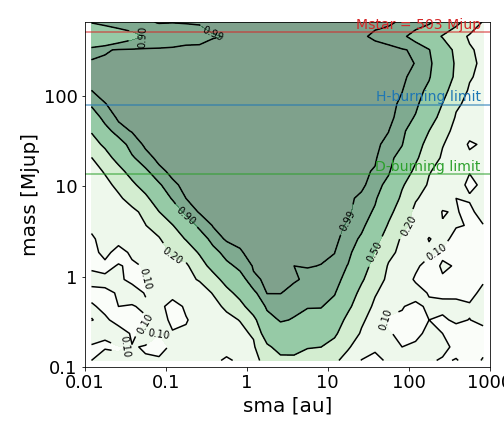}}
    \put(37,68){\glsxtrshort{pma} $\bigcup$ \glsxtrshort{mse}}
    \end{picture}
    \caption{Theoretical \textit{Gaia} detection rates of companions with a given mass and \glsxtrshort{sma} around GJ\,832 as permitted by allowing a detection with \glsxtrshort{pma} $\bigcup$ \glsxtrshort{mse}. White regions show the mass and \glsxtrshort{sma} for which the probability of detection of the companion is less than 10\%; dark green regions show a probability of detection $>$99\%; light green regions have intermediate probabilities. Individual maps for \glsxtrshort{mse} and \glsxtrshort{pma} are shown in Fig.~G.3.}
    \label{fig:detec_rate_GJ832}
\end{figure}

Focusing now on close-by low mass stars, we invoked again GJ\,832, a 5--pc distant M-dwarf, and derived the maps of \textit{Gaia}'s sensitivity to companion detection around this star. We modeled the \glsxtrshort{mse} and \glsxtrshort{pma} on a grid of \glsxtrshort{sma} and $M_c$, the same that was used for the \glsxtrshort{pmex} constraints maps, and determined the detection rates $r$, as done above. The map for detection with \glsxtrshort{mse} $\bigcup$ \glsxtrshort{pma} is shown in Fig.~\ref{fig:detec_rate_GJ832} and the individuals maps for detection with either \glsxtrshort{mse} or \glsxtrshort{pma} are shown in Fig.~G.3   . Around such a star, super-Jupiter and brown dwarfs with mass $>$10\,M$_{\rm J}$ and with an \glsxtrshort{sma} within 0.2--20\,au are detected in $>$99\% of the simulations. Lighter planetary companions whose mass is within 2--10\,M$_{\rm J}$ may lead to a significant astrometric signal in $>$99\% of the cases, provided their \glsxtrshort{sma} is contained within a narrower range of 1--10\,au. At masses in the Saturn-to-Jupiter regime (0.2--1\,M$_{\rm J}$) planetary companions have a lower detection rate, with a $>$50\% chance of being detected for an \glsxtrshort{sma} within 2--10\,au.  Planets of mass $<$0.2\,M$_{\rm J}$ are much less frequently detected ($<$50\%) whatever their \glsxtrshort{sma}. The detection rates obtained when considering only \glsxtrshort{pma} or only \glsxtrshort{mse} (Fig.~G.3) show that the \glsxtrshort{pma} is more sensitive to planets orbiting in the 2--20\,au range, while the \glsxtrshort{mse} tends to perform the most efficiently at shorter separations from 0.2 to 2\,au. 

In summary, \glsxtrshort{g3} currently performs best at detecting sub-stellar companions that are more massive than Jupiter on Earth-to-Saturn orbits (1--10\,au) around any star closer than 100\,pc from the Sun. 
%
\section{Conclusion}
We have introduced \glsxtrshort{pmex}, a tool that allows for characterization of the possible mass and \glsxtrshort{sma} of companions to stars observed with \textit{Gaia}. It uses the \glsxtrlong{pma} (\glsxtrshort{pma}) and the excess of the five-parameter model residuals, the \glsxtrlong{mse} (\glsxtrshort{mse}), indicated by the values of \glsxtrlong{ruwe} (\glsxtrshort{ruwe}) or astrometric excess noise (\glsxtrshort{aen}). \glsxtrshort{pmex} determines their significance with respect to the null hypothesis of a single star by calculating an astrometric signature for each indicator, $\alpha_{\rm \glsxtrshort{pma}}$ and $\alpha_{\rm \glsxtrshort{mse}}$, and then models them by the star's reflex motion due to a companion with ranges of possible mass and \glsxtrshort{sma}. 

Being otherwise highly degenerate in mass and \glsxtrshort{sma}, by combining the use of \glsxtrshort{aen} or \glsxtrshort{ruwe} and \glsxtrshort{pma}, the space of solutions is significantly reduced. This has never been done before and is highly promising for characterizing systems also observed by other means (radial velocities mostly). 
When $\alpha_{\rm \glsxtrshort{mse}}$ or $\alpha_{\rm \glsxtrshort{pma}}$ are significant, the posterior maps follow mass--\glsxtrshort{sma} relationships that are given in Eqs.~\ref{eq:mass_sma_AEN} and~\ref{eq:mass_sma_PMa}
 and that vary with orbital periods. Thereby, the \glsxtrshort{pma} could only be considered as a proxy for orbital acceleration when the orbital period of the companion is longer than 49\,yr, while the \glsxtrshort{mse} varies positively with the photocenter semi-major axis only when the orbital period of the companion is shorter than 3\,yr.

We determined companion detection rates and \textit{Gaia}'s sensitivity to exoplanet detection by comparing modeled \glsxtrshort{mse} and \glsxtrshort{pma} for companions with a hypothesized mass and \glsxtrshort{sma} to expectations for single stars. We show that in the exoplanet domain, the best detection rates are obtained for M-dwarf sources that are closest to the Sun ($<$10\,pc) and for exoplanet's \glsxtrshort{sma} within Earth-to-Saturn orbits, that is, 1--10\,au with masses down to 0.1\,M$_{\rm J}$. 

In the present article, we introduced in greater detail the \glsxtrshort{pmex} tool and the needed estimation of calibration, attitude and measurement noise levels. We plan on extensively exploiting \glsxtrshort{pmex} for the search and characterization of exoplanets in further studies.

\section*{Data availability}
Appendix E--H are available at the following url at \href{https://zenodo.org/records/13737120?token=eyJhbGciOiJIUzUxMiJ9.eyJpZCI6ImFkYWVlNDg1LTQzMmQtNDczZC1hNzk3LWYzYzUwNmYxZjE1MSIsImRhdGEiOnt9LCJyYW5kb20iOiIyNjRlN2QyYjczOWRhODVlMjc1NmU1NjlhNjQ2MmFlMyJ9.n8FLZELuZ8Oifu1hqDkJ6svxBxeBxmsDauL3LmlGmDVv5OJk3FxKYF4KuYMO_2ExMNdHpB4LkSpvN_urjrCLuw}{https://zenodo.org}.

\begin{acknowledgements}
We are very thankful to the anonymous referee for her/his thorough and courageous reading that led to significant improvements of this article. This work has made use of data from the European Space Agency (ESA) mission
{\it Gaia} (\url{https://www.cosmos.esa.int/gaia}), processed by the {\it Gaia}
\glsxtrlong{dpac} (\glsxtrshort{dpac},
\url{https://www.cosmos.esa.int/web/gaia/dpac/consortium}). Funding for the \glsxtrshort{dpac}
has been provided by national institutions, in particular the institutions
participating in the {\it Gaia} Multilateral Agreement. This work was granted access to the HPC resources of MesoPSL financed by the Region Ile de France and the project Equip\@Meso (reference ANR-10-EQPX-29-01) of the programme Investissements d’Avenir supervised by the Agence Nationale pour la Recherche. This project has received funding from the European Research Council (ERC) under the European Union's Horizon 2020 research and innovation programme (COBREX; grant agreement n° 885593). F.K. acknowledges funding from the initiative de recherches interdisciplinaires et stratégiques (IRIS) of Université PSL "Origines et Conditions d’Apparition de la Vie (OCAV)", as well as from the Action Pluriannuelle Incitative Exoplanètes from the Observatoire de Paris - Universit\'e PSL. F.K. also acknowledges funding from the American University of Paris.
\end{acknowledgements}

\bibliographystyle{aa}
\bibliography{main}

\begin{thebibliography}{68}
\expandafter\ifx\csname natexlab\endcsname\relax\def\natexlab#1{#1}\fi

\bibitem[{{Babusiaux} {et~al.}(2023){Babusiaux}, {Fabricius}, {Khanna}, {Muraveva}, {Reyl{\'e}}, {Spoto}, {Vallenari}, {Luri}, {Arenou}, {{\'A}lvarez}, {Anders}, {Antoja}, {Balbinot}, {Barache}, {Bauchet}, {Bossini}, {Busonero}, {Cantat-Gaudin}, {Carrasco}, {Dafonte}, {Diakit{\'e}}, {Figueras}, {Garcia-Gutierrez}, {Garofalo}, {Helmi}, {Jim{\'e}nez-Arranz}, {Jordi}, {Kervella}, {Kostrzewa-Rutkowska}, {Leclerc}, {Licata}, {Manteiga}, {Masip}, {Mongui{\'o}}, {Ramos}, {Robichon}, {Robin}, {Romero-G{\'o}mez}, {S{\'a}ez}, {Santove{\~n}a}, {Spina}, {Torralba Elipe}, \& {Weiler}}]{Babusiaux2023}
{Babusiaux}, C., {Fabricius}, C., {Khanna}, S., {et~al.} 2023, \aap, 674, A32

\bibitem[{{Bailey} {et~al.}(2009){Bailey}, {Butler}, {Tinney}, {Jones}, {O'Toole}, {Carter}, \& {Marcy}}]{Bailey2009}
{Bailey}, J., {Butler}, R.~P., {Tinney}, C.~G., {et~al.} 2009, \apj, 690, 743

\bibitem[{{Benest} \& {Duvent}(1995)}]{Benest1995}
{Benest}, D. \& {Duvent}, J.~L. 1995, \aap, 299, 621

\bibitem[{{Bessel}(1844)}]{Bessel1844}
{Bessel}, F.~W. 1844, \mnras, 6, 136

\bibitem[{{Bonnet-Bidaud} \& {Pantin}(2008)}]{Bonnet2008}
{Bonnet-Bidaud}, J.~M. \& {Pantin}, E. 2008, \aap, 489, 651

\bibitem[{{Brandt}(2021)}]{Brandt2021}
{Brandt}, T.~D. 2021, \apjs, 254, 42

\bibitem[{{Brandt} {et~al.}(2019){Brandt}, {Dupuy}, \& {Bowler}}]{Brandt2019}
{Brandt}, T.~D., {Dupuy}, T.~J., \& {Bowler}, B.~P. 2019, \aj, 158, 140

\bibitem[{{Calabretta} \& {Greisen}(2002)}]{Calabretta2002}
{Calabretta}, M.~R. \& {Greisen}, E.~W. 2002, \aap, 395, 1077

\bibitem[{Canal(2005)}]{canal2005}
Canal, L. 2005, Computational Statistics \& Data Analysis, 48, 803

\bibitem[{{Cochran} {et~al.}(1991){Cochran}, {Hatzes}, \& {Hancock}}]{Cochran1991}
{Cochran}, W.~D., {Hatzes}, A.~P., \& {Hancock}, T.~J. 1991, \apjl, 380, L35

\bibitem[{{Dalal} {et~al.}(2021){Dalal}, {Kiefer}, {H{\'e}brard}, {Sahlmann}, {Sousa}, {Forveille}, {Delfosse}, {Arnold}, {Astudillo-Defru}, {Bonfils}, {Boisse}, {Bouchy}, {Bourrier}, {Brugger}, {Cort{\'e}s-Zuleta}, {Deleuil}, {Demangeon}, {D{\'\i}az}, {Hara}, {Heidari}, {Hobson}, {Lopez}, {Lovis}, {Martioli}, {Mignon}, {Mousis}, {Moutou}, {Rey}, {Santerne}, {Santos}, {S{\'e}gransan}, {Str{\o}m}, \& {Udry}}]{Dalal2021}
{Dalal}, S., {Kiefer}, F., {H{\'e}brard}, G., {et~al.} 2021, \aap, 651, A11

\bibitem[{{De Rosa} {et~al.}(2023){De Rosa}, {Nielsen}, {Wahhaj}, {Ruffio}, {Kalas}, {Peck}, {Hirsch}, \& {Roberson}}]{DeRosa2023}
{De Rosa}, R.~J., {Nielsen}, E.~L., {Wahhaj}, Z., {et~al.} 2023, \aap, 672, A94

\bibitem[{{Fabricius} {et~al.}(2021){Fabricius}, {Luri}, {Arenou}, {Babusiaux}, {Helmi}, {Muraveva}, {Reyl{\'e}}, {Spoto}, {Vallenari}, {Antoja}, {Balbinot}, {Barache}, {Bauchet}, {Bragaglia}, {Busonero}, {Cantat-Gaudin}, {Carrasco}, {Diakit{\'e}}, {Fabrizio}, {Figueras}, {Garcia-Gutierrez}, {Garofalo}, {Jordi}, {Kervella}, {Khanna}, {Leclerc}, {Licata}, {Lambert}, {Marrese}, {Masip}, {Ramos}, {Robichon}, {Robin}, {Romero-G{\'o}mez}, {Rubele}, \& {Weiler}}]{Fabricius2021}
{Fabricius}, C., {Luri}, X., {Arenou}, F., {et~al.} 2021, \aap, 649, A5

\bibitem[{{Feng} {et~al.}(2021){Feng}, {Butler}, {Jones}, {Phillips}, {Vogt}, {Oppenheimer}, {Holden}, {Burt}, \& {Boss}}]{Feng2021}
{Feng}, F., {Butler}, R.~P., {Jones}, H. R.~A., {et~al.} 2021, \mnras, 507, 2856

\bibitem[{{Feng} {et~al.}(2022){Feng}, {Butler}, {Vogt}, {Clement}, {Tinney}, {Cui}, {Aizawa}, {Jones}, {Bailey}, {Burt}, {Carter}, {Crane}, {Flammini Dotti}, {Holden}, {Ma}, {Ogihara}, {Oppenheimer}, {O'Toole}, {Shectman}, {Wittenmyer}, {Wang}, {Wright}, \& {Xuan}}]{Feng2022}
{Feng}, F., {Butler}, R.~P., {Vogt}, S.~S., {et~al.} 2022, \apjs, 262, 21

\bibitem[{{Flammarion}(1877)}]{Flammarion1877}
{Flammarion}, C. 1877, Astronomical register, 15, 186

\bibitem[{{Franson} {et~al.}(2023){Franson}, {Bowler}, {Zhou}, {Pearce}, {Bardalez Gagliuffi}, {Biddle}, {Brandt}, {Crepp}, {Dupuy}, {Faherty}, {Jensen-Clem}, {Morgan}, {Sanghi}, {Theissen}, {Tran}, \& {Wolf}}]{Franson2023}
{Franson}, K., {Bowler}, B.~P., {Zhou}, Y., {et~al.} 2023, \apjl, 950, L19

\bibitem[{{Gaia collaboration}(2022)}]{vizierI357}
{Gaia collaboration}. 2022, {Gaia DR3 Part 3. Non-single stars}

\bibitem[{{Gaia Collaboration} {et~al.}(2023{\natexlab{a}}){Gaia Collaboration}, {Arenou}, {Babusiaux}, {Barstow}, {Faigler}, {Jorissen}, {Kervella}, {Mazeh}, {Mowlavi}, {Panuzzo}, {Sahlmann}, {Shahaf}, {Sozzetti}, {Bauchet}, {Damerdji}, {Gavras}, {Giacobbe}, {Gosset}, {Halbwachs}, {Holl}, {Lattanzi}, {Leclerc}, {Morel}, {Pourbaix}, {Re Fiorentin}, {Sadowski}, {S{\'e}gransan}, {Siopis}, {Teyssier}, {Zwitter}, {Planquart}, {Brown}, {Vallenari}, {Prusti}, {de Bruijne}, {Biermann}, {Creevey}, {Ducourant}, {Evans}, {Eyer}, {Guerra}, {Hutton}, {Jordi}, {Klioner}, {Lammers}, {Lindegren}, {Luri}, {Mignard}, {Panem}, {Randich}, {Sartoretti}, {Soubiran}, {Tanga}, {Walton}, {Bailer-Jones}, {Bastian}, {Drimmel}, {Jansen}, {Katz}, {van Leeuwen}, {Bakker}, {Cacciari}, {Casta{\~n}eda}, {De Angeli}, {Fabricius}, {Fouesneau}, {Fr{\'e}mat}, {Galluccio}, {Guerrier}, {Heiter}, {Masana}, {Messineo}, {Nicolas}, {Nienartowicz}, {Pailler}, {Riclet}, {Roux}, {Seabroke}, {Sordo}, {Th{\'e}venin}, {Gracia-Abril}, {Portell}, {Altmann},
  {Andrae}, {Audard}, {Bellas-Velidis}, {Benson}, {Berthier}, {Blomme}, {Burgess}, {Busonero}, {Busso}, {C{\'a}novas}, {Carry}, {Cellino}, {Cheek}, {Clementini}, {Davidson}, {de Teodoro}, {Nu{\~n}ez Campos}, {Delchambre}, {Dell'Oro}, {Esquej}, {Fern{\'a}ndez-Hern{\'a}ndez}, {Fraile}, {Garabato}, {Garc{\'\i}a-Lario}, {Haigron}, {Hambly}, {Harrison}, {Hern{\'a}ndez}, {Hestroffer}, {Hodgkin}, {Jan{\ss}en}, {Jevardat de Fombelle}, {Jordan}, {Krone-Martins}, {Lanzafame}, {L{\"o}ffler}, {Marchal}, {Marrese}, {Moitinho}, {Muinonen}, {Osborne}, {Pancino}, {Pauwels}, {Recio-Blanco}, {Reyl{\'e}}, {Riello}, {Rimoldini}, {Roegiers}, {Rybizki}, {Sarro}, {Smith}, {Utrilla}, {van Leeuwen}, {Abbas}, {{\'A}brah{\'a}m}, {Abreu Aramburu}, {Aerts}, {Aguado}, {Ajaj}, {Aldea-Montero}, {Altavilla}, {{\'A}lvarez}, {Alves}, {Anders}, {Anderson}, {Anglada Varela}, {Antoja}, {Baines}, {Baker}, {Balaguer-N{\'u}{\~n}ez}, {Balbinot}, {Balog}, {Barache}, {Barbato}, {Barros}, {Bartolom{\'e}}, {Bassilana}, {Becciani}, {Bellazzini},
  {Berihuete}, {Bernet}, {Bertone}, {Bianchi}, {Binnenfeld}, {Blanco-Cuaresma}, {Blazere}, {Boch}, {Bombrun}, {Bossini}, {Bouquillon}, {Bragaglia}, {Bramante}, {Breedt}, {Bressan}, {Brouillet}, {Brugaletta}, {Bucciarelli}, {Burlacu}, {Butkevich}, {Buzzi}, {Caffau}, {Cancelliere}, {Cantat-Gaudin}, {Carballo}, {Carlucci}, {Carnerero}, {Carrasco}, {Casamiquela}, {Castellani}, {Castro-Ginard}, {Chaoul}, {Charlot}, {Chemin}, {Chiaramida}, {Chiavassa}, {Chornay}, {Comoretto}, {Contursi}, {Cooper}, {Cornez}, {Cowell}, {Crifo}, {Cropper}, {Crosta}, {Crowley}, {Dafonte}, {Dapergolas}, {David}, {de Laverny}, {De Luise}, {De March}, {De Ridder}, {de Souza}, {de Torres}, {del Peloso}, {del Pozo}, {Delbo}, {Delgado}, {Delisle}, {Demouchy}, {Dharmawardena}, {Diakite}, {Diener}, {Distefano}, {Dolding}, {Enke}, {Fabre}, {Fabrizio}, {Fedorets}, {Fernique}, {Figueras}, {Fournier}, {Fouron}, {Fragkoudi}, {Gai}, {Garcia-Gutierrez}, {Garcia-Reinaldos}, {Garc{\'\i}a-Torres}, {Garofalo}, {Gavel}, {Gerlach}, {Geyer}, {Gilmore},
  {Girona}, {Giuffrida}, {Gomel}, {Gomez}, {Gonz{\'a}lez-N{\'u}{\~n}ez}, {Gonz{\'a}lez-Santamar{\'\i}a}, {Gonz{\'a}lez-Vidal}, {Granvik}, {Guillout}, {Guiraud}, {Guti{\'e}rrez-S{\'a}nchez}, {Guy}, {Hatzidimitriou}, {Hauser}, {Haywood}, {Helmer}, {Helmi}, {Sarmiento}, {Hidalgo}, {Hilger}, {H{\l}adczuk}, {Hobbs}, {Holland}, {Huckle}, {Jardine}, {Jasniewicz}, {Jean-Antoine Piccolo}, {Jim{\'e}nez-Arranz}, {Juaristi Campillo}, {Julbe}, {Karbevska}, {Khanna}, {Kordopatis}, {Korn}, {K{\'o}sp{\'a}l}, {Kostrzewa-Rutkowska}, {Kruszy{\'n}ska}, {Kun}, {Laizeau}, {Lambert}, {Lanza}, {Lasne}, {Le Campion}, {Lebreton}, {Lebzelter}, {Leccia}, {Lecoeur-Taibi}, {Liao}, {Licata}, {Lindstr{\o}m}, {Lister}, {Livanou}, {Lobel}, {Lorca}, {Loup}, {Madrero Pardo}, {Magdaleno Romeo}, {Managau}, {Mann}, {Manteiga}, {Marchant}, {Marconi}, {Marcos}, {Marcos Santos}, {Mar{\'\i}n Pina}, {Marinoni}, {Marocco}, {Marshall}, {Martin Polo}, {Mart{\'\i}n-Fleitas}, {Marton}, {Mary}, {Masip}, {Massari}, {Mastrobuono-Battisti}, {McMillan},
  {Messina}, {Michalik}, {Millar}, {Mints}, {Molina}, {Molinaro}, {Moln{\'a}r}, {Monari}, {Mongui{\'o}}, {Montegriffo}, {Montero}, {Mor}, {Mora}, {Morbidelli}, {Morris}, {Muraveva}, {Murphy}, {Musella}, {Nagy}, {Noval}, {Oca{\~n}a}, {Ogden}, {Ordenovic}, {Osinde}, {Pagani}, {Pagano}, {Palaversa}, {Palicio}, {Pallas-Quintela}, {Panahi}, {Payne-Wardenaar}, {Pe{\~n}alosa Esteller}, {Penttil{\"a}}, {Pichon}, {Piersimoni}, {Pineau}, {Plachy}, {Plum}, {Poggio}, {Pr{\v{s}}a}, {Pulone}, {Racero}, {Ragaini}, {Rainer}, {Raiteri}, {Ramos}, {Ramos-Lerate}, {Regibo}, {Richards}, {Rios Diaz}, {Ripepi}, {Riva}, {Rix}, {Rixon}, {Robichon}, {Robin}, {Robin}, {Roelens}, {Rogues}, {Rohrbasser}, {Romero-G{\'o}mez}, {Rowell}, {Royer}, {Ruz Mieres}, {Rybicki}, {S{\'a}ez N{\'u}{\~n}ez}, {Sagrist{\`a} Sell{\'e}s}, {Salguero}, {Samaras}, {Sanchez Gimenez}, {Sanna}, {Santove{\~n}a}, {Sarasso}, {Schultheis}, {Sciacca}, {Segol}, {Segovia}, {Semeux}, {Siddiqui}, {Siebert}, {Siltala}, {Silvelo}, {Slezak}, {Slezak}, {Smart}, {Snaith},
  {Solano}, {Solitro}, {Souami}, {Souchay}, {Spagna}, {Spina}, {Spoto}, {Steele}, {Steidelm{\"u}ller}, {Stephenson}, {S{\"u}veges}, {Surdej}, {Szabados}, {Szegedi-Elek}, {Taris}, {Taylor}, {Teixeira}, {Tolomei}, {Tonello}, {Torra}, {Torra}, {Torralba Elipe}, {Trabucchi}, {Tsounis}, {Turon}, {Ulla}, {Unger}, {Vaillant}, {van Dillen}, {van Reeven}, {Vanel}, {Vecchiato}, {Viala}, {Vicente}, {Voutsinas}, {Weiler}, {Wevers}, {Wyrzykowski}, {Yoldas}, {Yvard}, {Zhao}, {Zorec}, \& {Zucker}}]{Arenou2023}
{Gaia Collaboration}, {Arenou}, F., {Babusiaux}, C., {et~al.} 2023{\natexlab{a}}, \aap, 674, A34

\bibitem[{{Gaia Collaboration} {et~al.}(2021){Gaia Collaboration}, {Brown}, {Vallenari}, {Prusti}, {de Bruijne}, {Babusiaux}, {Biermann}, {Creevey}, {Evans}, {Eyer}, {Hutton}, {Jansen}, {Jordi}, {Klioner}, {Lammers}, {Lindegren}, {Luri}, {Mignard}, {Panem}, {Pourbaix}, {Randich}, {Sartoretti}, {Soubiran}, {Walton}, {Arenou}, {Bailer-Jones}, {Bastian}, {Cropper}, {Drimmel}, {Katz}, {Lattanzi}, {van Leeuwen}, {Bakker}, {Cacciari}, {Casta{\~n}eda}, {De Angeli}, {Ducourant}, {Fabricius}, {Fouesneau}, {Fr{\'e}mat}, {Guerra}, {Guerrier}, {Guiraud}, {Jean-Antoine Piccolo}, {Masana}, {Messineo}, {Mowlavi}, {Nicolas}, {Nienartowicz}, {Pailler}, {Panuzzo}, {Riclet}, {Roux}, {Seabroke}, {Sordo}, {Tanga}, {Th{\'e}venin}, {Gracia-Abril}, {Portell}, {Teyssier}, {Altmann}, {Andrae}, {Bellas-Velidis}, {Benson}, {Berthier}, {Blomme}, {Brugaletta}, {Burgess}, {Busso}, {Carry}, {Cellino}, {Cheek}, {Clementini}, {Damerdji}, {Davidson}, {Delchambre}, {Dell'Oro}, {Fern{\'a}ndez-Hern{\'a}ndez}, {Galluccio}, {Garc{\'\i}a-Lario},
  {Garcia-Reinaldos}, {Gonz{\'a}lez-N{\'u}{\~n}ez}, {Gosset}, {Haigron}, {Halbwachs}, {Hambly}, {Harrison}, {Hatzidimitriou}, {Heiter}, {Hern{\'a}ndez}, {Hestroffer}, {Hodgkin}, {Holl}, {Jan{\ss}en}, {Jevardat de Fombelle}, {Jordan}, {Krone-Martins}, {Lanzafame}, {L{\"o}ffler}, {Lorca}, {Manteiga}, {Marchal}, {Marrese}, {Moitinho}, {Mora}, {Muinonen}, {Osborne}, {Pancino}, {Pauwels}, {Petit}, {Recio-Blanco}, {Richards}, {Riello}, {Rimoldini}, {Robin}, {Roegiers}, {Rybizki}, {Sarro}, {Siopis}, {Smith}, {Sozzetti}, {Ulla}, {Utrilla}, {van Leeuwen}, {van Reeven}, {Abbas}, {Abreu Aramburu}, {Accart}, {Aerts}, {Aguado}, {Ajaj}, {Altavilla}, {{\'A}lvarez}, {{\'A}lvarez Cid-Fuentes}, {Alves}, {Anderson}, {Anglada Varela}, {Antoja}, {Audard}, {Baines}, {Baker}, {Balaguer-N{\'u}{\~n}ez}, {Balbinot}, {Balog}, {Barache}, {Barbato}, {Barros}, {Barstow}, {Bartolom{\'e}}, {Bassilana}, {Bauchet}, {Baudesson-Stella}, {Becciani}, {Bellazzini}, {Bernet}, {Bertone}, {Bianchi}, {Blanco-Cuaresma}, {Boch}, {Bombrun}, {Bossini},
  {Bouquillon}, {Bragaglia}, {Bramante}, {Breedt}, {Bressan}, {Brouillet}, {Bucciarelli}, {Burlacu}, {Busonero}, {Butkevich}, {Buzzi}, {Caffau}, {Cancelliere}, {C{\'a}novas}, {Cantat-Gaudin}, {Carballo}, {Carlucci}, {Carnerero}, {Carrasco}, {Casamiquela}, {Castellani}, {Castro-Ginard}, {Castro Sampol}, {Chaoul}, {Charlot}, {Chemin}, {Chiavassa}, {Cioni}, {Comoretto}, {Cooper}, {Cornez}, {Cowell}, {Crifo}, {Crosta}, {Crowley}, {Dafonte}, {Dapergolas}, {David}, {David}, {de Laverny}, {De Luise}, {De March}, {De Ridder}, {de Souza}, {de Teodoro}, {de Torres}, {del Peloso}, {del Pozo}, {Delbo}, {Delgado}, {Delgado}, {Delisle}, {Di Matteo}, {Diakite}, {Diener}, {Distefano}, {Dolding}, {Eappachen}, {Edvardsson}, {Enke}, {Esquej}, {Fabre}, {Fabrizio}, {Faigler}, {Fedorets}, {Fernique}, {Fienga}, {Figueras}, {Fouron}, {Fragkoudi}, {Fraile}, {Franke}, {Gai}, {Garabato}, {Garcia-Gutierrez}, {Garc{\'\i}a-Torres}, {Garofalo}, {Gavras}, {Gerlach}, {Geyer}, {Giacobbe}, {Gilmore}, {Girona}, {Giuffrida}, {Gomel}, {Gomez},
  {Gonzalez-Santamaria}, {Gonz{\'a}lez-Vidal}, {Granvik}, {Guti{\'e}rrez-S{\'a}nchez}, {Guy}, {Hauser}, {Haywood}, {Helmi}, {Hidalgo}, {Hilger}, {H{\l}adczuk}, {Hobbs}, {Holland}, {Huckle}, {Jasniewicz}, {Jonker}, {Juaristi Campillo}, {Julbe}, {Karbevska}, {Kervella}, {Khanna}, {Kochoska}, {Kontizas}, {Kordopatis}, {Korn}, {Kostrzewa-Rutkowska}, {Kruszy{\'n}ska}, {Lambert}, {Lanza}, {Lasne}, {Le Campion}, {Le Fustec}, {Lebreton}, {Lebzelter}, {Leccia}, {Leclerc}, {Lecoeur-Taibi}, {Liao}, {Licata}, {Lindstr{\o}m}, {Lister}, {Livanou}, {Lobel}, {Madrero Pardo}, {Managau}, {Mann}, {Marchant}, {Marconi}, {Marcos Santos}, {Marinoni}, {Marocco}, {Marshall}, {Martin Polo}, {Mart{\'\i}n-Fleitas}, {Masip}, {Massari}, {Mastrobuono-Battisti}, {Mazeh}, {McMillan}, {Messina}, {Michalik}, {Millar}, {Mints}, {Molina}, {Molinaro}, {Moln{\'a}r}, {Montegriffo}, {Mor}, {Morbidelli}, {Morel}, {Morris}, {Mulone}, {Munoz}, {Muraveva}, {Murphy}, {Musella}, {Noval}, {Ord{\'e}novic}, {Orr{\`u}}, {Osinde}, {Pagani}, {Pagano},
  {Palaversa}, {Palicio}, {Panahi}, {Pawlak}, {Pe{\~n}alosa Esteller}, {Penttil{\"a}}, {Piersimoni}, {Pineau}, {Plachy}, {Plum}, {Poggio}, {Poretti}, {Poujoulet}, {Pr{\v{s}}a}, {Pulone}, {Racero}, {Ragaini}, {Rainer}, {Raiteri}, {Rambaux}, {Ramos}, {Ramos-Lerate}, {Re Fiorentin}, {Regibo}, {Reyl{\'e}}, {Ripepi}, {Riva}, {Rixon}, {Robichon}, {Robin}, {Roelens}, {Rohrbasser}, {Romero-G{\'o}mez}, {Rowell}, {Royer}, {Rybicki}, {Sadowski}, {Sagrist{\`a} Sell{\'e}s}, {Sahlmann}, {Salgado}, {Salguero}, {Samaras}, {Sanchez Gimenez}, {Sanna}, {Santove{\~n}a}, {Sarasso}, {Schultheis}, {Sciacca}, {Segol}, {Segovia}, {S{\'e}gransan}, {Semeux}, {Shahaf}, {Siddiqui}, {Siebert}, {Siltala}, {Slezak}, {Smart}, {Solano}, {Solitro}, {Souami}, {Souchay}, {Spagna}, {Spoto}, {Steele}, {Steidelm{\"u}ller}, {Stephenson}, {S{\"u}veges}, {Szabados}, {Szegedi-Elek}, {Taris}, {Tauran}, {Taylor}, {Teixeira}, {Thuillot}, {Tonello}, {Torra}, {Torra}, {Turon}, {Unger}, {Vaillant}, {van Dillen}, {Vanel}, {Vecchiato}, {Viala}, {Vicente},
  {Voutsinas}, {Weiler}, {Wevers}, {Wyrzykowski}, {Yoldas}, {Yvard}, {Zhao}, {Zorec}, {Zucker}, {Zurbach}, \& {Zwitter}}]{Gaia2021}
{Gaia Collaboration}, {Brown}, A.~G.~A., {Vallenari}, A., {et~al.} 2021, \aap, 649, A1

\bibitem[{{Gaia Collaboration} {et~al.}(2023{\natexlab{b}}){Gaia Collaboration}, {Creevey}, {Sarro}, {Lobel}, {Pancino}, {Andrae}, {Smart}, {Clementini}, {Heiter}, {Korn}, {Fouesneau}, {Fr{\'e}mat}, {De Angeli}, {Vallenari}, {Harrison}, {Th{\'e}venin}, {Reyl{\'e}}, {Sordo}, {Garofalo}, {Brown}, {Eyer}, {Prusti}, {de Bruijne}, {Arenou}, {Babusiaux}, {Biermann}, {Ducourant}, {Evans}, {Guerra}, {Hutton}, {Jordi}, {Klioner}, {Lammers}, {Lindegren}, {Luri}, {Mignard}, {Panem}, {Pourbaix}, {Randich}, {Sartoretti}, {Soubiran}, {Tanga}, {Walton}, {Bailer-Jones}, {Bastian}, {Drimmel}, {Jansen}, {Katz}, {Lattanzi}, {van Leeuwen}, {Bakker}, {Cacciari}, {Casta{\~n}eda}, {Fabricius}, {Galluccio}, {Guerrier}, {Masana}, {Messineo}, {Mowlavi}, {Nicolas}, {Nienartowicz}, {Pailler}, {Panuzzo}, {Riclet}, {Roux}, {Seabroke}, {Gracia-Abril}, {Portell}, {Teyssier}, {Altmann}, {Audard}, {Bellas-Velidis}, {Benson}, {Berthier}, {Blomme}, {Burgess}, {Busonero}, {Busso}, {C{\'a}novas}, {Carry}, {Cellino}, {Cheek}, {Damerdji},
  {Davidson}, {de Teodoro}, {Nu{\~n}ez Campos}, {Delchambre}, {Dell'Oro}, {Esquej}, {Fern{\'a}ndez-Hern{\'a}ndez}, {Fraile}, {Garabato}, {Garc{\'\i}a-Lario}, {Gosset}, {Haigron}, {Halbwachs}, {Hambly}, {Hern{\'a}ndez}, {Hestroffer}, {Hodgkin}, {Holl}, {Jan{\ss}en}, {Jevardat de Fombelle}, {Jordan}, {Krone-Martins}, {Lanzafame}, {L{\"o}ffler}, {Marchal}, {Marrese}, {Moitinho}, {Muinonen}, {Osborne}, {Pauwels}, {Recio-Blanco}, {Riello}, {Rimoldini}, {Roegiers}, {Rybizki}, {Siopis}, {Smith}, {Sozzetti}, {Utrilla}, {van Leeuwen}, {Abbas}, {{\'A}brah{\'a}m}, {Abreu Aramburu}, {Aerts}, {Aguado}, {Ajaj}, {Aldea-Montero}, {Altavilla}, {{\'A}lvarez}, {Alves}, {Anders}, {Anderson}, {Anglada Varela}, {Antoja}, {Baines}, {Baker}, {Balaguer-N{\'u}{\~n}ez}, {Balbinot}, {Balog}, {Barache}, {Barbato}, {Barros}, {Barstow}, {Bartolom{\'e}}, {Bassilana}, {Bauchet}, {Becciani}, {Bellazzini}, {Berihuete}, {Bernet}, {Bertone}, {Bianchi}, {Binnenfeld}, {Blanco-Cuaresma}, {Boch}, {Bombrun}, {Bossini}, {Bouquillon}, {Bragaglia},
  {Bramante}, {Breedt}, {Bressan}, {Brouillet}, {Brugaletta}, {Bucciarelli}, {Burlacu}, {Butkevich}, {Buzzi}, {Caffau}, {Cancelliere}, {Cantat-Gaudin}, {Carballo}, {Carlucci}, {Carnerero}, {Carrasco}, {Casamiquela}, {Castellani}, {Castro-Ginard}, {Chaoul}, {Charlot}, {Chemin}, {Chiaramida}, {Chiavassa}, {Chornay}, {Comoretto}, {Contursi}, {Cooper}, {Cornez}, {Cowell}, {Crifo}, {Cropper}, {Crosta}, {Crowley}, {Dafonte}, {Dapergolas}, {David}, {de Laverny}, {De Luise}, {De March}, {De Ridder}, {de Souza}, {de Torres}, {del Peloso}, {del Pozo}, {Delbo}, {Delgado}, {Delisle}, {Demouchy}, {Dharmawardena}, {Di Matteo}, {Diakite}, {Diener}, {Distefano}, {Dolding}, {Enke}, {Fabre}, {Fabrizio}, {Faigler}, {Fedorets}, {Fernique}, {Figueras}, {Fournier}, {Fouron}, {Fragkoudi}, {Gai}, {Garcia-Gutierrez}, {Garcia-Reinaldos}, {Garc{\'\i}a-Torres}, {Gavel}, {Gavras}, {Gerlach}, {Geyer}, {Giacobbe}, {Gilmore}, {Girona}, {Giuffrida}, {Gomel}, {Gomez}, {Gonz{\'a}lez-N{\'u}{\~n}ez}, {Gonz{\'a}lez-Santamar{\'\i}a},
  {Gonz{\'a}lez-Vidal}, {Granvik}, {Guillout}, {Guiraud}, {Guti{\'e}rrez-S{\'a}nchez}, {Guy}, {Hatzidimitriou}, {Hauser}, {Haywood}, {Helmer}, {Helmi}, {Hilger}, {Sarmiento}, {Hidalgo}, {H{\l}adczuk}, {Hobbs}, {Holland}, {Huckle}, {Jardine}, {Jasniewicz}, {Jean-Antoine Piccolo}, {Jim{\'e}nez-Arranz}, {Juaristi Campillo}, {Julbe}, {Karbevska}, {Kervella}, {Khanna}, {Kordopatis}, {K{\'o}sp{\'a}l}, {Kostrzewa-Rutkowska}, {Kruszy{\'n}ska}, {Kun}, {Laizeau}, {Lambert}, {Lanza}, {Lasne}, {Le Campion}, {Lebreton}, {Lebzelter}, {Leccia}, {Leclerc}, {Lecoeur-Taibi}, {Liao}, {Licata}, {Lindstr{\o}m}, {Lister}, {Livanou}, {Lorca}, {Loup}, {Madrero Pardo}, {Magdaleno Romeo}, {Managau}, {Mann}, {Manteiga}, {Marchant}, {Marconi}, {Marcos}, {Marcos Santos}, {Mar{\'\i}n Pina}, {Marinoni}, {Marocco}, {Marshall}, {Martin Polo}, {Mart{\'\i}n-Fleitas}, {Marton}, {Mary}, {Masip}, {Massari}, {Mastrobuono-Battisti}, {Mazeh}, {McMillan}, {Messina}, {Michalik}, {Millar}, {Mints}, {Molina}, {Molinaro}, {Moln{\'a}r}, {Monari},
  {Mongui{\'o}}, {Montegriffo}, {Montero}, {Mor}, {Mora}, {Morbidelli}, {Morel}, {Morris}, {Muraveva}, {Murphy}, {Musella}, {Nagy}, {Noval}, {Oca{\~n}a}, {Ogden}, {Ordenovic}, {Osinde}, {Pagani}, {Pagano}, {Palaversa}, {Palicio}, {Pallas-Quintela}, {Panahi}, {Payne-Wardenaar}, {Pe{\~n}alosa Esteller}, {Penttil{\"a}}, {Pichon}, {Piersimoni}, {Pineau}, {Plachy}, {Plum}, {Poggio}, {Pr{\v{s}}a}, {Pulone}, {Racero}, {Ragaini}, {Rainer}, {Raiteri}, {Ramos}, {Ramos-Lerate}, {Re Fiorentin}, {Regibo}, {Richards}, {Rios Diaz}, {Ripepi}, {Riva}, {Rix}, {Rixon}, {Robichon}, {Robin}, {Robin}, {Roelens}, {Rogues}, {Rohrbasser}, {Romero-G{\'o}mez}, {Rowell}, {Royer}, {Ruz Mieres}, {Rybicki}, {Sadowski}, {S{\'a}ez N{\'u}{\~n}ez}, {Sagrist{\`a} Sell{\'e}s}, {Sahlmann}, {Salguero}, {Samaras}, {Sanchez Gimenez}, {Sanna}, {Santove{\~n}a}, {Sarasso}, {Schultheis}, {Sciacca}, {Segol}, {Segovia}, {S{\'e}gransan}, {Semeux}, {Shahaf}, {Siddiqui}, {Siebert}, {Siltala}, {Silvelo}, {Slezak}, {Slezak}, {Snaith}, {Solano}, {Solitro},
  {Souami}, {Souchay}, {Spagna}, {Spina}, {Spoto}, {Steele}, {Steidelm{\"u}ller}, {Stephenson}, {S{\"u}veges}, {Surdej}, {Szabados}, {Szegedi-Elek}, {Taris}, {Taylor}, {Teixeira}, {Tolomei}, {Tonello}, {Torra}, {Torra}, {Torralba Elipe}, {Trabucchi}, {Tsounis}, {Turon}, {Ulla}, {Unger}, {Vaillant}, {van Dillen}, {van Reeven}, {Vanel}, {Vecchiato}, {Viala}, {Vicente}, {Voutsinas}, {Weiler}, {Wevers}, {Wyrzykowski}, {Yoldas}, {Yvard}, {Zhao}, {Zorec}, {Zucker}, \& {Zwitter}}]{Creevey2023}
{Gaia Collaboration}, {Creevey}, O.~L., {Sarro}, L.~M., {et~al.} 2023{\natexlab{b}}, \aap, 674, A39

\bibitem[{{Gaia Collaboration} {et~al.}(2016){Gaia Collaboration}, {Prusti}, {de Bruijne}, {Brown}, {Vallenari}, {Babusiaux}, {Bailer-Jones}, {Bastian}, {Biermann}, {Evans}, {Eyer}, {Jansen}, {Jordi}, {Klioner}, {Lammers}, {Lindegren}, {Luri}, {Mignard}, {Milligan}, {Panem}, {Poinsignon}, {Pourbaix}, {Randich}, {Sarri}, {Sartoretti}, {Siddiqui}, {Soubiran}, {Valette}, {van Leeuwen}, {Walton}, {Aerts}, {Arenou}, {Cropper}, {Drimmel}, {H{\o}g}, {Katz}, {Lattanzi}, {O'Mullane}, {Grebel}, {Holland}, {Huc}, {Passot}, {Bramante}, {Cacciari}, {Casta{\~n}eda}, {Chaoul}, {Cheek}, {De Angeli}, {Fabricius}, {Guerra}, {Hern{\'a}ndez}, {Jean-Antoine-Piccolo}, {Masana}, {Messineo}, {Mowlavi}, {Nienartowicz}, {Ord{\'o}{\~n}ez-Blanco}, {Panuzzo}, {Portell}, {Richards}, {Riello}, {Seabroke}, {Tanga}, {Th{\'e}venin}, {Torra}, {Els}, {Gracia-Abril}, {Comoretto}, {Garcia-Reinaldos}, {Lock}, {Mercier}, {Altmann}, {Andrae}, {Astraatmadja}, {Bellas-Velidis}, {Benson}, {Berthier}, {Blomme}, {Busso}, {Carry}, {Cellino}, {Clementini},
  {Cowell}, {Creevey}, {Cuypers}, {Davidson}, {De Ridder}, {de Torres}, {Delchambre}, {Dell'Oro}, {Ducourant}, {Fr{\'e}mat}, {Garc{\'\i}a-Torres}, {Gosset}, {Halbwachs}, {Hambly}, {Harrison}, {Hauser}, {Hestroffer}, {Hodgkin}, {Huckle}, {Hutton}, {Jasniewicz}, {Jordan}, {Kontizas}, {Korn}, {Lanzafame}, {Manteiga}, {Moitinho}, {Muinonen}, {Osinde}, {Pancino}, {Pauwels}, {Petit}, {Recio-Blanco}, {Robin}, {Sarro}, {Siopis}, {Smith}, {Smith}, {Sozzetti}, {Thuillot}, {van Reeven}, {Viala}, {Abbas}, {Abreu Aramburu}, {Accart}, {Aguado}, {Allan}, {Allasia}, {Altavilla}, {{\'A}lvarez}, {Alves}, {Anderson}, {Andrei}, {Anglada Varela}, {Antiche}, {Antoja}, {Ant{\'o}n}, {Arcay}, {Atzei}, {Ayache}, {Bach}, {Baker}, {Balaguer-N{\'u}{\~n}ez}, {Barache}, {Barata}, {Barbier}, {Barblan}, {Baroni}, {Barrado y Navascu{\'e}s}, {Barros}, {Barstow}, {Becciani}, {Bellazzini}, {Bellei}, {Bello Garc{\'\i}a}, {Belokurov}, {Bendjoya}, {Berihuete}, {Bianchi}, {Bienaym{\'e}}, {Billebaud}, {Blagorodnova}, {Blanco-Cuaresma}, {Boch},
  {Bombrun}, {Borrachero}, {Bouquillon}, {Bourda}, {Bouy}, {Bragaglia}, {Breddels}, {Brouillet}, {Br{\"u}semeister}, {Bucciarelli}, {Budnik}, {Burgess}, {Burgon}, {Burlacu}, {Busonero}, {Buzzi}, {Caffau}, {Cambras}, {Campbell}, {Cancelliere}, {Cantat-Gaudin}, {Carlucci}, {Carrasco}, {Castellani}, {Charlot}, {Charnas}, {Charvet}, {Chassat}, {Chiavassa}, {Clotet}, {Cocozza}, {Collins}, {Collins}, {Costigan}, {Crifo}, {Cross}, {Crosta}, {Crowley}, {Dafonte}, {Damerdji}, {Dapergolas}, {David}, {David}, {De Cat}, {de Felice}, {de Laverny}, {De Luise}, {De March}, {de Martino}, {de Souza}, {Debosscher}, {del Pozo}, {Delbo}, {Delgado}, {Delgado}, {di Marco}, {Di Matteo}, {Diakite}, {Distefano}, {Dolding}, {Dos Anjos}, {Drazinos}, {Dur{\'a}n}, {Dzigan}, {Ecale}, {Edvardsson}, {Enke}, {Erdmann}, {Escolar}, {Espina}, {Evans}, {Eynard Bontemps}, {Fabre}, {Fabrizio}, {Faigler}, {Falc{\~a}o}, {Farr{\`a}s Casas}, {Faye}, {Federici}, {Fedorets}, {Fern{\'a}ndez-Hern{\'a}ndez}, {Fernique}, {Fienga}, {Figueras}, {Filippi},
  {Findeisen}, {Fonti}, {Fouesneau}, {Fraile}, {Fraser}, {Fuchs}, {Furnell}, {Gai}, {Galleti}, {Galluccio}, {Garabato}, {Garc{\'\i}a-Sedano}, {Gar{\'e}}, {Garofalo}, {Garralda}, {Gavras}, {Gerssen}, {Geyer}, {Gilmore}, {Girona}, {Giuffrida}, {Gomes}, {Gonz{\'a}lez-Marcos}, {Gonz{\'a}lez-N{\'u}{\~n}ez}, {Gonz{\'a}lez-Vidal}, {Granvik}, {Guerrier}, {Guillout}, {Guiraud}, {G{\'u}rpide}, {Guti{\'e}rrez-S{\'a}nchez}, {Guy}, {Haigron}, {Hatzidimitriou}, {Haywood}, {Heiter}, {Helmi}, {Hobbs}, {Hofmann}, {Holl}, {Holland}, {Hunt}, {Hypki}, {Icardi}, {Irwin}, {Jevardat de Fombelle}, {Jofr{\'e}}, {Jonker}, {Jorissen}, {Julbe}, {Karampelas}, {Kochoska}, {Kohley}, {Kolenberg}, {Kontizas}, {Koposov}, {Kordopatis}, {Koubsky}, {Kowalczyk}, {Krone-Martins}, {Kudryashova}, {Kull}, {Bachchan}, {Lacoste-Seris}, {Lanza}, {Lavigne}, {Le Poncin-Lafitte}, {Lebreton}, {Lebzelter}, {Leccia}, {Leclerc}, {Lecoeur-Taibi}, {Lemaitre}, {Lenhardt}, {Leroux}, {Liao}, {Licata}, {Lindstr{\o}m}, {Lister}, {Livanou}, {Lobel}, {L{\"o}ffler},
  {L{\'o}pez}, {Lopez-Lozano}, {Lorenz}, {Loureiro}, {MacDonald}, {Magalh{\~a}es Fernandes}, {Managau}, {Mann}, {Mantelet}, {Marchal}, {Marchant}, {Marconi}, {Marie}, {Marinoni}, {Marrese}, {Marschalk{\'o}}, {Marshall}, {Mart{\'\i}n-Fleitas}, {Martino}, {Mary}, {Matijevi{\v{c}}}, {Mazeh}, {McMillan}, {Messina}, {Mestre}, {Michalik}, {Millar}, {Miranda}, {Molina}, {Molinaro}, {Molinaro}, {Moln{\'a}r}, {Moniez}, {Montegriffo}, {Monteiro}, {Mor}, {Mora}, {Morbidelli}, {Morel}, {Morgenthaler}, {Morley}, {Morris}, {Mulone}, {Muraveva}, {Musella}, {Narbonne}, {Nelemans}, {Nicastro}, {Noval}, {Ord{\'e}novic}, {Ordieres-Mer{\'e}}, {Osborne}, {Pagani}, {Pagano}, {Pailler}, {Palacin}, {Palaversa}, {Parsons}, {Paulsen}, {Pecoraro}, {Pedrosa}, {Pentik{\"a}inen}, {Pereira}, {Pichon}, {Piersimoni}, {Pineau}, {Plachy}, {Plum}, {Poujoulet}, {Pr{\v{s}}a}, {Pulone}, {Ragaini}, {Rago}, {Rambaux}, {Ramos-Lerate}, {Ranalli}, {Rauw}, {Read}, {Regibo}, {Renk}, {Reyl{\'e}}, {Ribeiro}, {Rimoldini}, {Ripepi}, {Riva}, {Rixon},
  {Roelens}, {Romero-G{\'o}mez}, {Rowell}, {Royer}, {Rudolph}, {Ruiz-Dern}, {Sadowski}, {Sagrist{\`a} Sell{\'e}s}, {Sahlmann}, {Salgado}, {Salguero}, {Sarasso}, {Savietto}, {Schnorhk}, {Schultheis}, {Sciacca}, {Segol}, {Segovia}, {Segransan}, {Serpell}, {Shih}, {Smareglia}, {Smart}, {Smith}, {Solano}, {Solitro}, {Sordo}, {Soria Nieto}, {Souchay}, {Spagna}, {Spoto}, {Stampa}, {Steele}, {Steidelm{\"u}ller}, {Stephenson}, {Stoev}, {Suess}, {S{\"u}veges}, {Surdej}, {Szabados}, {Szegedi-Elek}, {Tapiador}, {Taris}, {Tauran}, {Taylor}, {Teixeira}, {Terrett}, {Tingley}, {Trager}, {Turon}, {Ulla}, {Utrilla}, {Valentini}, {van Elteren}, {Van Hemelryck}, {van Leeuwen}, {Varadi}, {Vecchiato}, {Veljanoski}, {Via}, {Vicente}, {Vogt}, {Voss}, {Votruba}, {Voutsinas}, {Walmsley}, {Weiler}, {Weingrill}, {Werner}, {Wevers}, {Whitehead}, {Wyrzykowski}, {Yoldas}, {{\v{Z}}erjal}, {Zucker}, {Zurbach}, {Zwitter}, {Alecu}, {Allen}, {Allende Prieto}, {Amorim}, {Anglada-Escud{\'e}}, {Arsenijevic}, {Azaz}, {Balm}, {Beck}, {Bernstein},
  {Bigot}, {Bijaoui}, {Blasco}, {Bonfigli}, {Bono}, {Boudreault}, {Bressan}, {Brown}, {Brunet}, {Bunclark}, {Buonanno}, {Butkevich}, {Carret}, {Carrion}, {Chemin}, {Ch{\'e}reau}, {Corcione}, {Darmigny}, {de Boer}, {de Teodoro}, {de Zeeuw}, {Delle Luche}, {Domingues}, {Dubath}, {Fodor}, {Fr{\'e}zouls}, {Fries}, {Fustes}, {Fyfe}, {Gallardo}, {Gallegos}, {Gardiol}, {Gebran}, {Gomboc}, {G{\'o}mez}, {Grux}, {Gueguen}, {Heyrovsky}, {Hoar}, {Iannicola}, {Isasi Parache}, {Janotto}, {Joliet}, {Jonckheere}, {Keil}, {Kim}, {Klagyivik}, {Klar}, {Knude}, {Kochukhov}, {Kolka}, {Kos}, {Kutka}, {Lainey}, {LeBouquin}, {Liu}, {Loreggia}, {Makarov}, {Marseille}, {Martayan}, {Martinez-Rubi}, {Massart}, {Meynadier}, {Mignot}, {Munari}, {Nguyen}, {Nordlander}, {Ocvirk}, {O'Flaherty}, {Olias Sanz}, {Ortiz}, {Osorio}, {Oszkiewicz}, {Ouzounis}, {Palmer}, {Park}, {Pasquato}, {Peltzer}, {Peralta}, {P{\'e}turaud}, {Pieniluoma}, {Pigozzi}, {Poels}, {Prat}, {Prod'homme}, {Raison}, {Rebordao}, {Risquez}, {Rocca-Volmerange}, {Rosen},
  {Ruiz-Fuertes}, {Russo}, {Sembay}, {Serraller Vizcaino}, {Short}, {Siebert}, {Silva}, {Sinachopoulos}, {Slezak}, {Soffel}, {Sosnowska}, {Strai{\v{z}}ys}, {ter Linden}, {Terrell}, {Theil}, {Tiede}, {Troisi}, {Tsalmantza}, {Tur}, {Vaccari}, {Vachier}, {Valles}, {Van Hamme}, {Veltz}, {Virtanen}, {Wallut}, {Wichmann}, {Wilkinson}, {Ziaeepour}, \& {Zschocke}}]{Gaia2016}
{Gaia Collaboration}, {Prusti}, T., {de Bruijne}, J.~H.~J., {et~al.} 2016, \aap, 595, A1

\bibitem[{{Gatewood} \& {Gatewood}(1978)}]{Gatewood1978}
{Gatewood}, G.~D. \& {Gatewood}, C.~V. 1978, \apj, 225, 191

\bibitem[{{Gorrini} {et~al.}(2022){Gorrini}, {Astudillo-Defru}, {Dreizler}, {Damasso}, {D{\'\i}az}, {Bonfils}, {Jeffers}, {Barnes}, {Del Sordo}, {Almenara}, {Artigau}, {Bouchy}, {Charbonneau}, {Delfosse}, {Doyon}, {Figueira}, {Forveille}, {Haswell}, {L{\'o}pez-Gonz{\'a}lez}, {Melo}, {Mennickent}, {Gaisn{\'e}}, {Morales Morales}, {Murgas}, {Pepe}, {Rodr{\'\i}guez}, {Santos}, {Tal-Or}, {Tsapras}, \& {Udry}}]{Gorrini2022}
{Gorrini}, P., {Astudillo-Defru}, N., {Dreizler}, S., {et~al.} 2022, \aap, 664, A64

\bibitem[{{Halbwachs} {et~al.}(2000){Halbwachs}, {Arenou}, {Mayor}, {Udry}, \& {Queloz}}]{Halbwachs2000}
{Halbwachs}, J.~L., {Arenou}, F., {Mayor}, M., {Udry}, S., \& {Queloz}, D. 2000, \aap, 355, 581

\bibitem[{{Halbwachs} {et~al.}(2023){Halbwachs}, {Pourbaix}, {Arenou}, {Galluccio}, {Guillout}, {Bauchet}, {Marchal}, {Sadowski}, \& {Teyssier}}]{Halbwachs2023}
{Halbwachs}, J.-L., {Pourbaix}, D., {Arenou}, F., {et~al.} 2023, \aap, 674, A9

\bibitem[{{Hale}(1995)}]{Hale1995}
{Hale}, A. 1995, \pasp, 107, 22

\bibitem[{{Holl} {et~al.}(2022){Holl}, {Perryman}, {Lindegren}, {Segransan}, \& {Raimbault}}]{Holl2022}
{Holl}, B., {Perryman}, M., {Lindegren}, L., {Segransan}, D., \& {Raimbault}, M. 2022, \aap, 661, A151

\bibitem[{{Holl} {et~al.}(2023){Holl}, {Sozzetti}, {Sahlmann}, {Giacobbe}, {S{\'e}gransan}, {Unger}, {Delisle}, {Barbato}, {Lattanzi}, {Morbidelli}, \& {Sosnowska}}]{Holl2023}
{Holl}, B., {Sozzetti}, A., {Sahlmann}, J., {et~al.} 2023, \aap, 674, A10

\bibitem[{{Kane} {et~al.}(2011){Kane}, {Henry}, {Dragomir}, {Fischer}, {Howard}, {Wang}, \& {Wright}}]{Kane2011}
{Kane}, S.~R., {Henry}, G.~W., {Dragomir}, D., {et~al.} 2011, \apjl, 735, L41

\bibitem[{{Kervella} {et~al.}(2019){Kervella}, {Arenou}, {Mignard}, \& {Th{\'e}venin}}]{Kervella2019}
{Kervella}, P., {Arenou}, F., {Mignard}, F., \& {Th{\'e}venin}, F. 2019, \aap, 623, A72

\bibitem[{{Kervella} {et~al.}(2022){Kervella}, {Arenou}, \& {Th{\'e}venin}}]{Kervella2022}
{Kervella}, P., {Arenou}, F., \& {Th{\'e}venin}, F. 2022, \aap, 657, A7

\bibitem[{{Kiefer}(2019)}]{Kiefer2019b}
{Kiefer}, F. 2019, \aap, 632, L9

\bibitem[{{Kiefer} {et~al.}(2021){Kiefer}, {H{\'e}brard}, {Lecavelier des Etangs}, {Martioli}, {Dalal}, \& {Vidal-Madjar}}]{Kiefer2021}
{Kiefer}, F., {H{\'e}brard}, G., {Lecavelier des Etangs}, A., {et~al.} 2021, \aap, 645, A7

\bibitem[{{Kiefer} {et~al.}(2019){Kiefer}, {H{\'e}brard}, {Sahlmann}, {Sousa}, {Forveille}, {Santos}, {Mayor}, {Deleuil}, {Wilson}, {Dalal}, {D{\'\i}az}, {Henry}, {Hagelberg}, {Hobson}, {Demangeon}, {Bourrier}, {Delfosse}, {Arnold}, {Astudillo-Defru}, {Beuzit}, {Boisse}, {Bonfils}, {Borgniet}, {Bouchy}, {Courcol}, {Ehrenreich}, {Hara}, {Lagrange}, {Lovis}, {Montagnier}, {Moutou}, {Pepe}, {Perrier}, {Rey}, {Santerne}, {S{\'e}gransan}, {Udry}, \& {Vidal-Madjar}}]{Kiefer2019a}
{Kiefer}, F., {H{\'e}brard}, G., {Sahlmann}, J., {et~al.} 2019, \aap, 631, A125

\bibitem[{{Kiefer} {et~al.}(2024){Kiefer}, {Lagrange}, {Rubini}, \& {Philipot}}]{Kiefer2024c}
{Kiefer}, F., {Lagrange}, A.-M., {Rubini}, P., \& {Philipot}, F. 2024, \aap, submitted

\bibitem[{{Lagrange} {et~al.}(2024){Lagrange}, {Kiefer}, {Rubini}, {Squicciarini}, {Chomez}, {Milli}, {Zurlo}, {Bouvier}, {Delorme}, {Beust}, {Mazoyer}, {Flasseur}, {Chauvin}, \& {Palma-Bifani}}]{Lagrange2024a}
{Lagrange}, A.-M., {Kiefer}, F., {Rubini}, P., {et~al.} 2024, \aap, submitted

\bibitem[{{Lagrange} {et~al.}(2020){Lagrange}, {Rubini}, {Nowak}, {Lacour}, {Grandjean}, {Boccaletti}, {Langlois}, {Delorme}, {Gratton}, {Wang}, {Flasseur}, {Galicher}, {Kral}, {Meunier}, {Beust}, {Babusiaux}, {Le Coroller}, {Thebault}, {Kervella}, {Zurlo}, {Maire}, {Wahhaj}, {Amorim}, {Asensio-Torres}, {Benisty}, {Berger}, {Bonnefoy}, {Brandner}, {Cantalloube}, {Charnay}, {Chauvin}, {Choquet}, {Cl{\'e}net}, {Christiaens}, {Coud{\'e} Du Foresto}, {de Zeeuw}, {Desidera}, {Duvert}, {Eckart}, {Eisenhauer}, {Galland}, {Gao}, {Garcia}, {Garcia Lopez}, {Gendron}, {Genzel}, {Gillessen}, {Girard}, {Hagelberg}, {Haubois}, {Henning}, {Heissel}, {Hippler}, {Horrobin}, {Janson}, {Kammerer}, {Kenworthy}, {Keppler}, {Kreidberg}, {Lapeyr{\`e}re}, {Le Bouquin}, {L{\'e}na}, {M{\'e}rand}, {Messina}, {Molli{\`e}re}, {Monnier}, {Ott}, {Otten}, {Paumard}, {Paladini}, {Perraut}, {Perrin}, {Pueyo}, {Pfuhl}, {Rodet}, {Rodriguez-Coira}, {Rousset}, {Samland}, {Shangguan}, {Schmidt}, {Straub}, {Straubmeier}, {Stolker}, {Vigan},
  {Vincent}, {Widmann}, {Woillez}, \& {GRAVITY Collaboration}}]{Lagrange2020}
{Lagrange}, A.~M., {Rubini}, P., {Nowak}, M., {et~al.} 2020, \aap, 642, A18

\bibitem[{{Latham} {et~al.}(1989){Latham}, {Mazeh}, {Stefanik}, {Mayor}, \& {Burki}}]{Latham1989}
{Latham}, D.~W., {Mazeh}, T., {Stefanik}, R.~P., {Mayor}, M., \& {Burki}, G. 1989, \nat, 339, 38

\bibitem[{{Li} {et~al.}(2021){Li}, {Brandt}, {Brandt}, {Dupuy}, {Michalik}, {Jensen-Clem}, {Zeng}, {Faherty}, \& {Mitra}}]{Li2021}
{Li}, Y., {Brandt}, T.~D., {Brandt}, G.~M., {et~al.} 2021, \aj, 162, 266

\bibitem[{{Lindegren} {et~al.}(2018){Lindegren}, {Hern{\'a}ndez}, {Bombrun}, {Klioner}, {Bastian}, {Ramos-Lerate}, {de Torres}, {Steidelm{\"u}ller}, {Stephenson}, {Hobbs}, {Lammers}, {Biermann}, {Geyer}, {Hilger}, {Michalik}, {Stampa}, {McMillan}, {Casta{\~n}eda}, {Clotet}, {Comoretto}, {Davidson}, {Fabricius}, {Gracia}, {Hambly}, {Hutton}, {Mora}, {Portell}, {van Leeuwen}, {Abbas}, {Abreu}, {Altmann}, {Andrei}, {Anglada}, {Balaguer-N{\'u}{\~n}ez}, {Barache}, {Becciani}, {Bertone}, {Bianchi}, {Bouquillon}, {Bourda}, {Br{\"u}semeister}, {Bucciarelli}, {Busonero}, {Buzzi}, {Cancelliere}, {Carlucci}, {Charlot}, {Cheek}, {Crosta}, {Crowley}, {de Bruijne}, {de Felice}, {Drimmel}, {Esquej}, {Fienga}, {Fraile}, {Gai}, {Garralda}, {Gonz{\'a}lez-Vidal}, {Guerra}, {Hauser}, {Hofmann}, {Holl}, {Jordan}, {Lattanzi}, {Lenhardt}, {Liao}, {Licata}, {Lister}, {L{\"o}ffler}, {Marchant}, {Martin-Fleitas}, {Messineo}, {Mignard}, {Morbidelli}, {Poggio}, {Riva}, {Rowell}, {Salguero}, {Sarasso}, {Sciacca}, {Siddiqui}, {Smart},
  {Spagna}, {Steele}, {Taris}, {Torra}, {van Elteren}, {van Reeven}, \& {Vecchiato}}]{Lindegren2018}
{Lindegren}, L., {Hern{\'a}ndez}, J., {Bombrun}, A., {et~al.} 2018, \aap, 616, A2

\bibitem[{{Lindegren} {et~al.}(2021){Lindegren}, {Klioner}, {Hern{\'a}ndez}, {Bombrun}, {Ramos-Lerate}, {Steidelm{\"u}ller}, {Bastian}, {Biermann}, {de Torres}, {Gerlach}, {Geyer}, {Hilger}, {Hobbs}, {Lammers}, {McMillan}, {Stephenson}, {Casta{\~n}eda}, {Davidson}, {Fabricius}, {Gracia-Abril}, {Portell}, {Rowell}, {Teyssier}, {Torra}, {Bartolom{\'e}}, {Clotet}, {Garralda}, {Gonz{\'a}lez-Vidal}, {Torra}, {Abbas}, {Altmann}, {Anglada Varela}, {Balaguer-N{\'u}{\~n}ez}, {Balog}, {Barache}, {Becciani}, {Bernet}, {Bertone}, {Bianchi}, {Bouquillon}, {Brown}, {Bucciarelli}, {Busonero}, {Butkevich}, {Buzzi}, {Cancelliere}, {Carlucci}, {Charlot}, {Cioni}, {Crosta}, {Crowley}, {del Peloso}, {del Pozo}, {Drimmel}, {Esquej}, {Fienga}, {Fraile}, {Gai}, {Garcia-Reinaldos}, {Guerra}, {Hambly}, {Hauser}, {Jan{\ss}en}, {Jordan}, {Kostrzewa-Rutkowska}, {Lattanzi}, {Liao}, {Licata}, {Lister}, {L{\"o}ffler}, {Marchant}, {Masip}, {Mignard}, {Mints}, {Molina}, {Mora}, {Morbidelli}, {Murphy}, {Pagani}, {Panuzzo}, {Pe{\~n}alosa
  Esteller}, {Poggio}, {Re Fiorentin}, {Riva}, {Sagrist{\`a} Sell{\'e}s}, {Sanchez Gimenez}, {Sarasso}, {Sciacca}, {Siddiqui}, {Smart}, {Souami}, {Spagna}, {Steele}, {Taris}, {Utrilla}, {van Reeven}, \& {Vecchiato}}]{Lindegren2021}
{Lindegren}, L., {Klioner}, S.~A., {Hern{\'a}ndez}, J., {et~al.} 2021, \aap, 649, A2

\bibitem[{{Lindegren} {et~al.}(2016){Lindegren}, {Lammers}, {Bastian}, {Hern{\'a}ndez}, {Klioner}, {Hobbs}, {Bombrun}, {Michalik}, {Ramos-Lerate}, {Butkevich}, {Comoretto}, {Joliet}, {Holl}, {Hutton}, {Parsons}, {Steidelm{\"u}ller}, {Abbas}, {Altmann}, {Andrei}, {Anton}, {Bach}, {Barache}, {Becciani}, {Berthier}, {Bianchi}, {Biermann}, {Bouquillon}, {Bourda}, {Br{\"u}semeister}, {Bucciarelli}, {Busonero}, {Carlucci}, {Casta{\~n}eda}, {Charlot}, {Clotet}, {Crosta}, {Davidson}, {de Felice}, {Drimmel}, {Fabricius}, {Fienga}, {Figueras}, {Fraile}, {Gai}, {Garralda}, {Geyer}, {Gonz{\'a}lez-Vidal}, {Guerra}, {Hambly}, {Hauser}, {Jordan}, {Lattanzi}, {Lenhardt}, {Liao}, {L{\"o}ffler}, {McMillan}, {Mignard}, {Mora}, {Morbidelli}, {Portell}, {Riva}, {Sarasso}, {Serraller}, {Siddiqui}, {Smart}, {Spagna}, {Stampa}, {Steele}, {Taris}, {Torra}, {van Reeven}, {Vecchiato}, {Zschocke}, {de Bruijne}, {Gracia}, {Raison}, {Lister}, {Marchant}, {Messineo}, {Soffel}, {Osorio}, {de Torres}, \& {O'Mullane}}]{Lindegren2016}
{Lindegren}, L., {Lammers}, U., {Bastian}, U., {et~al.} 2016, \aap, 595, A4

\bibitem[{{Lindegren} {et~al.}(2012){Lindegren}, {Lammers}, {Hobbs}, {O'Mullane}, {Bastian}, \& {Hern{\'a}ndez}}]{Lindegren2012}
{Lindegren}, L., {Lammers}, U., {Hobbs}, D., {et~al.} 2012, \aap, 538, A78

\bibitem[{{Lucas} {et~al.}(2022){Lucas}, {Bottom}, {Ruane}, \& {Ragland}}]{Lucas2022}
{Lucas}, M., {Bottom}, M., {Ruane}, G., \& {Ragland}, S. 2022, \aj, 163, 81

\bibitem[{{Mamajek} \& {Bell}(2014)}]{Mamajek2014}
{Mamajek}, E.~E. \& {Bell}, C. P.~M. 2014, \mnras, 445, 2169

\bibitem[{{Mesa} {et~al.}(2023){Mesa}, {Gratton}, {Kervella}, {Bonavita}, {Desidera}, {D'Orazi}, {Marino}, {Zurlo}, \& {Rigliaco}}]{Mesa2023}
{Mesa}, D., {Gratton}, R., {Kervella}, P., {et~al.} 2023, \aap, 672, A93

\bibitem[{{Michalik} {et~al.}(2014){Michalik}, {Lindegren}, {Hobbs}, \& {Lammers}}]{Michalik2014}
{Michalik}, D., {Lindegren}, L., {Hobbs}, D., \& {Lammers}, U. 2014, \aap, 571, A85

\bibitem[{{Pecaut} \& {Mamajek}(2013)}]{Pecaut2013}
{Pecaut}, M.~J. \& {Mamajek}, E.~E. 2013, \apjs, 208, 9

\bibitem[{{Pecaut} {et~al.}(2012){Pecaut}, {Mamajek}, \& {Bubar}}]{Pecaut2012}
{Pecaut}, M.~J., {Mamajek}, E.~E., \& {Bubar}, E.~J. 2012, \apj, 746, 154

\bibitem[{{Perrier} {et~al.}(2003){Perrier}, {Sivan}, {Naef}, {Beuzit}, {Mayor}, {Queloz}, \& {Udry}}]{Perrier2003}
{Perrier}, C., {Sivan}, J.~P., {Naef}, D., {et~al.} 2003, \aap, 410, 1039

\bibitem[{{Perryman} {et~al.}(2014){Perryman}, {Hartman}, {Bakos}, \& {Lindegren}}]{Perryman2014}
{Perryman}, M., {Hartman}, J., {Bakos}, G.~{\'A}., \& {Lindegren}, L. 2014, \apj, 797, 14

\bibitem[{{Philipot} {et~al.}(2023{\natexlab{a}}){Philipot}, {Lagrange}, {Kiefer}, {Rubini}, {Delorme}, \& {Chomez}}]{Philipot2023b}
{Philipot}, F., {Lagrange}, A.~M., {Kiefer}, F., {et~al.} 2023{\natexlab{a}}, \aap, 678, A107

\bibitem[{{Philipot} {et~al.}(2023{\natexlab{b}}){Philipot}, {Lagrange}, {Rubini}, {Kiefer}, \& {Chomez}}]{Philipot2023a}
{Philipot}, F., {Lagrange}, A.~M., {Rubini}, P., {Kiefer}, F., \& {Chomez}, A. 2023{\natexlab{b}}, \aap, 670, A65

\bibitem[{{Press} {et~al.}(2002){Press}, {Teukolsky}, {Vetterling}, \& {Flannery}}]{NumericalRecipes}
{Press}, W.~H., {Teukolsky}, S.~A., {Vetterling}, W.~T., \& {Flannery}, B.~P. 2002, {Numerical recipes in C++ : the art of scientific computing} ({Cambridge University Press})

\bibitem[{{Sahlmann} {et~al.}(2015){Sahlmann}, {Triaud}, \& {Martin}}]{Sahlmann2015}
{Sahlmann}, J., {Triaud}, A.~H.~M.~J., \& {Martin}, D.~V. 2015, \mnras, 447, 287

\bibitem[{Silvey(1970)}]{Silvey1970}
Silvey, S.~D. 1970, Statistical inference, by S. D. Silvey (Penguin Harmondsworth), 3--192 p.

\bibitem[{{Sozzetti} {et~al.}(2006){Sozzetti}, {Udry}, {Zucker}, {Torres}, {Beuzit}, {Latham}, {Mayor}, {Mazeh}, {Naef}, {Perrier}, {Queloz}, \& {Sivan}}]{Sozzetti2006}
{Sozzetti}, A., {Udry}, S., {Zucker}, S., {et~al.} 2006, \aap, 449, 417

\bibitem[{{Stassun} {et~al.}(2017){Stassun}, {Collins}, \& {Gaudi}}]{Stassun2017}
{Stassun}, K.~G., {Collins}, K.~A., \& {Gaudi}, B.~S. 2017, \aj, 153, 136

\bibitem[{{van den Bos}(1960)}]{vandenbos1960}
{van den Bos}, W.~H. 1960, Journal des Observateurs, 43, 145

\bibitem[{{van Leeuwen}(2007)}]{vanLeeuwen2007}
{van Leeuwen}, F. 2007, \aap, 474, 653

\bibitem[{{Volet}(1932)}]{Volet1932}
{Volet}, C. 1932, Bulletin Astronomique, 8, 51

\bibitem[{Wilks(1938)}]{Wilks1938}
Wilks, S.~S. 1938, Annals Math. Statist., 9, 60

\bibitem[{{Wilson} \& {Hilferty}(1931)}]{Wilson1931}
{Wilson}, E.~B. \& {Hilferty}, M.~M. 1931, Proceedings of the National Academy of Science, 17, 684

\bibitem[{{Winn}(2022)}]{Winn2022}
{Winn}, J.~N. 2022, \aj, 164, 196

\bibitem[{{Wittenmyer} {et~al.}(2009){Wittenmyer}, {Endl}, {Cochran}, {Levison}, \& {Henry}}]{Wittenmyer2009}
{Wittenmyer}, R.~A., {Endl}, M., {Cochran}, W.~D., {Levison}, H.~F., \& {Henry}, G.~W. 2009, \apjs, 182, 97

\bibitem[{{Wittenmyer} {et~al.}(2014){Wittenmyer}, {Tuomi}, {Butler}, {Jones}, {Anglada-Escud{\'e}}, {Horner}, {Tinney}, {Marshall}, {Carter}, {Bailey}, {Salter}, {O'Toole}, {Wright}, {Crane}, {Schectman}, {Arriagada}, {Thompson}, {Minniti}, {Jenkins}, \& {Diaz}}]{Wittenmyer2014}
{Wittenmyer}, R.~A., {Tuomi}, M., {Butler}, R.~P., {et~al.} 2014, \apj, 791, 114

\bibitem[{{Xiao} {et~al.}(2023){Xiao}, {Liu}, {Teng}, {Wang}, {Brandt}, {Zhao}, {Zhao}, {Zhai}, \& {Gao}}]{Xiao2023}
{Xiao}, G.-Y., {Liu}, Y.-J., {Teng}, H.-Y., {et~al.} 2023, Research in Astronomy and Astrophysics, 23, 055022

\end{thebibliography}

\begin{appendix}

\onecolumn
\section{Table of acronyms used in the text with their definitions and page references}
\label{sec:acronyms}
\vspace{-1cm}
\setglossarystyle{longragged3col-booktabs}
\printglossary[type=\acronymtype,title=]
\FloatBarrier

\newpage
\section{Additional table}
\begin{table*}[hbt]
\centering
\caption{\label{tab:examples_param}Parameters for the illustrative cases discussed in this paper.}
\resizebox*{!}{0.90\textheight}{%
\begin{tabular}{lcccccccc}
Parameters                                           & Units                   & \multicolumn{6}{c}{Targets} \\
\hline
Target name                                          &                         & HD114762            & GJ832               & HD81040            & AF Lep              & HD23596            & Alf Cma B           & Beta Pic            \\
\hline
\multicolumn{8}{l}{\textit{Aliases}} \\ \\
HIP name                                             &                         & HIP 64426           & HIP 106440          & HIP 46076          & HIP 25486           & HIP 17747          &                     & HIP 27321           \\
Gaia DR3 ID                                          &                         & 3937211745-         & 6562924609-         & 637329067-         & 3009908378-         & 224870885-         & 2947050466-         & 4792774797-         \\
                                                     &                         & 905473024           & 150908416           & 477530368          & 049913216           & 460646016          & 531873024           & 545800832           \\
\hline
\multicolumn{8}{l}{\textit{Main parameters}} \\ \\
$M_\star$                                            & $\mathrm{M_{\odot}}$    & 1.047               & 0.480               & 1.070              & 1.200               & 1.320              & 1.050               & 1.670               \\
$\sigma_{M\star}$                                    & $\mathrm{M_{\odot}}$    & 0.105               & 0.050               & 0.107              & 0.060               & 0.020              & 0.100               & 0.167               \\
V                                                    & $\mathrm{}$             & ---                 & 8.67                & ---                & 6.30                & 7.24               & 8.44                & 3.86                \\
RA                                                   &                         & 13:12:19.0912       & 21:33:33.9004       & 09:23:46.9152      & 05:27:04.7817       & 03:48:00.4487      & 06:45:08.7901       & 05:47:17.0964       \\
DEC                                                  &                         & +17:31:01.647       & -49:00:45.468       & +20:21:52.606      & -11:54:04.255       & +40:31:50.641      & -16:43:15.357       & -51:03:58.096       \\
\hline
\multicolumn{8}{l}{\textit{\glsxtrshort{g3} data}} \\ \\
\texttt{ra}                                          & $\mathrm{{}^{\circ}}$   & 198.080             & 323.391             & 140.945            & 81.770              & 57.002             & 101.287             & 86.821              \\
\texttt{ra\_error}                                   & $\mathrm{{}^{\circ}}$   & 0.078               & 0.018               & 0.033              & 0.012               & 0.026              & 0.165               & 0.137               \\
\texttt{dec}                                         & $\mathrm{{}^{\circ}}$   & 17.517              & -49.013             & 20.365             & -11.901             & 40.531             & -16.721             & -51.066             \\
\texttt{dec\_error}                                  & $\mathrm{{}^{\circ}}$   & 0.067               & 0.014               & 0.026              & 0.010               & 0.018              & 0.227               & 0.131               \\
\texttt{pmra}                                        & $\mathrm{mas\,yr^{-1}}$ & -580.999            & -45.917             & -151.265           & 16.915              & 52.742             & -461.571            & 5.160               \\
\texttt{pmra\_error}                                 & $\mathrm{mas\,yr^{-1}}$ & 0.126               & 0.023               & 0.045              & 0.018               & 0.039              & 0.278               & 0.202               \\
\texttt{pmdec}                                       & $\mathrm{mas\,yr^{-1}}$ & 1.062               & -816.875            & 35.708             & -49.318             & 21.740             & -914.520            & 84.041              \\
\texttt{pmdec\_error}                                & $\mathrm{mas\,yr^{-1}}$ & 0.142               & 0.018               & 0.036              & 0.016               & 0.026              & 0.332               & 0.187               \\
\texttt{parallax}                                    & $\mathrm{mas}$          & 26.20               & 201.33              & 29.06              & 37.25               & 19.32              & 374.49              & 50.93               \\
\texttt{parallax\_error}                             & $\mathrm{mas}$          & 0.11                & 0.02                & 0.04               & 0.02                & 0.03               & 0.23                & 0.15                \\
\texttt{phot\_g\_mean\_mag}                          & $\mathrm{}$             & 7.15                & 7.74                & 7.57               & 6.21                & 7.12               & 8.52                & 3.82                \\
\texttt{bp\_rp}                                      & $\mathrm{}$             & 0.733               & 2.240               & 0.814              & 0.736               & 0.745              & -0.278              & 0.261               \\
\texttt{astrometric\_matched\_transits}              & $\mathrm{}$             & 37                  & 47                  & 41                 & 72                  & 42                 & 22                  & 27                  \\
\texttt{astrometric\_n\_good\_obs\_al}               & $\mathrm{}$             & 327                 & 414                 & 367                & 627                 & 370                & 195                 & 231                 \\
\texttt{astrometric\_params\_solved}                 & $\mathrm{}$             & 95                  & 31                  & 31                 & 31                  & 31                 & 31                  & 95                  \\
\texttt{ipd\_frac\_multi\_peak}                      & $\mathrm{}$             & 0.0                 & 0.0                 & 0.0                & 0.0                 & 0.0                & 18.0                & 0.0                 \\
\texttt{ipd\_gof\_harmonic\_amplitude}               & $\mathrm{}$             & 0.036               & 0.013               & 0.016              & 0.019               & 0.003              & 0.056               & 0.013               \\
\texttt{ipd\_frac\_odd\_win}                         & $\mathrm{}$             & 0.000               & 0.000               & 0.000              & 0.000               & 0.000              & 0.000               & 0.000               \\
\texttt{astrometric\_chi2\_al}                       & $\mathrm{}$             & 15999               & 987                 & 2088               & 2105                & 1726               & 12128               & 66642               \\
\texttt{astrometric\_excess\_noise}                  & $\mathrm{mas}$          & 0.708               & 0.160               & 0.267              & 0.127               & 0.211              & 1.475               & 1.386               \\
\texttt{ruwe}                                        & $\mathrm{}$             & 3.161               & 1.097               & 1.598              & 0.918               & 1.345              & 2.419               & 3.072               \\
UWE factor  $u_0$                                    & $\mathrm{}$             & 2.230               & 1.417               & 1.503              & 2.003               & 1.616              & 3.303               & 5.590               \\
\hline
\multicolumn{8}{l}{\textit{\textsc{Hipparcos}-2 data}} \\ \\
e$_\mathrm{RA} \cos \mathrm{DEC}$                    & $\mathrm{mas}$          & 0.515               & 0.420               & 0.769              & 0.362               & 0.319              & ---                 & 0.063               \\
e$_\mathrm{DEC}$                                     & $\mathrm{mas}$          & 0.540               & 0.600               & 0.490              & 0.290               & 0.280              & ---                 & 0.110               \\
$\sigma_\mathrm{pos}$                                & $\mathrm{mas}$          & 0.746               & 0.732               & 0.912              & 0.464               & 0.425              & ---                 & 0.127               \\
\hline
\multicolumn{8}{l}{\textit{Data from~\citet{Kervella2022}}} \\ \\
pmRAH2EG3b                                           & $\mathrm{mas\,yr^{-1}}$ & -582.576            & -46.046             & -151.162           & 17.125              & 53.354             & ---                 & 4.918               \\
e\_pmRAH2EG3b                                        & $\mathrm{mas\,yr^{-1}}$ & 0.024               & 0.018               & 0.019              & 0.011               & 0.017              & ---                 & 0.006               \\
pmDEH2EG3b                                           & $\mathrm{mas\,yr^{-1}}$ & -0.462              & -816.289            & 35.792             & -49.179             & 21.842             & ---                 & 83.947              \\
e\_pmDEH2EG3b                                        & $\mathrm{mas\,yr^{-1}}$ & 0.025               & 0.020               & 0.023              & 0.010               & 0.011              & ---                 & 0.006               \\
PMaRAH2EG3b                                          & $\mathrm{mas\,yr^{-1}}$ & 1.579               & 0.140               & -0.090             & -0.206              & -0.578             & ---                 & 0.223               \\
e\_PMaRAH2EG3b                                       & $\mathrm{mas\,yr^{-1}}$ & 0.128               & 0.029               & 0.049              & 0.021               & 0.043              & ---                 & 0.202               \\
PMaDEH2EG3b                                          & $\mathrm{mas\,yr^{-1}}$ & 1.498               & -0.547              & -0.121             & -0.152              & -0.099             & ---                 & 0.077               \\
e\_PMaDEH2EG3b                                       & $\mathrm{mas\,yr^{-1}}$ & 0.144               & 0.027               & 0.043              & 0.019               & 0.028              & ---                 & 0.187               \\
$\Vert \mathrm{PMa} \Vert$                           & $\mathrm{mas\,yr^{-1}}$ & 2.177               & 0.565               & 0.151              & 0.256               & 0.586              & ---                 & 0.236               \\
$\sigma_\mathrm{PMa}$                                & $\mathrm{mas\,yr^{-1}}$ & 0.136               & 0.027               & 0.045              & 0.020               & 0.043              & ---                 & 0.200               \\
\hline 
\multicolumn{7}{l}{\textit{\glsxtrshort{g3} noise estimations}} \\ \\
$\sigma_\mathrm{AL}$                                 & $\mathrm{mas}$          & 0.081               & 0.095               & 0.085              & 0.039               & 0.068              & 0.073               & 0.012               \\
$\sigma_\mathrm{att}$                                & $\mathrm{mas}$          & 0.074               & 0.077               & 0.078              & 0.072               & 0.073              & 0.072               & 0.074               \\
$\sigma_\mathrm{calib}$                              & $\mathrm{mas}$          & 0.281               & 0.150               & 0.153              & 0.176               & 0.156              & 0.348               & 1.548               \\
\hline 
\multicolumn{8}{l}{\textit{Modeled distributions for single stars}} \\ \\
AEN$_\mathrm{simu,single}$                           & $\mathrm{mas}$          & 0.250               & 0.119               & 0.120              & 0.154               & 0.127              & 0.295               & 1.388               \\
$\sigma_\mathrm{AEN,simu,single}$                   & $\mathrm{mas}$          & 0.034               & 0.020               & 0.021              & 0.016               & 0.020              & 0.055               & 0.212               \\
RUWE$_\mathrm{simu,single}$                          & $\mathrm{}$             & 1.112               & 0.986               & 0.959              & 1.063               & 1.003              & 0.924               & 3.319               \\
$\sigma_\mathrm{RUWE,simu,single}$                  & $\mathrm{}$             & 0.128               & 0.079               & 0.088              & 0.088               & 0.098              & 0.153               & 0.506               \\
UEVA$_\mathrm{simu,single}$                          & $\mathrm{mas^{2}}$      & 0.076               & 0.030               & 0.028              & 0.031               & 0.027              & 0.102               & 1.993               \\
$\sigma_\mathrm{UEVA,simu,single}$                  & $\mathrm{mas^{2}}$      & 0.018               & 0.005               & 0.005              & 0.005               & 0.005              & 0.033               & 0.601               \\
PMa$_\mathrm{simu,single}$                           & $\mathrm{mas\,yr^{-1}}$ & 0.140               & 0.046               & 0.078              & 0.060               & 0.076              & ---                 & 0.689               \\
$\sigma_\mathrm{PMa,simu,single}$                   & $\mathrm{mas\,yr^{-1}}$ & 0.083               & 0.025               & 0.044              & 0.033               & 0.043              & ---                 & 0.383               \\
\hline 
\multicolumn{8}{l}{\textit{Astrometric signatures}} \\ \\
UEVA$_\mathrm{AEN}$                                  & $\mathrm{mas^{2}}$      & 0.514               & 0.041               & 0.085              & 0.023               & 0.054              & 2.187               & 1.927               \\
UEVA$_\mathrm{RUWE}$                                 & $\mathrm{mas^{2}}$      & 0.605               & 0.036               & 0.077              & 0.023               & 0.047              & 0.668               & 1.654               \\
$\alpha_\mathrm{UEVA,AEN}$                           & $\mathrm{mas}$          & 0.661               & 0.105               & 0.237              & ---                 & 0.167              & 1.444               & ---                 \\
$\alpha_\mathrm{UEVA,RUWE}$                          & $\mathrm{mas}$          & 0.727               & 0.081               & 0.221              & ---                 & 0.143              & 0.753               & ---                 \\
$\alpha_\mathrm{PMa}$                                & $\mathrm{mas\,yr^{-1}}$ & 2.172               & 0.563               & 0.129              & 0.249               & 0.582              & ---                 & ---                 \\
significance AEN                                     & $N-\sigma$              & $>$9                & 2.436               & 7.472              & 0.066               & 4.389              & $>$9                & 0.670               \\
significance RUWE                                    & $N-\sigma$              & $>$9                & 1.726               & 6.767              & 0.059               & 3.491              & 8.210               & 0.399               \\
significance PMa                                     & $N-\sigma$              & $>$9                & $>$9                & 1.513              & 4.285               & 7.035              & ---                 & 0.170         \\     
\hline
\end{tabular}%
}
\end{table*}

\newpage
\twocolumn
\section{The \textit{Gaia} point and line spread function variations}
\label{sec:IPD}
The \glsxtrlong{ipd} (\glsxtrshort{ipd}) of \textit{Gaia} makes a fit by a single \glsxtrlong{psf} (\glsxtrshort{psf}; \glsxtrlong{lsf} or \glsxtrshort{lsf}, if $G$$>$13) of the flux distribution within some defined window around any source transiting the detector. When polluting light is present and not masked out, it periodically affects the location of the measured photocenter compared to the theoretical photocenter at the light-barycenter of the main targeted source. 
The presence of unresolved sources (background or wide-orbit companions) nearby may thus cause undesirable variable shifts of the photocenter of the main source. The amplitude of the shifts is correlated with the angle between the \glsxtrshort{al} scan direction (with position angle $\psi$) and the lines joining the polluting sources and the main source (with position angle $\theta$). 

The \verb+IPD_frac_multi_peak+ indicator published in the \glsxtrshort{g3} archive gives the information on the fraction of exposures for which multiple modes have been detected. The window is then recalculated by masking out the parts containing the identified peaks~\citep{Holl2023}. In theory, since some peaks may not be detected and masked out at all epochs, this can be a source of supplementary spurious variations in the \glsxtrshort{5p}-fit residuals. 
These spurious variations are partly removed by the \textit{Gaia}'s reduction software. It fits a sinusoidal function of the angle of the \glsxtrshort{al} scan direction $\psi$ (a.k.a subpixel phase;~\citealt{Lindegren2021}) to the astrometric time series of the targeted source ($c+a\sin\psi + b\cos\psi$; Eq. 9 in~\citealt{Lindegren2021}). 
The shift function is not a perfect sinusoid due to the non-axisymetric shape of the \glsxtrshort{psf} on the detector~\citep{Holl2023}. Thus, a residual \glsxtrshort{al} angle $\Delta\eta$ remains that may vary through time and mimick the signature of true orbital astrometric motion.
The \glsxtrshort{ipd} produced another important indicator of \glsxtrshort{psf} distortion, the \verb+IPD_gof_harmonic_amplitude+. It measures the amplitude of the sinusoidal variation of the goodness-of-fit (GOF=$\ln \chi^2_{\rm reduced}$) of the \glsxtrshort{psf} (\glsxtrshort{lsf}) fit.
\citet{Fabricius2021} showed that sources with a \glsxtrshort{ruwe} indicative of possible binarity, and \verb+IPD_frac_multi_peak+$>$2 or the \verb+IPD_gof_harmonic_amplitude+$>$0.1, must be considered as resolved doubles.  

\begin{figure}
    \centering
    \includegraphics[width=60mm,clip=true,trim=0 0 0 10]{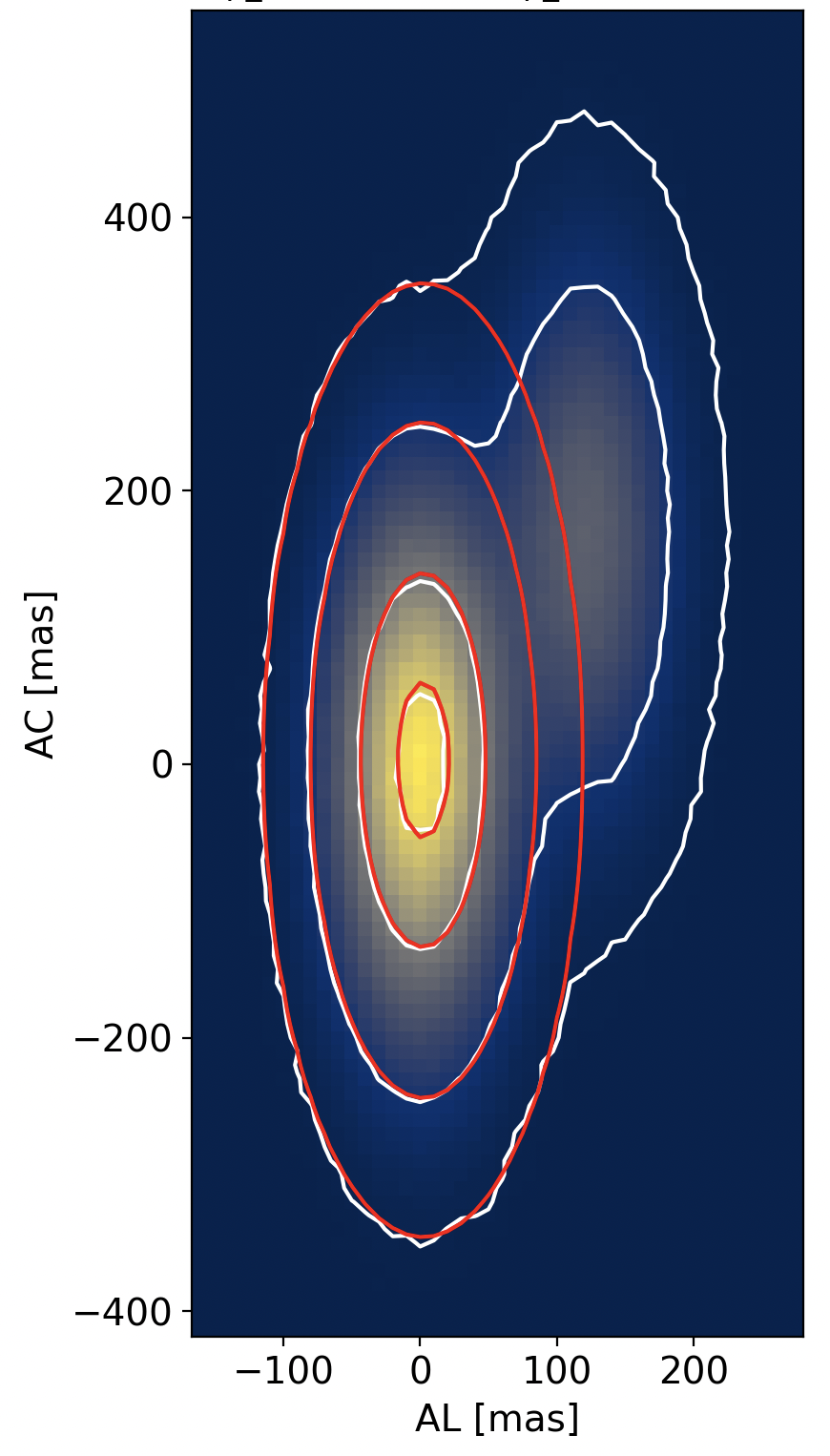}
    \caption{Simulation of the \glsxtrshort{ipd} fit of a 2D Gaussian \glsxtrshort{psf} on a wide-orbit binary with separation of $\sim$200\,mas and a flux ratio of 0.42.}
    \label{fig:IPD_simu}
\end{figure}

\begin{figure}
    \centering
    \includegraphics[width=85mm]{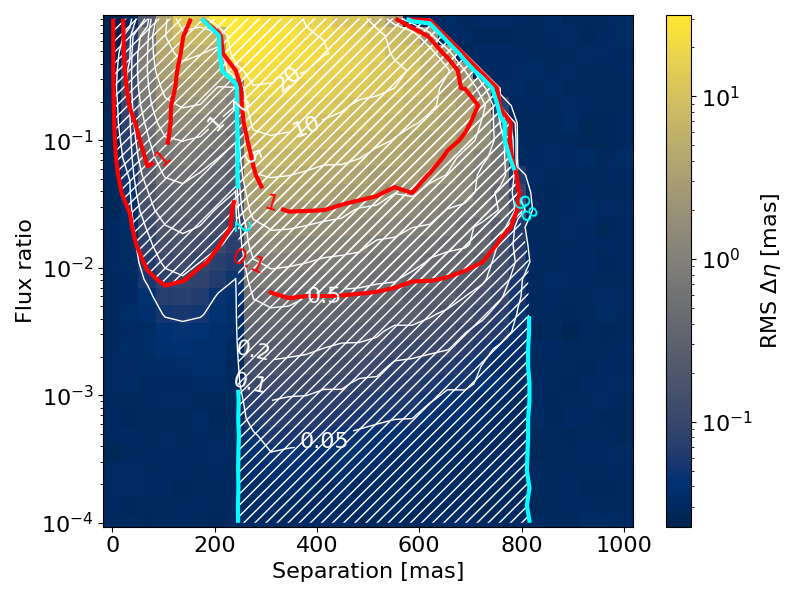}
    \caption{Photocenter centroid variations \glsxtrshort{rse} with respect to binary separation and flux ratio delineated with white contours. The red contours delineate the GOF amplitude (at levels 0.1 and 1) and the cyan contours delineate the 2 and 98\% levels of the \texttt{IPD\_frac\_multi\_peak}. The white hatched region shows, as discussed in the text, where the \glsxtrshort{rms} of the $\Delta \eta$ residuals could be larger than 0.1\,mas, while outside this region it is always $<$0.1\,mas.}
    \label{fig:IPD_binary}
\end{figure}

We modeled the observation by \textit{Gaia}, through time, of two Gaussian \glsxtrshort{psf} of a wide-orbit binary, that is, with virtually no orbital motion during \textit{Gaia}'s 3-yr monitoring. We assumed separations ranging from 0 to 1000\,mas and flux ratio of $0.001$--$1$. An example is shown in Fig.~\ref{fig:IPD_simu}. We fixed the CCD noise to 1\% of the flux, corresponding to \glsxtrshort{al} measurement uncertainty of~$\sim$0.03\,mas. We arbitrarily fixed the flux at the tip of the \glsxtrshort{psf} of the main source to $1$. We fit the double star \glsxtrshort{psf} by a single 2D-Gaussian varying only the scale and the centroid locus. A fit result is also shown in red in Fig.~\ref{fig:IPD_simu}. We obtained series of GOF and photocenter \glsxtrshort{al} angle that we fit by a sinusoidal function following the \glsxtrshort{al} scan rotation law, leading to GOF variation amplitude and residual $\Delta \eta$ beyond the five-parameter model. Figure~\ref{fig:IPD_binary} shows the \glsxtrlong{rms} (\glsxtrshort{rms}) of $\Delta \eta$ with respect to separation and flux ratio, as well as some isocontours of \verb+IPD_frac_multi_peak+ (at 2 and 98\%) and \verb+IPD_gof_harmonic_amplitude+ (at 0.1 and 1\%). 

We found that significant residuals are associated with \verb+IPD_frac_multi_peak+ being different than 0 or 100\% or with \verb+IPD_gof_harmonic_amplitude+ being larger than 0.1, while, insignificant variations on the order of the \glsxtrshort{al} measurement uncertainty were mostly found when \verb+IPD_frac_multi_peak+ is close to 0 or 100\% and \verb+IPD_gof_harmonic_amplitude+$<$0.1. This agrees well with \citet{Fabricius2021} conclusions, but shows also that for fully resolved binary the \glsxtrshort{ipd} \glsxtrshort{psf} fit remains unperturbed. As long as \verb+IPD_gof_harmonic_amplitude+$<$0.1 and \verb+IPD_frac_multi_peak+ $<$2 or $>$98\%, the astrometric signature $\alpha_{\rm \glsxtrshort{mse}}$ is only lightly affected by the binary astrometric bias and correctly measures supplementary astrometric motion. However, for the sources that do not comply to these conditions, the interpretation of $\alpha_{\rm \glsxtrshort{mse}}$, as indicative of orbital motion of the photocenter, is hazardous.

\newpage
\FloatBarrier
\section{Detailed calculation of the approximate probability density function followed by \texorpdfstring{$\chi^2_{\rm astro}$}{the astrometric chi-square} }
\label{sec:details_chi2}

We show here that for a single star, the $\chi^2_{\rm astro}$ that can be found in the \glsxtrshort{g3} archive does not actually follow a $\chi^2$ distribution with $N-5$ degree of freedom. Instead we show that it more accurately follows a linear combination of normal distributions. The $\chi^2_{\rm astro}$ is expressed with respect to the residuals of the five-parameter astrometric fit: 
\begin{align}\label{eq:chi2_astro}
    \chi^2_{\rm astro} = \sum_\ell^N\frac{R_\ell^2}{\sigma_\ell^2},
\end{align}
where as assumed in the rest of this paper, the formal error $\sigma_\ell\equiv\sigma_f$ is approximately constant throughout all data points. The formal error is the quadratic sum $\sigma_f=\sqrt{\sigma_{\rm AL}^2+\sigma_{\rm att}^2}$~\citep{Lindegren2012} with $\sigma_{\rm att}$ the attitude excess noise (see Appendix~\ref{sec:attitude_error}) and $\sigma_{\rm AL}$ the along-scan measurement error (see Appendix~\ref{sec:AL_error}). The formal error is explicitly calculated for each \glsxtrshort{g3} source in Appendix~\ref{sec:formal_error}. One residual $R_\ell$ is the sum of two contributions $r_i$ and $\xi^{(i)}_j$, with $r_i$ that varies randomly from one \glsxtrshort{fov} transit ($i$) to another with normal distribution ${\mathcal N}(0,\sigma_{\rm calib})$, and $\xi^{(i)}_j$ that varies for a given \glsxtrshort{fov} transit ($i$) from one \glsxtrshort{al} measurement ($j$) to another with normal distribution ${\mathcal N}(0,\sigma_{\rm AL}^2)$. It follows that
\begin{align}
\sum_\ell^N R_\ell^2 &\approx N_{\glsxtrshort{al}} \sum_i^{N_{\glsxtrshort{fov}}} r_i^2 + \sum^{N}_\ell \xi_\ell^2 + 2\,\sum_i^{N_{\glsxtrshort{fov}}} \left( r_i \sum_j^{N_{\glsxtrshort{al}}} \xi^{(i)}_j \right).
\end{align}

\noindent
We recall that $N_{\rm \glsxtrshort{fov}}$ is the number of \glsxtrshort{fov} transits on the detector, and $N_{\glsxtrshort{al}}$ is the average number of \glsxtrshort{al} angle measurements per transit, that is, $N_{\glsxtrshort{al}}={\rm int}(N/N_{\rm \glsxtrshort{fov}})$, given $N$ is the \verb+astrometric_n_good_obs_AL+. And thus $\chi^2_{\rm astro}$ is the combination of three terms:
\begin{align}\label{eq:chi2_normal}
\chi^2_{\rm astro} &\approx  N_{\glsxtrshort{al}} \frac{\sigma^2_{\rm calib}}{\sigma_f^2} \,X + \frac{\sigma^2_{\rm AL}}{\sigma_f^2} \,Y + \frac{2}{\sigma_f^2}\,Z,
\end{align}
with
\begin{align}
    X&\sim \chi^2(N_{\rm \glsxtrshort{fov}}-5)\sim {\mathcal N}\left(N_{\rm \glsxtrshort{fov}}-5,\sqrt{2(N_{\rm \glsxtrshort{fov}}-5)}\right) \\ 
    Y&\sim \chi^2(N)\sim {\mathcal N}\left(N,\sqrt{2N}\right) \\ 
    Z&\sim {\mathcal N}\left(0,\sqrt{N}\,\sigma_{\rm AL}\,\sigma_{\rm calib} \right)  \label{eq:Z}.
\end{align} 

\noindent
The last term $Z$ is obtained as the sum of the product of two normally distributed variables:
\begin{align}
    Z &= \sum_i^{N_{\rm \glsxtrshort{fov}}}\sum_j^{N_{\glsxtrshort{al}}} r_i \xi^{(i)}_j  \approx \sum_i^{N_{\rm \glsxtrshort{fov}}} r_i W_i \nonumber \\ & {\rm with} \quad W_i = {\mathcal N}\left(0,\sqrt{N_{\glsxtrshort{al}}}\,\sigma_{\rm AL}\right) \quad {\rm and} \quad
    r_i = {\mathcal N}\left(0,\sigma_{\rm calib}\right).
\end{align}

\noindent
The standard deviation of the product of two normally distributed variables centered on zero is the product of their standard deviation, and the standard deviation of the sum of normally distributed variables is the root sum square of their standard deviation. This leads us to the formula of $Z$ expressed in Eq.~\ref{eq:Z}, given moreover that the total number of astrometric points is $N\approx N_{\glsxtrshort{al}}N_{\rm \glsxtrshort{fov}}$. The two first terms $X$ and $Y$ dominate the spread, thus $\chi^2_{\rm astro}$ should mainly be distributed according to a skewed $\chi^2$ law.
Thus, $\chi^2_{\rm astro}$ follows a quasi-normal distribution ${\mathcal N}\left(\mu,\sigma \right)$ with 
\begin{align}
    \mu =& \frac{N_{\glsxtrshort{al}}}{\sigma^2_{\rm att}+\sigma_{\rm AL}^2}\,\left[(N_{\rm \glsxtrshort{fov}}-5) \, \sigma_{\rm calib}^2 + N_{\rm \glsxtrshort{fov}}\,\sigma^2_{\rm AL}\right] 
    \label{eq:chi2_mean}\\ 
    \sigma^2=&  \frac{2 N_{\glsxtrshort{al}}}{\left(\sigma^2_{\rm att}+\sigma_{\rm AL}^2\right)^2}\,\Bigg[ N_{\glsxtrshort{al}} \left(N_{\rm \glsxtrshort{fov}}-5\right)\,\sigma_{\rm calib}^4 \nonumber \\ & \qquad\qquad\qquad +  N_{\rm \glsxtrshort{fov}}\,\sigma^4_{\rm AL} 
    \nonumber \\ & \qquad\qquad\qquad + 2\,N_{\rm \glsxtrshort{fov}} \,\sigma_{\rm AL}^2\,\sigma_{\rm calib}^2\Bigg] \label{eq:chi2_sigma}.
\end{align}

\begin{figure}[hbt]
    \centering
    \includegraphics[width=89.3mm,clip=true]{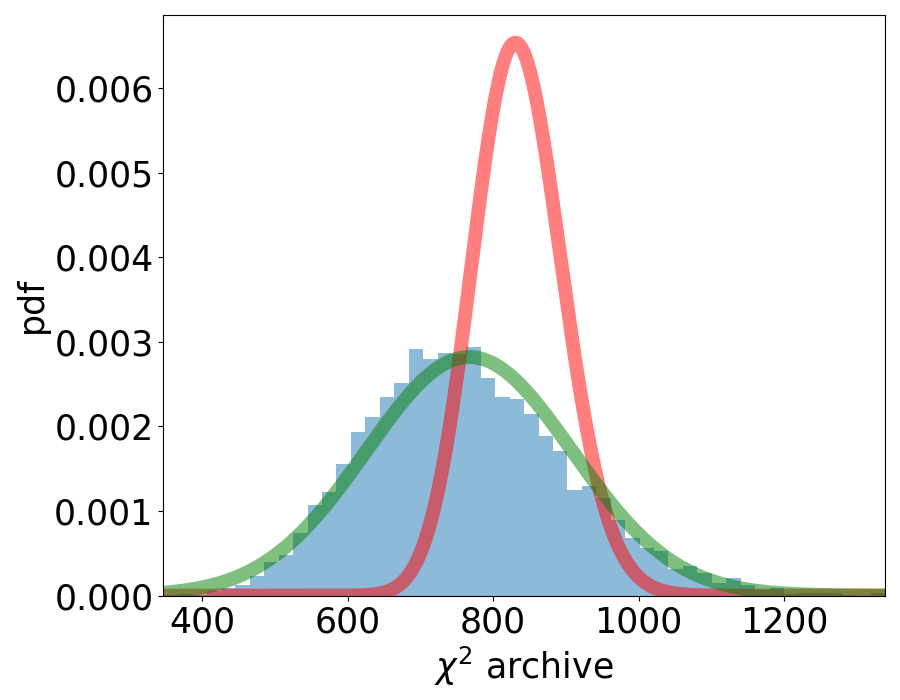}
    \caption{Distribution of $\chi^2_{\rm astro}$ calculated from 10,000 simulations of GJ\,832 astrometric data as a single star. They are compared to a theoretical $\chi^2$ probability density function with respectively $N-5$ degrees of freedom (red line) and to the normal distribution expressed in  Eqs.~\ref{eq:chi2_mean} and~\ref{eq:chi2_sigma} (green line).}
    \label{fig:compared_chi2}
\end{figure}

\noindent
We compared the simple $\chi^2$ distribution with $N-5$ degrees of freedom and this non-trivial normal model to the distribution of 10,000 $\chi^2_{\rm astro}$ values modeled for a single source from the \glsxtrshort{5p} dataset in Fig.~\ref{fig:compared_chi2}. Models are obtained by simulation, as explained in Sect.~\ref{sec:simu} by generating noisy astrometric measurements of GJ\,832 by \textit{Gaia}, assuming zero orbital motion, and fitting-out a five-parameter model, leading to residuals $R_\ell$ and a $\chi^2$ as defined in Eq.~\ref{eq:chi2_astro}.
We indeed found that the non-trivial normal distribution is a more accurate model of the true probability law that is followed by $\chi^2_{\rm astro}$, than a na\"ive $\chi^2$--law.

\end{appendix}

\end{document}